\documentclass[fleqn,usenatbib]{mnras}

\usepackage[T1]{fontenc}
\usepackage{ae,aecompl}

\usepackage[normalem]{ulem}
\usepackage{times}
\usepackage[fleqn]{amsmath}
\usepackage[pdftex]{graphicx}
\usepackage{lscape}
\usepackage{indentfirst}
\usepackage{enumitem}
\usepackage{xspace}

\usepackage{amssymb}
\usepackage{bm}
\usepackage[dvipsnames]{xcolor}
\usepackage{url}
\usepackage[flushleft]{threeparttable}
\usepackage[normalem]{ulem}

\newcommand{\alphaf}{\alpha_\mathrm{f}}
\newcommand{\beq}{\begin{equation}}
\newcommand{\dd}{\mathrm{d}}
\newcommand{\ee}{\mathrm{e}}
\newcommand{\eeq}{\end{equation}}
\newcommand{\ii}{\mathrm{i}}
\newcommand{\kB}{k_\mathrm{B}}
\newcommand{\mel}{m_\mathrm{e}}
\newcommand{\mion}{m_\mathrm{i}}
\newcommand{\nel}{n_\mathrm{e}}
\newcommand{\omB}[1]{\omega_{B\mathrm{,#1}}}
\newcommand{\ompe}{\omega_{\mathrm{pe}}}
\newcommand{\ompi}{\omega_{\mathrm{pi}}}

\newcommand{\req}[1]{equation~(\ref{#1})}

\date{\textbf{CORRECTED} version of: MNRAS \textbf{551} (3), stag1517
(2026)}

\title[Vacuum polarization in subcritical X-ray pulsars]
{Vacuum polarization and cyclotron resonance effects on radiative transfer and plasma deceleration in subcritical X-ray pulsars}
\author[I.D.~Markozov et al.] 
{
Ivan D. Markozov$^{1}$,
Alexander Y. Potekhin$^{1}$,
Alexander D. Kaminker$^{1}$,
Alexander A. Mushtukov$^{2,3}$\thanks{E-mail: 
al.mushtukov@gmail.com}
\\ 
$^1$ Ioffe Institute, Politekhnicheskaya 26, St Petersburg 194021, Russia\\
$^2$ Mullard Space Science Laboratory, University College London, Holmbury St. Mary, Surrey RH5 6NT, UK\\
$^3$ Astrophysics, Department of Physics, University of Oxford, Denys Wilkinson Building, 
Keble Road, Oxford OX1 3RH, UK \\
} 

\pubyear{2026}

\begin{document}
\label{firstpage}
\pagerange{\pageref{firstpage}--\pageref{lastpage}}
\maketitle

%%%%%%%%%%%%%%%%%%%%%%%%%%%%%%%%%%%%%%%%%%%

\begin{abstract} 
We investigate the spectrum and polarization of radiation emerging from a subcritical X-ray pulsar using 
self-consistent radiation-hydrodynamic simulations of an accretion channel in a strong magnetic field. 
The polarized radiative transfer in the channel above the hot spot is simulated for the two normal modes, 
taking into account resonant Compton scattering in a strongly magnetized plasma and the effects of vacuum polarization.
We show that the deceleration of the accreting matter in the subcritical regime is mainly governed by resonant scattering. 
Our simulations provide the velocity profiles of the plasma flow
and demonstrate that vacuum polarization dominates over plasma birefringence, 
enhancing both the cyclotron spectral feature and the radiative deceleration of the plasma. 
We also find that the
energy of the cyclotron
feature 
increases with accretion luminosity, indicating a positive correlation consistent 
with previous observational results and theoretical interpretation.
\end{abstract}


%%%%%%%%%%%%%%%%%%%%%%%%%%%%%%%%%%%%%%%%%%%
\section{Introduction}
\label{sec:Intro}
%%%%%%%%%%%%%%%%%%%%%%%%%%%%%%%%%%%%%%%%%%%%%%%%%
X-ray pulsars (XRPs) are strongly magnetized accreting neutron stars
(NSs) in close binary systems \citep[see][for
review]{2022arXiv220414185M}. A typical field strength at the surface
of an XRP is $B\sim10^{12}-10^{13}$~G. In such systems, matter from the
companion star falls onto the NS surface through an accretion channel
along the field lines around magnetic poles. The kinetic energy of the
accreting matter that hits the NS surface is mostly converted into
X-ray radiation. Because of a misalignment of the magnetic and rotation
axes of an NS, a remote observer detects a pulsating signal. 

X-ray luminosities $L_\mathrm{X}$ of the XRPs range over several orders of magnitude. 
XRPs powering some ultra-luminous X-ray sources can reach luminosities of 
$\gtrsim 10^{41}$ erg~s$^{-1}$ (see, e.g., \citealt{2017Sci...355..817I}).
At the other extreme, some XRPs are observed at $L_\mathrm{X}\sim10^{33}$ erg~s$^{-1}$, when accreting NSs
enter the propeller regime \citep{1975A&A....39..185I} -- emitting X-rays from cooling polar caps \citep{2016A&A...593A..16T,2025A&A...702A.216X} -- or, alternatively, when they sustain low-level accretion from a cold recombined disc \citep{2017A&A...608A..17T,2019A&A...621A.134T}
or from a hot quasi-static envelope around the NS magnetosphere
\citep{Shakura_12,Shakura_15}.

The XRP spectra usually resemble a power-law continuum with quasi-exponential cutoff at high energies
\citep[e.g.,][and references therein]{1992herm.book.....M,2022arXiv220414185M}. 
Spectra of many XRPs reveal cyclotron absorption lines superposed on the continuum 
(sometimes with one or a few harmonics), whose positions can vary with
varying $L_\mathrm{X}$ \citep[][and references
therein]{2019A&A...622A..61S}. 
Recently, due to the Imaging X-ray Polarimetry Explorer (IXPE, \citealt{2022JATIS...8b6002W}), 
it has become also possible to measure the polarization of the X-ray radiation of the XRPs 
in the band of 2--8 keV \citep{2022NatAs...6.1433D,2022ApJ...941L..14T}. 
The linear polarization degree of the XRPs turned out to be within 20\% (see \citealt{2024Galax..12...46P} for review), 
much below the expectations based on the previous theoretical studies
of X-ray polarization in hot magnetized plasmas
\citep*[e.g.,][]{1983Ap&SS..91..167K,1985ApJ...298..147M,1985ApJ...299..138M}, NS atmospheres \citep{Taverna_20}
and XRP accretion columns 
\citep{2021MNRAS.501..109C,2021MNRAS.501..129C}, although compatible
with the predictions based on the model of a condensed NS surface
(\citealt*{PotekhinHC16}; \citealt{Taverna_20}).

A variety of formation mechanisms and geometries of accretion flows
have been discussed in the literature \citep[e.g.,][and references
therein]{2022arXiv220414185M}. The higher the mass accretion rate
$\dot{M}$, the higher the luminosity. There is a critical value of
luminosity, at which  
radiation pressure becomes {sufficient to stop the plasma flow}
above the NS surface \citep{1973NPhS..246....1D}. 
XRPs with accretion rates above this value are called supercritical, while XRPs with lower accretion
rates are called subcritical. 
In the supercritical XRPs, a radiation-dominated shock in the accretion channel is expected
\citep{1976MNRAS.175..395B}. 
Below the shock front, there is a slow moving dense hot plasma zone that is called accretion column, 
where the observed radiation of the supercritical XRPs originates. 

The mechanism of the radiation production of the subcritical XRPs is
less certain. At relatively low accretion rates  $\dot{M} \la
(10^{15}-10^{16})$~g s$^{-1}$ two fundamentally different accretion regimes have been discussed.
One of them
is determined by Coulomb braking of the accreting plasma
in a relatively thin layer near the NS surface
\citep[e.g.,][and references therein]{2021MNRAS.503.5193M,2021A&A...651A..12S}.
This process
leads to heating of the deceleration layer and to the formation of hot
atmosphere regions around the magnetic poles, from which the observed 
X-ray radiation emerges. In the other regime
\citep[e.g.,][]{1975ApJ...198..671S,1982ApJ...257..733L,2004AstL...30..309B}
a collisionless shock wave appears, which plays the crucial role in
the accretion channel structure and radiation formation. In the present
paper we will focus on the first regime. So far it has been assumed that
the radiation that is formed in the atmosphere around the magnetic poles
freely escapes to infinity. In the present paper we study
the influence of radiative transfer in the accretion channel above the
hot spot on the spectra and polarization of the outgoing radiation of
the subcritical XRPs in the shockless regime.

The strong magnetic fields affect the formation of XRP radiation due to
a number of physical effects. First, a strongly magnetized plasma is a
birefringent medium \citep[e.g.,][]{Ginzburg70}. Under the conditions
established by \citet{GnedinPavlovModes}, electromagnetic waves can be
represented in terms of two normal modes (NMs) with separate
refractive indices and absorption coefficients.

Second,  strong magnetic fields change the wave functions of charged
particles, their motion transverse to the field  being quantized in the
Landau levels \citep[see, e.g.,][]{SokolovTernov}, which strongly
affects interaction of matter with radiation. Various radiation
processes in strong magnetic fields under different conditions were
studied by many authors \citep[see, e.g.,][for review and
references]{2006RPPh...69.2631H}. Under the conditions characteristic
of the XRPs, considered in the present work, the most important of these
processes is Compton scattering, which was treated in detail by
\citet{1996PhDT.......100S} and \citet*{2016PhRvD..93j5003M}.    

Third, formation of the XRP radiation can be strongly affected by vacuum
polarization, which is a specific effect of quantum electrodynamics
(QED) arising from interactions of virtual electron-positron pairs with
a magnetic field \citep[see, e.g.,][and references
therein]{PavlovGnedin84,HeylCaiazzo18}. This effect is important at
sufficiently strong magnetic fields and low plasma densities, where the
virtual pairs compete with real plasma particles in their influence on
the electromagnetic waves propagation.

A theoretical modelling of an XRP can be divided in two main stages. 
At the first stage, the hydrodynamic structure of the accretion channel
is determined. At the second stage, XRP radiation properties are
calculated.

The simulations of an accretion column structure were mostly performed 
for supercritical pulsars with a radiation-dominated shock wave. Both
stationary
(\citealt{1973NPhS..246....1D,1976MNRAS.175..395B,1975PASJ...27..311I,
1981A&A....93..255W,1984ApJ...278..349B,1984PASA....5..446K,1985A&A...142..430K,
1975ApJ...198..671S,1982ApJ...257..733L,1998ApJ...498..790B,
2015MNRAS.454.2539M,2015MNRAS.452.1601P};
\citealt*{2017ApJ...835..129W,2021MNRAS.501..564G}) and non-stationary
(\citealt{1989ESASP.296...89K,2004AstL...30..309B,2016PASJ...68...83K,2023MNRAS.524.4148A};
\citealt*{2022MNRAS.515.4371Z,2023MNRAS.520.1421Z,ZhangBW25};
\citealt{2023MNRAS.524.2431S}) regimes were explored, using
(semi)analytical and numerical models. Most of the works did not take
into account the effect of the resonant scattering of photons on the
deceleration of accreting plasma. \citet{2015MNRAS.454.2539M}
calculated the structure of an accretion column using fully
relativistic resonant cross sections, but neglecting plasma motion. 
\citet{2023MNRAS.524.2431S} and \citet{ZhangBW25} performed full
radiation-hydrodynamical calculations of an XRP accretion channel 
using an approximate treatment of the influence of a magnetic field on
the opacity in the case of supercritical pulsars. 

Hydrodynamical simulations of the subcritical accretion channel 
were carried out mainly in connection 
with studies of possible 
collisionless shock waves in such channels
\citep{1975ApJ...198..671S,1982ApJ...257..733L,
1984ApJ...278..349B,
2004AstL...30..309B}.
In particular, \citet{2004AstL...30..309B} 
considered in detail a number 
of resonant processes and non-stationary
 plasma evolution. 
However, their model did not include the recoil effect in the Compton scattering, 
which plays an important role 
in formation of accretion channel structure and 
outgoing spectrum \citep{2007ApJ...654..435B}. 
In a previous work \citep*[][hereafter Paper~I]{2023AstL...49..583M} 
we constructed a model of an accretion channel 
{taking resonant scattering into account}
in a strong magnetic field and 
calculated polarized radiation transfer. 
In Paper~I we used the electron plasma approximation, which did not 
take into account the damping, ion cyclotron motion and vacuum polarization 
effects in the dielectric tensor. 
Besides, we considered 
only the case of a completely filled accretion 
channel.
A
{disc accretion scenario}
allows formation of a hollow channel, 
only partially filled with plasma.  

Other theoretical studies of the XRPs rely on
radiation transfer simulations under 
predetermined physical conditions 
in the radiation subsystem. 
There are two prevailing techniques 
of solving the radiation transfer equations: 
the Feautrier method 
\citep[e.g.,][]{Mihalas78,1992herm.book.....M} 
and Monte Carlo method 
\citep[e.g.,][]{Marchuk_80,2019LRCA....5....1N}. 
{\citet{Nagel81a,Nagel81b} 
was the first to employ the Feautrier technique to study the XRP spectra, 
beaming and cyclotron line formation with Comptonization.}
\citet{2008ApJ...672.1127N,2015ApJ...807..164N} 
employed
{this}
technique to show that taking into account 
the dependence of the magnetic field, 
temperature and density on the height of  the 
accretion channel leads to broad and 
shallow line profiles, similar to those 
observed in the real XRP spectra. 
The Monte Carlo technique 
{has been used to model XRP spectra since 1970s \citep[e.g.,][]{Yahel79}.
This technique}
makes it possible 
to include the complex 
microphysics related to processes 
in a strong magnetic field. 
\citet{1999ApJ...517..334A} 
used the latter technique to include 
fully relativistic 
Compton scattering cross sections 
with a few cyclotron resonances. 
They found that a photon multiplication 
associated with relaxation
of excited electron states
makes the cyclotron 
lines broader and shallower even 
in the uniform medium. 
They also showed that the first 
cyclotron harmonic can be deeper 
than the fundamental one 
at certain angles of the line of sight. 
\citet{2007A&A...472..353S} and 
\citet{2017A&A...597A...3S,2017A&A...601A..99S} used 
the model by \citet{1999ApJ...517..334A} in Monte Carlo simulations 
of radiation in the accreting plasma.
\citet*{2022MNRAS.515..914K} 
included up to ten resonances 
and the general relativity (GR) effects to calculate 
observed pulsars spectra. 
\cite{Garasev_11,Garasev_16} studied the effects of 
frequency redistribution of the cyclotron resonant photons in the 
atmospheres of magnetized  compact stars and the impact of these
effects on the  spectral line shape and 
on dynamics of resonant photon transfer.
Recently, \citet{Fotiadis_26} simulated radiative transfer in an accretion channel of subcritical XRPs, 
using the polarization-averaged resonant Compton scattering cross section in a strong magnetic field, 
and obtained spectra of outgoing radiation with a pronounced cyclotron absorption line, 
which depends on the accretion luminosity and angles of observation. 
In particular, they demonstrated the increase of the centroid photon energy of this line 
with increasing the luminosity in agreement with observations.

Most of the above-mentioned works treated a hydrodynamical structure 
of the radiating region as predetermined, without consideration
{of}
its consistency with a
{radiative}
transfer. Coupling of the plasma
magnetohydrodynamics with 
{the}
radiative transfer has been included in
numerical simulations of accretion columns of supercritical XRPs in
several recent papers
\citep{2017ApJ...835..129W,2017ApJ...835..130W,2022MNRAS.515.4371Z, 
2023MNRAS.520.1421Z,ZhangBW25,2023MNRAS.524.2431S}.
{Two of them}
 \citep{2023MNRAS.524.2431S,ZhangBW25} have approximately
taken into account a resonant scattering in the strong magnetic field.

In this paper we present a self-consistent calculation of the
hydrodynamical structure of the accretion channels of the subcritical
XRPs together with their radiation, going beyond the electron plasma approximation 
by taking into account the ion cyclotron resonance and damping effects on the dielectric tensor, 
as well as 
the vacuum-polarization effects, which 
{constitute} the main advance {over} Paper~I.
The importance of vacuum polarization in the
simulations of subcritical pulsar radiation was shown in the recent
paper by \citet{2023A&A...674L...2S}, which was focused on the hot
polar cap radiation. Here we demonstrate that radiative transfer in the
accretion channel with allowance for the vacuum polarization
{has}
a substantial influence on the radiation of the subcritical XRPs. 
In addition, we consider not only the filled accretion channel, as in Paper~I, 
but also the case of a hollow accretion channel, which is considered 
as more realistic in the case of a disc accretion 
\citep[e.g.,][]{1976MNRAS.175..395B,2022arXiv220414185M}.

The paper is organized as follows{. I}n Section~\ref{sec:physics} 
we introduce definitions of the main physical quantities 
and their relationships
used in further calculations. In particular, we describe
polarization characteristics of the NMs in the
strongly magnetized plasma 
(Section~\ref{sec:rad}),
a set of cross-sections for 
 Compton scattering of the polarized photons
(Section~\ref{sec:compton_scat}),
 the general system of hydrodynamic  
and radiation transfer equations 
(Section~\ref{sec:rad_hydro}).
In Section~\ref{sec:num_mod} we describe
the employed numerical schemes:
the basic relationships used at the hydrodynamic substep
(Section~\ref{sec:hydro_rad}),
the algorithm of calculations at the radiation substep 
(Section~\ref{sec:rad_transfer}),
the energy-momentum exchange
(Section~\ref{sec:en_moment}) and boundary conditions
(Section~\ref{sec:bounds}).
In Section~\ref{sec:num_res} we discuss the results
of  numerical calculations, including 
the velocity distributions of  plasma flow along the accretion channel
(Section~\ref{sec:plasma_dec}),
spectra of the outgoing radiation (Section~\ref{sec:spectra}),
{dependence of the cyclotron feature on the accretion rate (Section~\ref{sec:cyclo})}
and {linear}
polarization {degree} (Section~\ref{sec:pol}).
The summary and discussion are given in Section~\ref{sec:summary}.
In Appendix~\ref{app:scattering} we give explicit expressions for the
resonant cross sections in NMs.
Appendix~\ref{app:code_verif} presents examples of  the code verification.

%%%%%%%%%%%%%%%%%%%%%%%%%%%%%%%%%%%%%%%%
\section{Physics input}
\label{sec:physics}
%%%%%%%%%%%%%%%%%%%%%%%%%%%%%%%%%%%%%%%%

\subsection{Radiation in a magnetized plasma}
\label{sec:rad} 

Propagation of electromagnetic waves in magnetized plasmas has been
comprehensively described in the monograph by \citet{Ginzburg70}. 
The waves propagate in the form of
two NMs at 
photon
circular frequencies $\omega$, which lie sufficiently far from resonances and 
are much larger than the electron plasma
frequency $\ompe=\left({4\pi e^2 \nel / \mel} \right)^{1/2}$, 
where $\mel$ is the electron mass, $\nel$ is the electron number
density and $e$ is the elementary charge.
The NM whose electric vector oscillates predominantly in
{directions close to}
the $\bm{k}-\bm{B}$ plane, where $\bm{k}$ is the wave vector, is called ordinary (O-mode), and the NM whose
electric vector oscillates predominantly in the perpendicular plane is
called extraordinary (X-mode). They have different polarization vectors
$\bm{e}_j$ and different absorption and scattering cross sections, which
depend on the angle $\theta_B$ between $\bm{k}$ and $\bm{B}$. The two modes interact with one another
through scattering. \citet{Ventura79} performed an analysis of the
polarization modes in application to the NSs
from the physics
point of view. \cite{GnedinPavlovModes} formulated the radiative
transfer problem in terms of these modes and specified the conditions of
their applicability. They showed that in the conditions typical for strongly
magnetized NSs, except narrow frequency ranges near resonances, a strong
Faraday depolarization occurs, which allows one to consider specific intensities of the two
NMs instead of the four components of the Stokes vector.

\subsubsection{Dielectric tensor}
\label{sec:RTEmag}

It is convenient to describe a monochromatic electromagnetic 
wave in the complex vector representation, in which the electric field
vector of a plane electromagnetic wave is written as
$\bm{\mathcal{E}}(\omega)\ee^{\ii\bm{k}\cdot\bm{r}-\ii\omega t}$, 
where $\bm{r}$ is the radius vector and $t$ is the time. In the following we will omit the
common factor $\ee^{\ii\bm{k}\cdot\bm{r}-\ii\omega t}$ for brevity.
In this representation and in the Cartesian coordinate system $(xyz)$
with the $z$-axis along $\bm{B}$, the dielectric tensor of a plasma has
the form \citep[e.g.,][]{Ginzburg70}
\beq
 \bm{\varepsilon} =
 \left( \begin{array}{ccc}
 \varepsilon_\perp & \ii \varepsilon_\wedge & 0 \\
 -\ii\varepsilon_\wedge & \varepsilon_\perp & 0 \\
 0 & 0 & \varepsilon_\| 
 \end{array} \right),
\label{eps-p}
\eeq
where\footnote{\label{i-omega-t}Here we have taken into account that $\ompi\ll\ompe$.
The
plus sign at the damping factor in the denominator corresponds 
to the chosen form $\ee^{\ii\bm{k}\cdot\bm{r}-\ii\omega t}$ of a monochromatic wave; it is opposite to
the sign in \citet{Ginzburg70} according to his convention $\ee^{\ii\omega t-\ii\bm{k}\cdot\bm{r}}$.} 
\beq
    \varepsilon_\| \approx 1 - \frac{\ompe^2}{\omega^2 + \ii \omega
  \nu_\|(\omega)}
\label{eps_long}
\eeq
and $\nu_\|(\omega)$ is an effective damping frequency.
Here and hereinafter we use the cold plasma approximation (that is, 
neglect thermal motion effects on the dielectric tensor) and assume that damping
frequencies are small compared to $\omega$.
For the other components of $\bm\varepsilon$, we will use
the approximation suggested by \citet{2003ApJ...588..962L}:
\beq
  \varepsilon_\perp \pm \varepsilon_\wedge
 \approx
  1 - \frac{\ompe^2(1+\ii \nu_\mathrm{ri}/\omega)
  + \ompi^2(1+\ii \nu_\mathrm{re}/\omega)
  }{
  (\omega \pm \omB{e} + \ii \nu_\mathrm{re})
  (\omega \mp \omB{i} + \ii \nu_\mathrm{ri})
   + \ii \omega\nu_\mathrm{ei}^\perp(\omega)
   },
\label{eps_perp}
\eeq
where $\ompe$ is the electron plasma frequency defined above,
$\ompi=\left({4\pi Z e^2 \nel / \mion} \right)^{1/2}$ is the ion plasma
frequency, $\mion$ is the ion mass, $Z$ is its effective
charge number,
\beq
  \omB{e} = \frac{eB}{\mel c},
\quad
  \omB{i} = \frac{ZeB}{\mion c}
\eeq
are the electron and ion cyclotron frequencies,
\beq
  \nu_\mathrm{re} = \frac{2}{3}\frac{e^2}{\mel c^3}\omega^2,
\quad
  \nu_\mathrm{ri} = \frac{2}{3}\frac{Z^2e^2}{\mion c^3}\omega^2
\label{nu_r}
\eeq
are natural radiative half-widths of the electron and ion
quantum states
and $\nu_\mathrm{ei}^\perp(\omega)$ is an effective damping
frequency due to the electron-ion collisions. 
Since $\nu_\mathrm{ri} \ll \nu_\mathrm{re}$, the damping frequency in \req{eps_long} is $\nu_\| \approx
\nu_\mathrm{ei}^\| + \nu_\mathrm{re}$.
At the high temperatures typical for the XRPs, $\nu_\mathrm{ei}$
is mainly determined by the free-free transitions. 
In this case, it is customary to write \citep[e.g.,][]{PavlovPanov76}
\beq
  \nu_\mathrm{ei}^{\|,\perp} = \frac{4}{3}\sqrt{\frac{2\pi}{\mel\kB T}}
  \frac{\nel e^4}{\hbar\omega}
  \left(1 - \ee^{-\hbar\omega/\kB T} \right)
  \Lambda^{\mathrm{ff}}_{\|,\perp}(\omega),
\label{nu_ei}
\eeq
where $\nel$ is the electron number density, 
$\kB$ is the Boltzmann constant and
$\Lambda^{\mathrm{ff}}_{\parallel,\perp}(\omega)$ is a Coulomb logarithm.
By definition, the Coulomb logarithm is related to the thermally
averaged Gaunt factor $g^\mathrm{ff}_{\parallel,\perp}$ as
$\Lambda^{\mathrm{ff}}_{\parallel,\perp}=(\pi/\sqrt{3})
g^{\mathrm{ff}}_{\parallel,\perp}$; symbols $\parallel$ and  $\perp$  
correspond to photon polarization along and transverse to $\bm{B}$,
respectively.

The most general explicit expressions for
$\Lambda^{\mathrm{ff}}_{\parallel,\perp}$ with allowance for
 quantizing magnetic
fields and ion motions are given in
\citet{Potekhin10cyclo}; they
generalize and correct earlier results presented by different authors
 for specific
particular cases (e.g., \citealt{PavlovPanov76} for $\omega\gg\omB{i}$
or \citealt{1992herm.book.....M} for
$\omB{i}\ll\omega\ll\omB{e}$).

The above results have been
obtained 
{within the framework} of the so-called elementary theory,
which is adequate at $\omega \gg \ompe$ \citep{Ginzburg70}.

%------------------------------------------------------
\subsubsection{Vacuum polarization}
\label{sec:vacpol}

The
influence of the vacuum polarization on the NS
radiation was
first evaluated by \citet{Novick_77}, whose results were revisited and
corrected by \citet*{GnedinPS78}; the effect was
studied in detail by \citet{PavlovGnedin84}. In the linear
approximation for the electromagnetic waves, the
dielectric tensor with allowance for both the magnetized plasma and vacuum
polarization can be written as
\beq
 \bm{\varepsilon}' = \bm{\varepsilon} + 4\pi\bm{\chi}_\mathrm{v},
\quad
\bm{\chi}_\mathrm{v} = \frac{1}{4\pi}
 \mathrm{diag}(a_\mathrm{v}, a_\mathrm{v}, a_\mathrm{v}+q_\mathrm{v} )
\eeq
where $\bm{\varepsilon}$ is given by \req{eps-p}, 
$\bm{\chi}_\mathrm{v}$ is the polarizability of vacuum,
diag(\ldots) denotes the diagonal matrix, $a_\mathrm{v}$ and
$q_\mathrm{v}$ are specified below.
Magnetic permeability of the QED vacuum $\bm{\mu}_\mathrm{v}$ is determined by 
\beq
 \bm{\mu}_\mathrm{v}^{-1} = \mathrm{diag}(1+a_\mathrm{v}, 1+a_\mathrm{v},
 1+a_\mathrm{v}+m_\mathrm{v}).
\eeq
{In these equations, $a_\mathrm{v}$, $q_\mathrm{v}$ and $m_\mathrm{v}$ are scalar functions of $B$,
which can be approximated by expressions}
\citep{2004ApJ...612.1034P}
\begin{align}
&
 a_\mathrm{v} = - \frac{2\alphaf}{9\pi} \ln\bigg(
 1 + \frac{b^2}{5}\,\frac{
 1+0.25487\,b^{3/4}
 }{
 1+0.75\,b^{5/4}}\bigg),
\label{fit-a}
\\&
 q_\mathrm{v} = \frac{7\alphaf}{45\pi}\,b^2\,\frac{
 1 + 1.2\,b
 }{
 1 + 1.33\,b + 0.56\,b^2
 },
\label{fit-q}
\\&
 m_\mathrm{v} = - \frac{\alphaf}{3\pi} \, \frac{ b^2 }{
 3.75 + 2.7\,b^{5/4} + b^2 },
\label{fit-m}
\end{align}
{where $\alphaf\equiv e^2/\hbar c \approx 1/137$ is the
fine-structure constant.
Equations (\ref{fit-a})--(\ref{fit-m})
agree with accurate analytical \citep{HeylHernquist97} and numerical \citep{KohriYamada02} results at any $B$.
They also reproduce the asymptotes derived by
\citet{1971AnPhy..67..599A} in the limit of small parameter
$
  b \equiv B/B_\mathrm{QED},
$
where
 $B_\mathrm{QED} \equiv \mel^2 c^3 /
e\hbar = 4.414\times10^{13}$~G is the natural magnetic field unit in the QED:
$
  a_\mathrm{v} = - 2 \delta_\mathrm{v}
$, $
  q_\mathrm{v} = 7\delta_\mathrm{v}
$ and $
  m_\mathrm{v} = -4 \delta_\mathrm{v}$,
where $\delta_\mathrm{v} \equiv (\alphaf/45\pi)
b^2 \approx 2.65\times10^{-8}B_{12}^2$ and $B_{12}\equiv{B/10^{12}}$~G.
These low-$b$ asymptotes are quite accurate in the case of $b\sim0.1$
considered below.}

The regions of predominant importance of either 
the plasma effects
or the vacuum polarization effects 
are separated by the vacuum resonance energy $E_{\mathrm{vac}}$ \citep[see][]{HoLai03}.
The concept of a vacuum resonance was introduced in the theory by   \citet{GnedinPS78}.
If 
{the photon energy $E\equiv\hbar\omega$ is small compared with $E_{\mathrm{vac}}$}, 
the dielectric tensor is determined by plasma, while at 
$E\gg E_{\mathrm{vac}}$ 
it is determined by vacuum polarization.
At $b\ll1$,
\begin{equation}
  E_{\mathrm{vac}} \approx \left(\frac{15\pi}{\alphaf}\right)^{1/2}
  \frac{\ompe}{\omB{e}}\,\mel c^2
  \approx 10^2 \left(\frac{Z}{A}\right)^{1/2}\frac{\sqrt{\rho_1}}{B_{12}}\text{ keV},
\label{Evac}
\end{equation}
where  $\rho_1 \equiv \rho /(1\ {\rm g}\  {\rm cm}^{-3} )$,\
$\rho$ is the plasma mass density
and $A$ is the ion mass number.
One can write
$E_{\mathrm{vac}}  \approx {E_B/\sqrt{W}}$,
where $E_B \equiv \hbar\omB{e} \approx 11.6B_{12}$ keV is the electron
cyclotron energy and
\begin{equation}
\label{eqW}
  W \simeq (3\times10^{28}\text{~cm}^{-3}/\nel)\,b^4,
\end{equation}
is the vacuum polarization 
parameter \citep{1979JETP...49..741P,1982Ap&SS..86..249K}.

%------------------------------------------------------
\subsubsection{Polarization vectors of the normal modes}
\label{sec:polarNM}

Let us consider
the right-handed Cartesian coordinate system $(xyz)$ with the $z$ axis
 directed along $\bm{k}$ and the $x$-axis
perpendicular to the $(\bm{k}$--$\bm{B})$ plane.\footnote{Note that this coordinate system is rotated 
by the angle $\theta_B$ relative to the
one used in Sections~\ref{sec:RTEmag} and~\ref{sec:vacpol} (see, e.g.,
\citealt{Ginzburg70}, \S\,10, for the
transformations of $\bm{\varepsilon}$ between the coordinate systems).}

In the NM approximation, the complex electric
vectors can be numbered by the
polarization index $j$, which takes two values for the two polarization
modes. We will use the transverse approximation
$\bm{\mathcal{E}}\cdot\bm{k}=0$. Then
\begin{equation}
\label{Ealpha}
\bm{\mathcal{E}}_j = \mathcal{E}_j \bm{e}_j
= \mathcal{E}_{j,x} \hat{\bm{e}}_x + 
    \mathcal{E}_{j,y} \hat{\bm{e}}_y ,
\end{equation}
where
$\mathcal{E}_j$ 
determines  the NM amplitude, $\bm{e}_j$ is the complex unit
vector of polarization, $\hat{\bm{e}}_x$ and $\hat{\bm{e}}_y$ 
are the unit vectors along the 
$x$- and $y$-axes, 
$\mathcal{E}_{j,x}$ and $\mathcal{E}_{j,y}$
are complex electric field amplitude components along these axes.
Following
\citet{HoLai01,HoLai03}, we adopt the values $j=1$ and 2 for the X-
and O-modes, respectively, and define the dimensionless
quantities
\beq
  u_\mathrm{e} = \frac{\omB{e}^2}{\omega^2},
\quad
  u_\mathrm{i} = \frac{\omB{i}^2}{\omega^2},
\quad
  v_\mathrm{e} = \frac{\ompe^2}{\omega^2},
\quad
  v_\mathrm{i} = \frac{\ompi^2}{\omega^2}.
\label{u_v_def}
\eeq
In the following we will assume that $v_\mathrm{e}\ll1$ (which also implies that $v_\mathrm{i}\ll1$).
It is also convenient to introduce the complex
ellipticity parameter \citep[e.g.][]{HoLai03}\footnote{The choice of sign in \req{xi_def} 
is opposite to that in \citet{2022PhRvD.105j3027M}, in order to make our definition of $K_j$ 
equivalent to that by \citet{HoLai03}, 
taking into account that the coordinate system with the $x$-axis lying in the ($\bm{k} - \bm{B}$) plane was chosen 
by these authors, while we have chosen the $y$-axis lying in this plane. 
Thus defined parameters $K_j$ are equivalent to 
$-\ii K_j$ in \S\,11 of \citet{Ginzburg70}.}
\beq
  {K}_{j} 
= \ii \mathcal{E}_{{j},y} / \mathcal{E}_{{j},x} 
= \ii e_{{j},y} / e_{{j},x},
\label{xi_def}
\eeq
where
$e_{{j},y} = 
\bm{e}_{j}\cdot \hat{\bm{e}}_y$ 
and $e_{{j}x,} = 
\bm{e}_{j} \cdot \hat{\bm{e}}_x$ are projections 
of $\bm{e}_{j}$ on the $y$ and $x$ axes.

In Paper~I we neglected the vacuum polarization, collisions between
plasma particles {and} ion cyclotron oscillations in a magnetic field.
In such approximations,
\begin{equation}
\label{xi0}
  {K}_j \approx \frac{- 2\cos\theta_B}
{\sqrt{u_\mathrm{e}}\sin^2\theta_B 
- (-1)^{j} \sqrt{u_\mathrm{e}\sin^4\theta_B +4\cos^2\theta_B} },
\end{equation}
where $\theta_B$ is the angle between $\bm{k}$
and $\bm{B}$, and we have neglected $v_\mathrm{e}$ in the factors $(1-v_\mathrm{e})$ 
\citep[cf.\ equation (11.26) of][]{Ginzburg70}.

%%%%%%%%%%%%%%%%%
\begin{figure}
	\centering 
	 \includegraphics[width=\columnwidth]{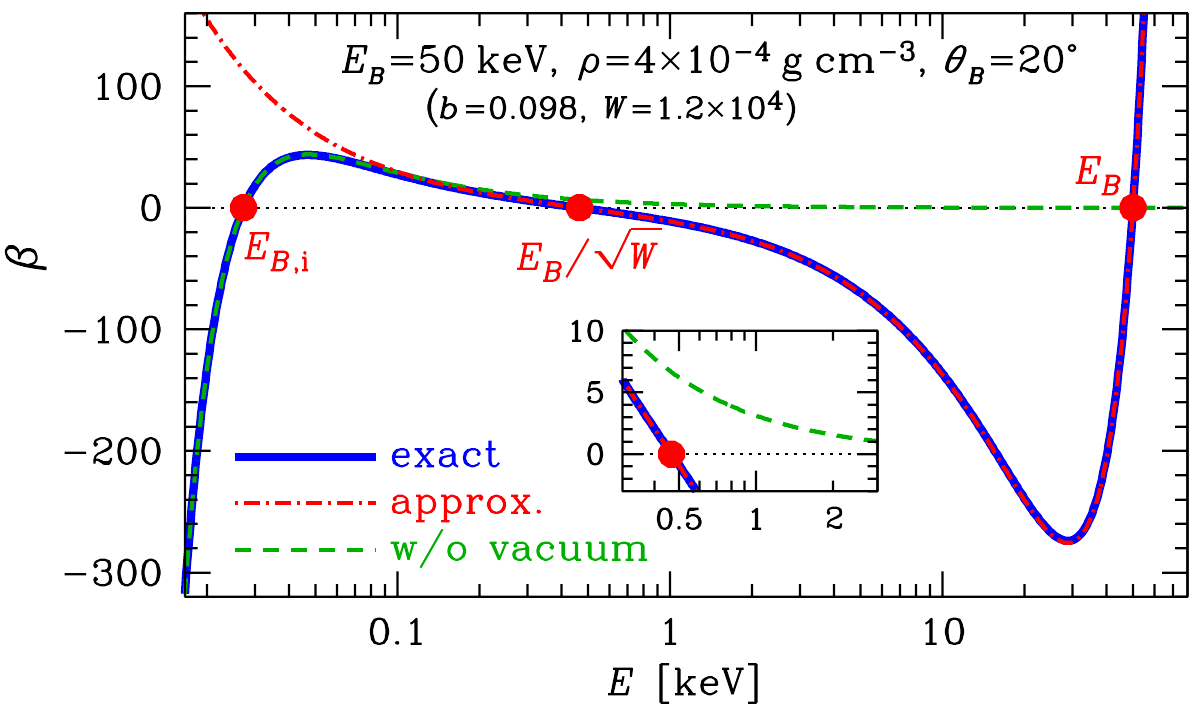}
 	\caption{Parameter $\beta$ according to equations (\ref{beta})
        (solid curve) and (\ref{beta-W}) (dot-dashed curve) 
        as functions of photon energy $E$ at magnetic field  $B=4.32\times10^{12}$~G, mass density
        $\rho=4\times10^{-4}$ g cm$^{-3}$ and angle $\theta_B=\pi/9$. The dashed curve shows
        this parameter without account of the QED vacuum, i.e.
        according to equation (\ref{beta}) with $\varepsilon'_{\|,\perp,\wedge}$ replaced by $\varepsilon_{\|,\perp,\wedge}$.
        Heavy dots mark the energy values $E_{B,\mathrm{i}}$, $E_B/\sqrt{W}$ and 
        $E_B$, which are approximations to the values of $E$ at which $\beta$ changes its sign.
        The inset shows a zoom to $E\sim E_B/\sqrt{W}$.
        }
\label{fig:beta}
\end{figure}
%%%%%%%%%%%%%%%%%%%%%%%%%%%%%%%%%%%%

In the present calculations we rely on a more general expression 
for the complex ellipticity parameter, which takes into account the finite ion mass,
damping at the resonances {and} vacuum polarization {\citep{HoLai03}:}
\beq
 {K}_j = \beta \left\{
  1 + (-1)^j \left[ 1 + \frac{1}{\beta^2} 
  + \frac{m_\mathrm{v}}{1+a_\mathrm{v}} \frac{\sin^2\theta_B}{\beta^2}\right]^{1/2}
  \right\},
\label{xi}
\eeq
where
\beq
 \beta = \frac{\varepsilon_\|' - \varepsilon_\perp' + \varepsilon_\wedge^2/\varepsilon_\perp' + \varepsilon_\|'
 \,m_\mathrm{v}/(1+a_\mathrm{v})
  }{
  2 \, \varepsilon_\wedge }
  \,\, \frac{ \varepsilon_\perp'}{\varepsilon_\|'}
  \,\,\frac{\sin^2\theta_B}{\cos\theta_B},
\label{beta}
\eeq
$\varepsilon_\perp' = \varepsilon_\perp + a_\mathrm{v}$
and
$\varepsilon_\|' = \varepsilon_\| + a_\mathrm{v} +
q_\mathrm{v}$.
The parameter $\beta$ is sometimes also called 
polarization parameter \citep[e.g.,][]{2003ApJ...588..962L}.
We use the transverse approximation ($\bm{e}_j\cdot\bm{k}=0$) and
therefore neglect the $z$-components of the polarization vectors
$e_{j,z} = \bm{e}_{j}\cdot \hat{\bm{e}}_z$
(see the explicit expression for 
${K}_{j,z} = - \ii e_{j,z} / e_{j,x} \propto v_\mathrm{e} \ll 1$
in \citealt{HoLai03}).
At $B\ll B_\mathrm{QED}$, 
using the relations
$v_\mathrm{i} \ll v_\mathrm{e} \ll 1$,
$\sqrt{u_\mathrm{i}} \ll \sqrt{u_\mathrm{e}}$
and $| u_\mathrm{e} - 1 | \gg v_\mathrm{e}$,
one can approximate \req{beta} by the expression 
\citep{1979JETP...49..741P,1982Ap&SS..86..249K}%
\begin{equation} 
\label{beta-W}
\beta = \sqrt{u_\mathrm{e}} \frac{\sin^2 \theta_B}{2 \cos \theta_B}
\left( 1 - W \frac{u_\mathrm{e} -1}{u_\mathrm{e}^2} \right),
\end{equation}
where $W$ is given by \req{eqW}.

{Figure \ref{fig:beta} demonstrates the typical behaviour of the parameter 
$\beta$ as a function of photon energy.
It crosses zero near the electron and ion cyclotron resonances and near the point 
$E=E_B/\sqrt{W}$;  the values of $\beta$ away from these points are large. 
In contrast, $\beta$ decreases with growing $E$ at
$E\gg E_{B,\mathrm{i}}$  as $\beta\propto E_B/E$},
{if the vacuum polarization is neglected.
We also see that the approximation (\ref{beta-W}) is excellent at $E\gg E_{B,\mathrm{i}}$, 
but becomes inapplicable at smaller energies, where the cyclotron motion of ions
becomes important. 
The dependence of $K_j$ on $E$ and $\rho$ reflects the dependence 
of $\beta$ according to equation (\ref{xi}): since the vacuum parameter 
$m_\mathrm{v}$ is small, $K_2\approx2\beta$ whenever $\beta^2\gg1$,
$K_2\approx\mathrm{sign}\beta$ if $\beta^2\ll1$, and $K_1\approx - K_2^{-1}$.
It means that away from the resonances, where
$\beta^2\gg1$,
the NM polarization is almost linear
\citep[as first noticed by][]{1979JETP...49..741P}.
}

Instead of the X- and O-modes with the ellipticities 
$K_j$, which are determined by
relation $|K_2| > |K_1|$ and given by \req{xi}, one can 
refer  to the `plus' and `minus' modes
\citep{2002ApJ...566..373L,HoLai03}, 
whose complex ellipticities 
are given by
\beq
  K_\pm = \beta \pm \sqrt{\beta^2+ 1 +
  m_\mathrm{v}\sin^2\theta_B/(1+a_\mathrm{v})}.
\label{xi+-}
\eeq
At $\beta>0$, the plus-mode is the O-mode and the minus-mode is the X-mode, 
and vice versa at $\beta < 0$.

Under the considered conditions, typical for the XRPs, the plasma
and damping frequencies are small compared to $\omega$. 
Then one can neglect the anti-Hermitian part 
of the dielectric tensor at $\omega$ outside the Doppler cores of 
cyclotron resonances \citep*[see][]{1979JETP...49..741P,PavlovSY80}. 
Thus the damping frequencies serve only to remove
divergencies near the resonances by replacing the
factors $(\omega-\omB{e})$ and $(\omega-\omB{i})$ by 
$(\omega-\omB{e}+\ii\nu_\mathrm{eff,e})$ and 
$(\omega-\omB{i}+\ii\nu_\mathrm{eff,i})$ at
$\omega\approx\omB{e}$ and $\omega\approx\omB{i}$,
respectively, where $\nu_\mathrm{eff,e}$ and $\nu_\mathrm{eff,i}$ are \emph{effective}
cumulative damping frequencies, which can be found from \req{eps_perp}.
Since $\omB{i}\ll\omB{e}$ and $\nu_\mathrm{ri}\ll\omB{e}$, 
the denominator in \req{eps_perp} for
$\varepsilon_\perp-\varepsilon_\wedge$
at the electron cyclotron resonance ($\omega\approx\omB{e}$) approaches
$\omB{e}(\omega-\omB{e}+\ii\nu_\mathrm{re} + \ii\nu_\mathrm{ei}^\perp)$,
which means
that
\beq
  \nu_\mathrm{eff,e} \approx \nu_\mathrm{re} + \nu_\mathrm{ei}^\perp(\omB{e}).
\label{nu_eff_e}
\eeq
In the same way, at the ion cyclotron resonance ($\omega\approx\omB{i}$)
the denominator in \req{eps_perp} for
$\varepsilon_\perp+\varepsilon_\wedge$ can be approximated by
$\omB{e}(\omega-\omB{i}+\ii\nu_\mathrm{ri})+\ii\omB{i}\nu_\mathrm{ei}^\perp$,
so that\footnote{The factor $(\mel/\mion)$ in
\req{nu_eff_i}, which also appears in equation~(53) of 
\citet{PotekhinChabrier03}, was
missed by \citet{2003ApJ...588..962L} and \citet{vanAdelsbergLai06}.}
\beq
  \nu_\mathrm{eff,i} \approx \nu_\mathrm{ri}
  + \frac{\omB{i}}{\omB{e}}\nu_\mathrm{ei}^\perp(\omB{i})
  = \nu_\mathrm{ri}
  + \frac{\mel}{\mion}\nu_\mathrm{ei}^\perp(\omB{i}).
\label{nu_eff_i}
\eeq
For estimations, 
it is convenient to rewrite equations~(\ref{nu_r}) and~(\ref{nu_ei}) in the
form \citep{2003ApJ...588..962L}
\begin{align}
\label{gam_ei}&
  \frac{\nu_{\mathrm{re}}}{\omega} \approx 9.5\times10^{-6}E_1,
\quad
  \frac{\nu_{\mathrm{ri}}}{\omega} \approx
  5.2\times10^{-9}\frac{Z^2}{A}E_1,
\\&
  \frac{\nu_{\mathrm{ei}}^{\parallel,\perp}}{\omega}
   \approx 9.2\times10^{-5}\frac{Z^2\rho_1}{AT_6^{1/2}E_1^2}
   \left(1-e^{-E/\kB T}\right)g^{\mathrm{ff}}_{\parallel,\perp},
\label{gam_r}
\end{align}
where
$E_1={E/(1~\text{keV})}$ and $T_6=T/(10^6~\text{K})\approx
11.6\,\kB T/(1\mbox{~keV})$. 

In the following we consider the
hydrogen plasma and set $Z=A=1$.
The Gaunt factor $g^{\mathrm{ff}}_{\|,\perp}$
depends on $E$, $\rho$, $T$ and $B$ in a non-trivial way, 
but typically remains in the range $\sim0.1$--10
\citep[see, e.g.,][]{Potekhin10cyclo}. Thus we see from
equations~(\ref{gam_ei}) and~(\ref{gam_r}) that
for the plasma parameters considered in this paper
($E_1\gtrsim 1$, $\rho_1 \lesssim10^{-2}$, $T_6 \gg 10$) we have
$\nu_\mathrm{re} \gg \nu_\mathrm{ei}^{\|,\perp}$
and $\nu_\mathrm{ri} \gg (\mel/\mion)\nu_\mathrm{ei}^{\|,\perp}$.
Then, according to equations~(\ref{nu_eff_e}) and~(\ref{nu_eff_i}),
$\nu_\mathrm{eff,e} \approx \nu_\mathrm{re}$ and $\nu_\mathrm{eff,i}
\approx \nu_\mathrm{ri}$, therefore we neglect the free-free damping.

As follows directly from the definition (\ref{xi_def}),
in the transverse approximation
the polarization vectors can be written in the form
\begin{equation} 
\label{ecyc}
\bm{e}_{j} = \frac{e^{\ii \phi_{j}}}{\sqrt{1 + |{K}_{j}|^2}} 
\left( \hat{\bm{e}}_x - \ii{K}_{j} \hat{\bm{e}}_y \right),
\end{equation}
where $\phi_{j}$ is an arbitrary phase.
If the vectors
$\bm{e}_{j}$ are mutually orthogonal,
$\bm{e}_1\cdot\bm{e}_2^\ast=0$
(the asterisk means complex conjugate), then \req{ecyc} provides the
relation
\beq
  {K}_1= - {K}_2^{-1}.
\label{K1K2}
\eeq
With allowance for the vacuum polarization, \req{K1K2}
is fulfilled only approximately, but with high accuracy for the
parameter values considered here. 
In this case, the
small inaccuracy of \req{K1K2} is equivalent to the small inaccuracy of
the  NM  orthogonality.
In general case the NM orthogonality 
was investigated by
\citet{GnedinPavlovModes,1979JETP...49..741P,PavlovSY80}.

Let us write \req{ecyc} in the matrix 
form by analogy with \citet{2022PhRvD.105j3027M},
%%%%%%%%%%%%%%%%%%%%%%%%%%%%%%%
\begin{equation}
  \label{eMe}
\begin{pmatrix}
 \bm{e}_1\\
 \bm{e}_2
\end{pmatrix}
=
\mathbfss{M}
\begin{pmatrix}
 \bm{e}_1^\ell \\
 \bm{e}_2^\ell 
\end{pmatrix},
\end{equation}
%%%%%%%%%%%%%%%%%%%%% 
where $\mathbfss{M}$ is composed of elements $M_{jl}$
with the first index (${j}=1,2$) numbering 
the complex polarization
 vectors $\bm{e}_{j}$ of the elliptical 
NMs
and the second one (${l}=1,2$) numbering the linear
 (indicated by the superscript $\ell$)
polarization vectors
$\bm{e}_1^\ell =\hat{\bm{e}}_x$ and
$\bm{e}_2^\ell =\hat{\bm{e}}_y$:
\beq
   M_{j1} = \frac{\ee^{\ii\phi_j}}{\sqrt{1+|K_j|^2}},
\quad
   M_{j2} = \frac{-\ii K_j \ee^{\ii\phi_j}}{\sqrt{1+|K_j|^2}}
\quad
   (j=1,2).
\label{M1}
\eeq

\citet{1979JETP...49..741P} represented the parameter $\beta$ [Eq.~(\ref{beta})] as
$\beta = q + \ii p$, where $q$ and $p$ are real, $q$ determines the ellipticity and $p$ 
is associated with collisional processes.
Under the conditions considered in this paper,
the inequality $q^2 \gg p^2$ is fulfilled and we may neglect the parameter $p$. 
Then $\bm{e}_1$ and $\bm{e}_2$ are mutually orthogonal 
(disregarding the above-mentioned minor inaccuracy)
and using \req{K1K2} we can write $\mathbfss{M}$ in the
form
\begin{align}
  \mathbfss{M} &= \frac{1}{\sqrt{1+{K}_1^2}}
  \left( \begin{array}{cc}
   \ee^{\ii\phi_1} & - \ii\ee^{\ii\phi_1} {K}_1
  \\
   |{K}_1|\,\ee^{\ii\phi_2}  &  \ii\ee^{\ii\phi_2}\,\mathrm{sign}\,{K}_1
   \end{array}
  \right)
\nonumber\\&
= \frac{1}{\sqrt{1+{K}_2^2}}
  \left( \begin{array}{cc}
  |{K}_2|\,\ee^{\ii\phi_1}  &  \ii\ee^{\ii\phi_1}\,\mathrm{sign}\,{K}_2
  \\
   \ee^{\ii\phi_2} & - \ii\ee^{\ii\phi_2} {K}_2
   \end{array}
  \right) ,
\label{M}
\end{align}
which 
generalizes%
\footnote{{The factors $\ee^{\ii\phi_j}$ are introduced for generality,
but the choice of the phases $\phi_j$ does not affect any results.
Indeed,
the cross sections do not depend on $\phi_j$, as can be seen from equations
(\ref{dsigma_a_fi}) and (\ref{MaM}).
In the calculations we set $\phi_1=0$ and $\phi_2=\pi$.}}
and corrects\footnote{Equations (4)--(6) in  
\citet{2022PhRvD.105j3027M}
mistakenly referred to the non-normalized vectors $\mathcal{E}$
instead of polarization vectors $\bm{e}$ and contained erroneous
signs, inconsistent with any choices of
$\phi_1$ and
$\phi_2$,
as well as equation (15) in \citet{2021MNRAS.503.5193M}.}
a similar matrix in \citet{2021MNRAS.503.5193M,2022PhRvD.105j3027M}.
This matrix is unitary:
$\mathbfss{M}^{-1}=\mathbfss{M}^\dag$,
where
the dagger sign means the Hermitian conjugate.

%%%%%%%%%%%%%%%%%%%%%%%%%%%%%%%%%%%%%%%%%%%
\subsection{Compton scattering in a strong magnetic field}
\label{sec:compton_scat}

%%%%%%%%%%%%%%%%%%%%%%%%%%%%%%%%%%%%%%%%%%%%%%
\begin{figure*}
\centering
\includegraphics[width=.33\textwidth]{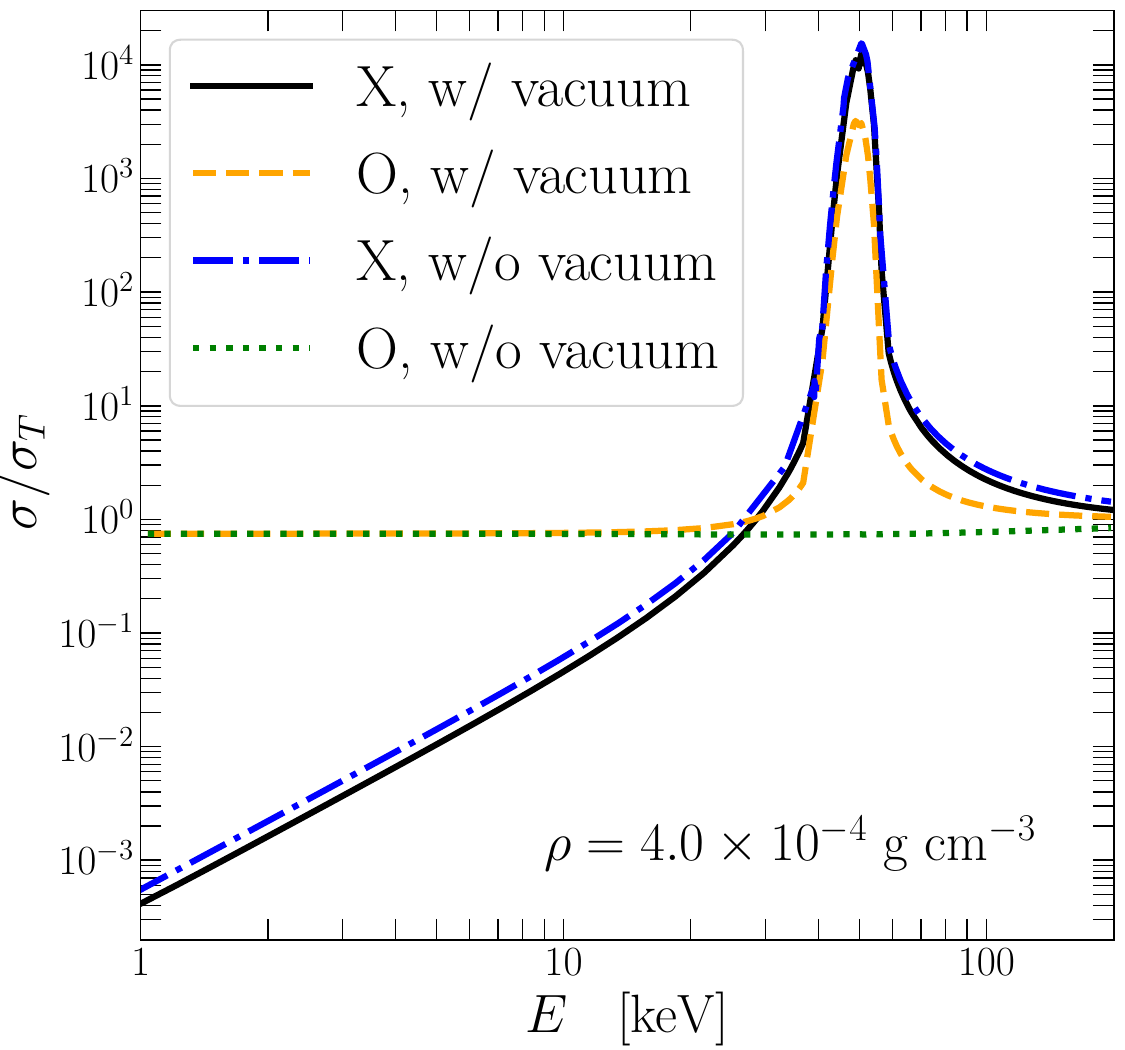}
\includegraphics[width=.33\textwidth]{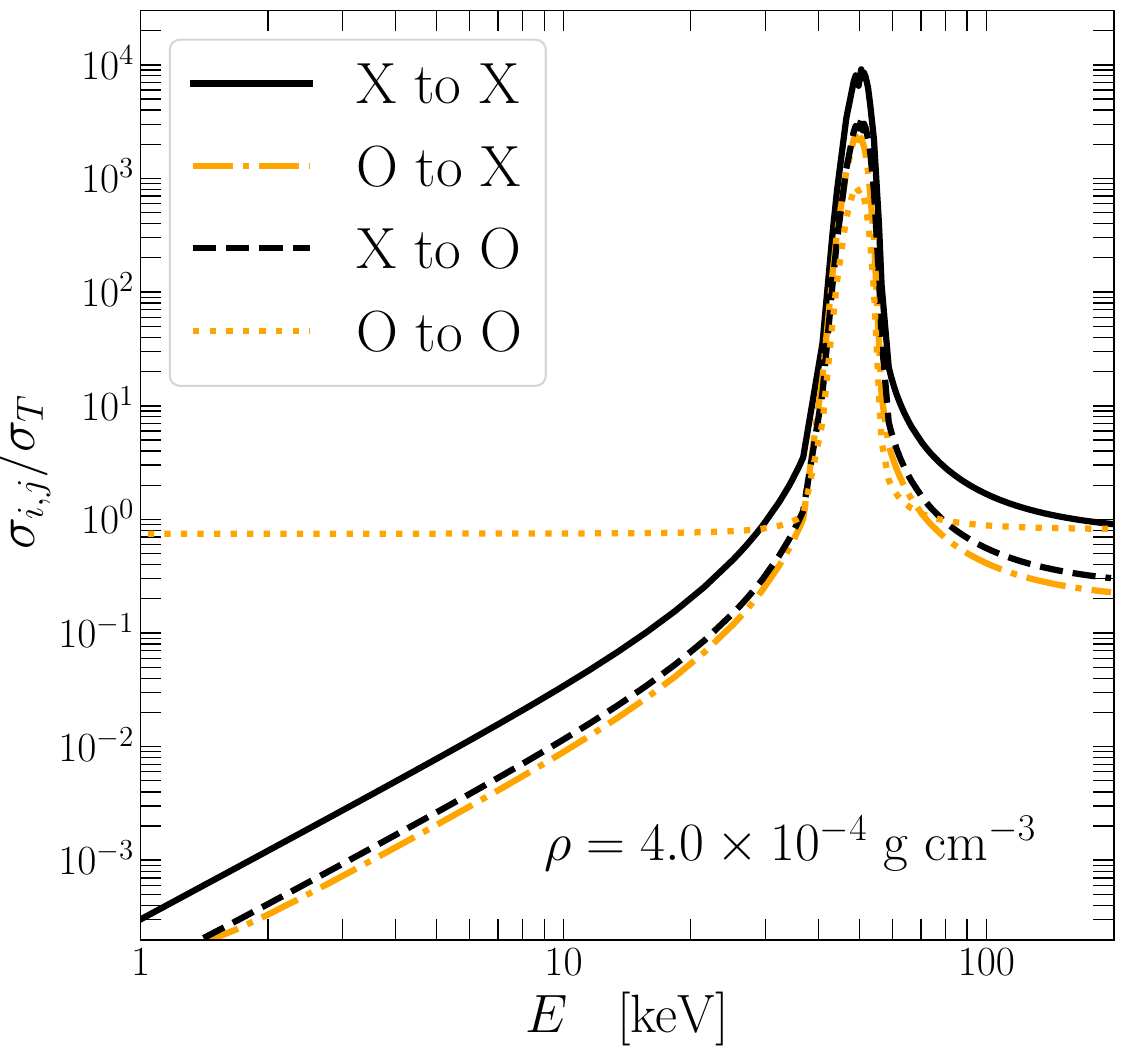}
\includegraphics[width=.33\textwidth]{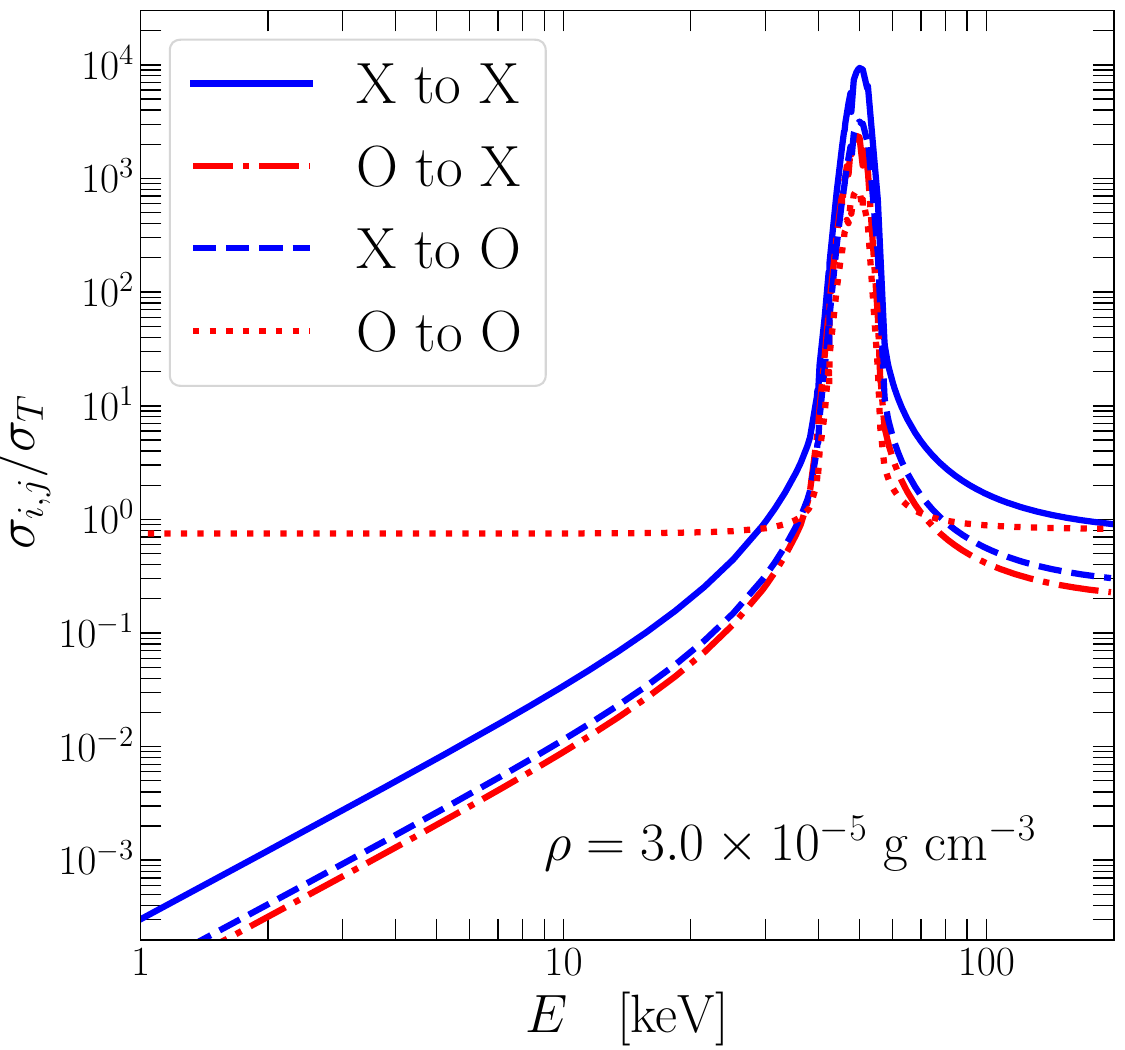}
   \caption{Scattering cross sections as functions
of photon energy in a
hydrogen plasma with temperature
$\kB T=5$ keV and cyclotron energy $E_B=50$ keV 
{for radiation propagating at angle}
$\theta_B=\pi/3$ {to the magnetic field.}
\textit{{Left} panel:} the total cross section (\ref{totcrsec})
for the X-mode  (solid and dot-dashed lines) and O-mode
(dashed and dotted lines) 
{at plasma mass density $\rho=4\times10^{-4}$ g cm$^{-3}$
(the vacuum polarization parameter (\ref{eqW}) $W \approx 1.2\times10^4$)}
{taking vacuum polarization into account}
(solid and dashed lines) and without it (dot-dashed and dotted lines). 
\textit{{Middle} panel:} 
cross sections (\ref{totcrsec1}) 
of
the scattering 
from X-  to X-mode (solid line), 
from O- to  X-mode (dot-dashed line),
from X-  to O-mode (dashed line) and
from O- to O-mode (dotted line)
{with account for the vacuum polarization at the same parameters.
\textit{Right panel:} the same as in the middle panel but at $\rho=3\times10^{-5}$ g cm$^{-3}$
($W \approx 1.5\times 10^5$).}
}
\label{f:cr_sec} 
\end{figure*}
%%%%%%%%%%%%%%%%%%%%%%%%%%%%%%%%%%%

Compton scattering in a strong magnetic field differs significantly
from the field-free one. In this subsection we outline cross sections of the process
by analogy with
\citet{2022PhRvD.105j3027M} and Paper~I. 

We will assume that an electron 
occupies the ground Landau level before and after scattering. 
This is generally a good approximation for the XRPs, 
because the cyclotron radiative de-excitation has a timescale much 
shorter than collisional processes
\citep[e.g.,][]{1992herm.book.....M}.\footnote{Note, however, that this
approximation does not allow one to reproduce the broad and shallow shapes
of the cyclotron
lines obtained by \citet{1999ApJ...517..334A}, which are due to the
multiple generation (\emph{spawning}) of photons by electrons 
at higher Landau levels in the process of scattering.}
Then the
energy and momentum conservation laws are
\begin{align}&
   \sqrt{m_\mathrm{e}^2c^4+p_\mathrm{i}^2c^2}+E_\mathrm{i} =
  \sqrt{m_\mathrm{e}^2c^4+p_\mathrm{f}^2c^2}+E_\mathrm{f} ,
\label{conserv_E}
\\&
   E_\mathrm{i}\cos\theta_\mathrm{i}+p_\mathrm{i}c=
   E_\mathrm{f}\cos\theta_\mathrm{f}+p_\mathrm{f}c,
\label{conserv_p}
\end{align}
where $p_\mathrm{i}$ ($p_\mathrm{f}$) is the projection of the electron
momentum on $\bm{B}$, $E_\mathrm{i}$ ($E_\mathrm{f}$) is the photon
energy and $\theta_\mathrm{i}$ ($\theta_\mathrm{f}$) is the value of
$\theta_B$ before (after) scattering.

The differential cross section of photon
scattering on an electron with $p_\mathrm{i}=0$ in a strong magnetic
field can be written as \citep[e.g.,][]{1979PhRvD..19.2868H}
\begin{equation}
  \label{dsigma_a_fi}
  \frac{\mathrm{d}\sigma_\mathrm{fi}^0}
  {\mathrm{d}\bm{\Omega}_\mathrm{f}^0}
 \left(E_\mathrm{i}^0,\bm{\Omega}_\mathrm{i}^0,
 \bm{\Omega}_\mathrm{f}^0\right)=
  \frac{3}{32\pi}\sigma_\mathrm{T}|a_\mathrm{f i}|^2,
\end{equation}
where $E_\mathrm{i}^0$ is the initial photon energy,
$\bm{\Omega}_\mathrm{i,f}^0$ are the unit vectors along $\bm{k}$ before
and after scattering, the superscript 0 indicates that they are measured
in the electron rest frame before the scattering, $\sigma_\mathrm{T}$ is
the Thomson cross section and $a_\mathrm{f i}$ is a complex scattering
amplitude. Hereinafter we use the non-relativistic approximation with a single cyclotron
resonance in $a_\mathrm{fi}$, which is valid at $B\ll B_\mathrm{QED}$ and 
${E_{B,\mathrm{i}}}\ll E_\mathrm{i}^0 \ll \mel c^2$,
{where ${E_{B,\mathrm{i}}\equiv \hbar\omB{i} = (\mel/\mion)E_B}$}. 
Then the scattering amplitudes for linear
polarizations orthogonal and
coplanar to the $\bm{k}-\bm{B}$ plane are \citep{1979PhRvD..19.2868H}
\begin{align}&
  a^\ell _\mathrm{11} = 
 A_+
 + A_-,
\label{Herold1}
\\&
  a^\ell _\mathrm{22} = 
2\sin\theta_\mathrm{i}\,\sin\theta_\mathrm{f} + 
(A_+ + A_-)
\cos\theta_\mathrm{i}\,\cos\theta_\mathrm{f},
\label{Herold2}
\\&
  \begin{pmatrix}
  a^\ell _\mathrm{12}
  \\
  a^\ell _\mathrm{21}
  \end{pmatrix}
 =
  \begin{pmatrix}
  -\mathrm{i}\cos\theta_\mathrm{i}
  \\
  \mathrm{i}\cos\theta_\mathrm{f}
  \end{pmatrix}
  (A_+ - A_-),
\label{Herold3}
\\&
\text{where}\quad
A_+ \equiv \frac{E_\mathrm{i}\mathrm{e}^{\mathrm{i}(\varphi_\mathrm{i}-\varphi_\mathrm{f})}}{E_\mathrm{i}+
  E_B},
\quad
 A_- \equiv \frac{E_\mathrm{i}\mathrm{e}^{-\mathrm{i}(\varphi_\mathrm{i}-
 \varphi_\mathrm{f})}}{E_\mathrm{i}-E_B + \ii\hbar\nu_\mathrm{eff,e}}.
\label{A+-}
\end{align}
Here the first and second
subscripts relate to the final and initial states of a photon,
respectively, $\varphi_\mathrm{i,f}$ are azimuthal angles in the plane
orthogonal to $\bm{B}$, the subscripts `1' and `2' indicate the linear
polarizations $\bm{e}_1^\ell $ and $\bm{e}_2^\ell $, and $\nu_\mathrm{eff,e}$ 
is the effective damping frequency [Eq.~(\ref{nu_eff_e})];  in our case 
$\nu_\mathrm{eff,e} \approx \nu_\mathrm{re}$.

Let us compose a $2\times2$ matrix
$\mathbfss{a}_\mathrm{fi}^\ell $ of the four scattering amplitudes for the linear polarizations $a^\ell _{l'l}$ in 
equations (\ref{Herold1})--(\ref{Herold3}). 
To transform it to the analogous matrix $\mathbfss{a}_\mathrm{fi}$ of the  amplitudes for 
the elliptical NM polarizations $a_{j'j}$, we convert the initial NM basis vectors $\bm{e}_j$ 
into the initial linearly polarized vectors $\bm{e}_l$ by applying the matrix 
$\mathbfss{M}_\mathrm{i}^{-1} = \mathbfss{M}_\mathrm{i}^\dag$ and the final 
linearly polarized vectors $\bm{e}_{l'}$ into the final NM basis $\bm{e}_{j'}$ 
by using the matrix $\mathbfss{M}_\mathrm{f}$, which gives
\beq
  \mathbfss{a}_\mathrm{fi} =
\mathbfss{M}_\mathrm{f}\,\mathbfss{a}_\mathrm{fi}^\ell\,
  \mathbfss{M}_\mathrm{i}^\dag.
\label{MaM}
\eeq
The rest-frame differential cross sections for 
NM scattering are given by equations (\ref{dsigma_a_fi})--(\ref{MaM})
with $a_\mathrm{fi}=a_{j'j}$,  where the subscripts $j$ and $j'$ enumerate the elliptical NMs
of initial and final photon states, respectively.
Explicit expressions for the resulting scattering amplitudes are given in Appendix~\ref{app:scattering}.

The differential cross sections in an arbitrary reference
frame, where the electron has a non-zero initial longitudinal momentum
$p_\mathrm{i}$, are given by the Lorentz transformation as
\begin{align}&
  \label{dsigma_transform}
  \frac{\mathrm{d}\sigma_{j'j}}
{\mathrm{d}\bm{\Omega}_\mathrm{f}}
\left(p_\mathrm{i},E_\mathrm{i},
\bm{\Omega}_\mathrm{i},\bm{\Omega}_\mathrm{f}\right)
  =
  \frac{\mathrm{d}\sigma_{j'j}^0}
{\mathrm{d}\bm{\Omega}_\mathrm{f}^0}
\left(E_\mathrm{i}^0,\bm{\Omega}_\mathrm{i}^0,
\bm{\Omega}_\mathrm{f}^0\right)
\frac{1-\beta_\mathrm{i}^2}
{(1-\beta_\mathrm{i}\cos\theta_\mathrm{f})^2},
\end{align}
where 
\beq
  \beta_\mathrm{i} =
  \frac{p_\mathrm{i}}{\sqrt{m_\mathrm{e}^2c^2+p_\mathrm{i}^2}}
\eeq
 is the dimensionless electron velocity before the scattering 
and the arguments of 
$\mathrm{d}\sigma_\mathrm{fi}/\mathrm{d}\bm{\Omega}_\mathrm{f}$ 
are related to their values in the electron rest frame by relativistic
 Doppler effect and aberration formulae (see, e.g., \citealt{1975ctf..book.....L})
\begin{equation}
  \label{Dopef}
  E_\mathrm{i}^0=E_\mathrm{i}\frac{1-
  \beta_\mathrm{i}\cos\theta_\mathrm{i}}{\sqrt{1-\beta_\mathrm{i}^2}},
  \quad 
  \cos\theta_\mathrm{i,f}^0=\frac{\cos\theta_\mathrm{i,f}-\beta_\mathrm{i}}{1-
\beta_\mathrm{i}\cos\theta_\mathrm{i,f}}.
\end{equation}

Since the electrons in the plasma have different longitudinal momenta,
a photon with a given energy and direction, scattered into another given
direction, can have different energies, depending on $p_\mathrm{i}$
according to equations~(\ref{conserv_E}) and~(\ref{conserv_p}),
\begin{align}&
  \hat{E}_\mathrm{f}(p_\mathrm{i}) = E_\mathrm{i} + 
    \left( \mel^2 c^4 + p_\mathrm{i}^2 c^2 \right)^{1/2}
\nonumber\\&\qquad
    - \left[ \mel^2 c^4 + (p_\mathrm{i} c +
    E_\mathrm{i}\cos\theta_\mathrm{i} -
     E_\mathrm{f}\cos\theta_\mathrm{f})^2
     \right]^{1/2}.
\label{E_f}
\end{align}
Therefore,
a differential cross section averaged over all electrons depends on both
the initial and final photon energies, so that
\begin{align}&
 \frac{ \dd \sigma_{j'j}
       }{
       \dd E_\mathrm{f}\,
       \dd\bm{\Omega}_\mathrm{f}}
      ([f_\mathrm{e}]; E_\mathrm{i},\bm{\Omega}_\mathrm{i},
            E_\mathrm{f},\bm{\Omega}_\mathrm{f})
=
\int_{-\infty}^{+\infty}
\frac{\mathrm{d}\sigma_{j'j}}
{\mathrm{d}\bm{\Omega}_\mathrm{f}}
\left(p_\mathrm{i},E_\mathrm{i},
\bm{\Omega}_\mathrm{i},\bm{\Omega}_\mathrm{f}\right)
\nonumber\\&\qquad
  \times\delta\big(E_\mathrm{f} - \hat{E}_\mathrm{f}(p_\mathrm{i}) \big)\,
   f_\mathrm{e}(p_\mathrm{i})\,
   \dd p_\mathrm{i}
\label{sigma_av}
\\&
=
\sum_{\hat{p}_\mathrm{i}} \left[
\frac{\mathrm{d}\sigma_{j'j}}
{\mathrm{d}\bm{\Omega}_\mathrm{f}}
\left(p,E_\mathrm{i},
\bm{\Omega}_\mathrm{i},\bm{\Omega}_\mathrm{f}\right)
 f_\mathrm{e}(p)
  \left|
  \frac{ \dd \hat{E}_\mathrm{f}(p)
  }{
  \dd p}
  \right|^{\!\!-1}
  \right]_{p=\hat{p}_\mathrm{i}}\!,
\label{avcrsec}
\end{align}
where
$f_\mathrm{e}(p)$ is the 
distribution function of the longitudinal momenta of the electrons and
$\hat{p}_\mathrm{i}$ is a value of $p_\mathrm{i}$ which solves the system
(\ref{conserv_E}), (\ref{conserv_p}) at given $E_\mathrm{i}$,
$E_\mathrm{f}$, $\theta_\mathrm{i}$ and $\theta_\mathrm{f}$. 
The sum in \req{avcrsec} is over all real solutions $\hat{p}_\mathrm{i}$; it is replaced by zero if they do not exist. 
We use a one-dimensional version of the Maxwell--J\"uttner distribution of electron momenta $f_\mathrm{e}(p)$ 
(e.g., \citealt*{2010PhRvE..81b1126C}; also see Appendix A in \citealt{2022PhRvD.105j3027M}).

In the simulations, we mostly use a simplified averaging, 
taking the delta function out of the integral 
in equation~(\ref{sigma_av}):
\beq
 \frac{ \dd \sigma_{j'j}
       }{
       \dd E_\mathrm{f}\,
       \dd\bm{\Omega}_\mathrm{f}}
      ([f_\mathrm{e}]; E_\mathrm{i},\bm{\Omega}_\mathrm{i},
            E_\mathrm{f},\bm{\Omega}_\mathrm{f})
\approx
\delta\big(E_\mathrm{f} - \hat{E}_\mathrm{f}(p_\mathrm{bulk}) \big)
\overline{\frac{\mathrm{d}\sigma_{j'j}}
{\mathrm{d}\bm{\Omega}_\mathrm{f}}},
\label{sigma_av_simple}
\eeq
where $p_\mathrm{bulk}$ is the bulk plasma momentum and
\beq
\overline{\frac{\mathrm{d}\sigma_{j'j}}
{\mathrm{d}\bm{\Omega}_\mathrm{f}}}
=
\int\limits_{-\infty}^{+\infty}
\frac{\mathrm{d}\sigma_{j'j}}
{\mathrm{d}\bm{\Omega}_\mathrm{f}}
\left(p_\mathrm{i},E_\mathrm{i},
\bm{\Omega}_\mathrm{i},\bm{\Omega}_\mathrm{f}\right)
   f_\mathrm{e}(p_\mathrm{i})\,
   \dd p_\mathrm{i}.
\label{sigma_mean}
\eeq
We have checked by a direct comparison that this simplification does not introduce 
any noticeable error in our results presented below.

The total scattering cross section for a photon of a NM $j$
is given by integration over the
final energies and angles, as well as
summation over the final modes,
\begin{align}&
  \label{totcrsec}
  \sigma_j([f_\mathrm{e}]; E_\mathrm{i},\bm{\Omega}_\mathrm{i})
  =
\sum_{j'=1}^2 \sigma_{j'j}([f_\mathrm{e}]; E_\mathrm{i},\bm{\Omega}_\mathrm{i}),
\\&
  \label{totcrsec1}
  \sigma_{j'j}([f_\mathrm{e}]; E_\mathrm{i},
   \bm{\Omega}_\mathrm{i})=
    \int_{(4\pi)} 
    \, 
    {\overline{\frac{\mathrm{d}\sigma_{j'j} }{ \dd\bm{\Omega}_\mathrm{f}}} \left([f_\mathrm{e}]; E_\mathrm{i},
\bm{\Omega}_\mathrm{i},\bm{\Omega}_\mathrm{f}\right)
\mathrm{d}\bm{\Omega}_\mathrm{f}
}
\,.
\end{align}

In the present work, we primarily use cross sections in which
the finite ion mass, damping and vacuum polarization effects are taken into account 
with 
the ellipticities given by 
equation (\ref{xi}). 
Figure~\ref{f:cr_sec} 
{shows examples} of cross sections 
{that include}
these effects.
For comparison, 
{in the left panel}
we also {show} the cross sections 
calculated with ellipticities ${K}_{j}$ 
in the electron plasma approximation (\ref{xi0}) that has been
used in Paper~I. A key difference of the more accurate 
cross sections is the increased strength
of the resonance in the O-mode, which becomes 
comparable with the resonance in the X-mode.
{This behaviour agrees with \citet[][cf. their figure 5]{1982Ap&SS..86..249K}
and is explained by the appearance of virtual positrons in the QED vacuum,
which interact with radiation similarly to the electrons, but for the
opposite rotation directions of the electric vector  in the
electromagnetic wave  (circular minus-mode), so that 
the scattering cross section of the O-mode  becomes 
resonant due to the interaction with virtual positrons.
Comparing the middle and right panels, 
we see that the density dependence of the cross sections at $W\gg1$ is insignificant. 
Since the the `plasma approximation' (\ref{xi0}) omits density dependence, it means that the impact of vacuum polarization on the cross sections is nearly the same for both considered densities.}

In the following sections we study the
impact of vacuum polarization on the radiation transfer and
hydrodynamics in the accretion channel of subcritical XRPs. 

%%%%%%%%%%%%%%%%%%%%%%%%%%%%%%%%%%%%%%%%%%%%%%%%%%%%%%%%%%%%%%%%%%%%%%%%%%
\subsection{Accretion channel hydrodynamics}
\label{sec:rad_hydro}

Accreting plasma moves toward the NS
surface 
along magnetic field lines 
and forms an
\emph{accretion channel} 
near the magnetic pole 
(see, e.g., \citealt{1976MNRAS.175..395B}). 
In the case where accreting matter 
is supplied by the disc,
the base of the accretion channel has a ring-shaped cross section
with a void in the centre,
while in the case of the wind accretion the channel is fully
filled with accreting matter
(see section~2 in \citealt{1976MNRAS.175..395B}). 
We consider both cases and approximately represent an
accretion channel as a hollow (with a ring-shaped cross section)
or filled cylinder 
(designated
as \emph{ring} or \emph{circle} geometry,
respectively), neglecting the curvature of the magnetic field.

{Due to the current limitation of the employed numerical code 
(specified in Section \ref{sec:hydro_rad} below)}, 
in the present work we do not account for the GR
effects and neglect the special relativity modifications 
of
{radiation hydrodynamics equations, which then read}
\citep[e.g.,][{\S2.2}]{2007rahy.book.....C}:
\begin{align}&
\label{rhov1}
\frac{\partial\rho}{\partial{t}}+\nabla\cdot(\rho\bm{v})=0, 
\\&
\frac{\partial}{\partial{t}} \left[ \rho\epsilon +
{
\frac{\rho v^2}{2} 
}
\right]
  +\nabla\cdot \left[
{
\rho\epsilon\bm{v}
  + \rho \frac{v^2}{2}\bm{v}
   + P\bm{v}
}
\right]
  =Q_g+Q_\mathrm{rad},
\label{rhov2}
\\&
\frac{\partial\rho\bm{v}}
{\partial{t}}+\nabla\cdot
(\rho\bm{v}\otimes\bm{v})+\nabla{P}
  = \bm{F}_\mathrm{rad} +\bm{F_{g}},
\label{rhov3}
\end{align}
where
\begin{align}&
Q_\mathrm{rad} = \int_0^\infty {\dd E\int_{(4\pi)}
  \dd\bm{\Omega}\,(\kappa_{E}I_{E} - J_{E})},
\label{Qrad}
\\&
\bm{F}_\mathrm{rad} = \frac{1}{c}\int_0^\infty
{\dd E\int_{(4\pi)} \dd 
\bm{\Omega}\,(\kappa_{E}I_{E} - J_{E})}\,\bm{n}.
\label{Frad}
\end{align}
Here, 
$\bm{v}$ is
the bulk velocity of the plasma, $\epsilon$ the mass energy density,
$P$ the pressure, 
$Q_g=\rho\bm{v}\cdot\bm{g}$ the gravity power density, 
$\bm{F}_g=\rho\bm{g}$ the gravity force density,
$\bm{g}=
-
({GM_\mathrm{NS}}/{R_\mathrm{NS}^2})\bm{n}$ 
the gravitational acceleration at the surface of an NS
with mass $M_\mathrm{NS}$ and radius $R_\mathrm{NS}$, $\bm{n}$ the outer
normal to the
surface, $\bm{\Omega}=\bm{k}/|\bm{k}|$ 
the unit vector of radiation
direction, 
$I_E^j$ the specific
intensity in the mode $j$ normalized to the photon energy $E$,
$I_E=\sum_{j=1}^2 I_E^j$,
$\kappa_E^j=\nel\sigma_j$ 
the absorption coefficient in the mode $j$, $\kappa_E=\sum_{j=1}^2 \kappa_E^j$,
$J_E^j$ the emission coefficient in the mode $j$ and $J_E=\sum_{j=1}^2
J_E^j$, while $Q_\mathrm{rad}$ and $\bm{F}_\mathrm{rad}$
describe, respectively, the energy 
and momentum 
 interchange between radiation and plasma. 

This system of equations is coupled to the stationary two-mode radiation transfer equation
\citep*[e.g.,][]{DolginovGS}
\begin{multline}
\label{IOm}
\bm{\Omega}\cdot\bm{\nabla}I^{j}_E=J_E^j-\kappa_E^j I_E^j
\\\quad
=\sum_{j'=1}^2 \int_0^\infty \dd E' \int_{(4\pi)}\dd\bm{\Omega}'
\big[R_{j'\rightarrow j}(E',\bm{\Omega}'\rightarrow 
E,\bm{\Omega})I_{E'}^{j'}(\bm{\Omega}')
\\
-R_{j\rightarrow j'}(E,\bm{\Omega}\rightarrow 
E',\bm{\Omega}')I_{E}^{j}(\bm{\Omega)}\big],
\end{multline}
where
\beq
  R_{j \rightarrow j'}(E,\bm{\Omega}\rightarrow E',\bm{\Omega}')
  = \nel\, \frac{\dd \sigma_{j'j}
       }{
       \dd E' \dd\bm{\Omega}'}
      ([f_\mathrm{e}]; E,\bm{\Omega},E',\bm{\Omega}')
\eeq
is the redistribution function of photons 
in the polarization state $j$, propagating with energy $E$ 
in the direction $\bm{\Omega}$, 
into the polarization state $j'$ with energy $E'$
 and propagation direction $\bm{\Omega}'$.

Deceleration of the accreting matter is determined by the radiative
(integral) terms in equations (\ref{rhov2}) and (\ref{rhov3}). In this
work we neglect all radiation processes except scattering,
hence the redistribution function $R_{j \rightarrow j'}
(E,\bm{\Omega}\rightarrow E',\bm{\Omega}')$ takes only this process
into account. Nevertheless it has a complex form, because the 
scattering cross sections depend strongly on the photon energy,
polarization and propagation direction.

The system of radiation hydrodynamics equations
(\ref{rhov1})--(\ref{IOm}) should be completed by the plasma equation of
state. We assume that electrons and protons have the same temperature
$T$ and apply an ideal gas equation of state, which is a good
approximation at the {considered} XRP conditions, because $\kB T\sim$ a few keV is
much greater than the hydrogen binding energy at $B\lesssim10^{13}$~G.
Since electrons at $E_B\gg\kB T$ constitute a one-dimensional gas with
the adiabatic index $\gamma=3$, while protons at
$E_{B,\mathrm{i}}\ll\kB T$
constitute the usual three-dimensional non-relativistic gas with
$\gamma={5/3}$, the total adiabatic index for a fully ionized hydrogen
plasma is $\gamma=2$. However, as mentioned in Paper~I, the adiabatic 
index as well as the gas pressure itself is unimportant for our model
calculations, because the accretion flow
is mainly controlled by the radiation pressure and
gravity, rather than by the plasma pressure.

%%%%%%%%%%%%%%%%%%%%%%%%%%%%%%%%%
\section{Numerical model}
\label{sec:num_mod}
 
\subsection{Numerical scheme}

To solve numerically the system of equations (\ref{rhov1})--(\ref{IOm}), 
we use time discretization and apply the operator 
splitting method 
(see, e.g, \citealt{LeVeque_2002}). 
Each time step is divided into three  substeps, described below.

%%%%%%%%%%%%%%%%%%%%%%%%%%%
\subsubsection{Hydrodynamical substep}
\label{sec:hydro_rad}

At the first substep we solve the hydrodynamical 
system of equations (\ref{rhov1})--(\ref{rhov3}), neglecting the 
radiative terms $Q_\mathrm{rad}$ and $\bm{F}_\mathrm{rad}$.
To solve this simplified system, we employ the open 
library VH-1 (`Virginia Hydrodynamics-1', 
\citealt{ColellaWoodward84}).%
\footnote{\href{http://wonka.physics.ncsu.edu/pub/VH-1/}{http://wonka.physics.ncsu.edu/pub/VH-1/};
it was also employed in the simulations of supercritical 
XRPs by \citet{2016PASJ...68...83K}.} 
The numerical grid 
is organized as a cylinder of radius $R_\mathrm{c}$ and height $H$, 
divided into $N_z$ 
slices of equal height ($N_z=100$ in the examples shown in Section~\ref{sec:num_res}).
The value of $H$ in each simulation is chosen sufficiently large 
to exceed the height of the zone where radiation pressure is significant, 
so that at the height $H$ above the NS surface the calculated plasma 
velocity $v$ almost coincides with the free-fall velocity.
Although our model allows two-dimensional hydrodynamical 
calculations, in the present work we neglect variations of all quantities 
in the hydrodynamic equations along the radius of the cylinder, 
because we have found in Paper~I that the profiles of these
quantities do not significantly vary along the radial coordinate.     
In the circle geometry there is only one 
cylindrical boundary with a radius $R_\mathrm{c}$. 
In the ring geometry, there are external 
and internal cylindrical boundaries, 
the latter having the radius 
$R_\mathrm{c}-d_\mathrm{w}$, 
where $d_\mathrm{w}$ is the width of the cylindrical wall.
Since the plasma moves along $\bm{B}$,
the hydrodynamical problem is reduced to one dimension.

\subsubsection{Radiative substep}
\label{sec:rad_transfer}

At  the second substep the system (\ref{IOm}) 
of polarized radiation transfer equations is solved using the Monte Carlo method
and the terms $Q_\mathrm{rad}$ (\ref{Qrad}) and 
$\bm{F}_\mathrm{rad}$ (\ref{Frad}) are computed. 
A detailed description of our Monte Carlo code 
can be found in \citet{2022PhRvD.105j3027M}.
We {summarize the main elements} here.

In the simulations, individual seed energy packets (`photons') 
are injected into the accretion channel at its bottom, 
which is associated with a hot polar cap at the NS
surface (details of their generation are described 
in Section \ref{sec:bounds}). 
Each photon packet moves in the accretion channel and scatters on the electrons. 
Density and velocity profiles in the channel 
are taken from the previous substep. 
The probability of a scattering event is computed 
according to the cross sections (\ref{totcrsec}),
which are
interpolated from precalculated tables. 

If a scattering has occurred, 
the probability that the photon 
goes into a certain direction 
is found from an interpolated 
cumulative distribution function 
$f_\mathrm{i \rightarrow f}$ 
of the probability that a photon with energy 
$E_\mathrm{i}$, 
polarization $j_\mathrm{i}$,
initial angle to the magnetic field 
$\theta_\mathrm{i}$ and initial azimuthal angle $\varphi_\mathrm{i}$
scatters 
to the angles
$\theta_\mathrm{f}$, $\varphi_\mathrm{f}$ 
and polarization $j_\mathrm{f}$, the  energy of the scattered photon being 
fixed to $E_\mathrm{f} = \hat{E}_\mathrm{f}(p_\mathrm{bulk})$ according to equations 
(\ref{E_f}) and (\ref{sigma_av_simple}).

The key ingredients of our Monte Carlo procedure are precalculated tables of scattering cross sections
$\sigma_{j'j}$ [equation (\ref{totcrsec1})],
converted into the distribution functions $f_\mathrm{i \rightarrow f}$, which depend on $\rho$, $T$ and $E_B$ as parameters. 
We treat these parameters as constants during 
each simulation, with values typical for accretion 
channels of subcritical XRPs. 
The tables are calculated for $p_\mathrm{bulk}=0$ (i.e., in the plasma reference frame) and used 
in the simulations together with the Lorentz transformation, analogous to equation (\ref{dsigma_transform}).
We work with the tables of $\sigma_{j'j}$ and
$f_\mathrm{i \rightarrow f}$ for all four combinations 
of the photon polarizations $(j,j')$ before and after scattering. 
The probability of scattering from mode $j$ into mode $j'$ equals $\sigma_{j'j}/\sigma_{j}$.

As can be seen from equations (\ref{Herold1})--(\ref{MaM}) and (\ref{M}), 
the scattering amplitude does not depend on the initial and final azimuthal angles separately, 
but only on their difference $\Delta\varphi = \varphi_\mathrm{f} - \varphi_\mathrm{i}$.
Moreover, using equations (\ref{dsigma_a_fi})--(\ref{dsigma_transform}) and (\ref{sigma_mean}), 
we can write the differential cross section in the general form
\beq
   \overline{\frac{\mathrm{d}\sigma_{j_\mathrm{f}j_\mathrm{i}}}
{\mathrm{d}\bm{\Omega}_\mathrm{f}}}
=
\sigma_\mathrm{d0} + \sigma_\mathrm{d1} \cos(\Delta\varphi) + \sigma_\mathrm{d2} \cos(2\Delta\varphi),
\label{sigma_dif}
\eeq
where the coefficients $\sigma_\mathrm{d0}$, $\sigma_\mathrm{d1}$ and 
$\sigma_\mathrm{d2}$ depend on $j_\mathrm{i,f}$, 
{$E_\mathrm{i}$} and $\theta_\mathrm{i,f}$, but not on $\varphi_\mathrm{i,f}$.
Therefore the integral cross section $\sigma_{j'j}$ 
{(\ref{totcrsec1})}
is independent of $\varphi_\mathrm{i}$ 
and the corresponding
tables are two-dimensional: one entry of each table is the initial angle $\theta_\mathrm{i}$ 
and the other is the initial photon energy $E_\mathrm{i}$. The full range of angles $\theta_\mathrm{i}\in[0,\pi]$ 
is divided in 100 equal steps. The 
{range of initial energies $E_\mathrm{i}$} is divided differently in three intervals
{with different arrays of $E_\mathrm{i}$ entries, adjusted empirically so as 
to reproduce with a sufficient accuracy the cross sections  near the resonance and at both its far wings.
In the benchmark case of $\kB T = 5$ keV, we set the intervals 
$[0.01E_B,0.5E_B]$, $[0.5E_B,1.5E_B]$ and $[1.5E_B,3E_B]$ with equidistant 250 table entries in 
$\log E_\mathrm{i}$ in the first and last of these intervals, and with} 500 entries in the middle, 
{distributed with a condensation towards $E_B$. For lower and higher $T$, 
we use analogous arrays of table entries with smaller and larger widths of the central interval, respectively.}

The cumulative distribution function
of photon scattering to directions defined by inequalities 
{$0 \leq \theta_\mathrm{f} \leq \theta$
(with $0 \leq \theta \leq \pi$; 
regardless of $\varphi_\mathrm{f}$)
can be written using \req{sigma_dif} as
\beq
f_\mathrm{j_\mathrm{f}j_\mathrm{i}}(E_\mathrm{i},\theta_\mathrm{i},\theta)
=
\frac{2\pi}{\sigma_{j'j}} \int\limits_0^{\theta}
\sigma_{\mathrm{d0},j_\mathrm{f}j_\mathrm{i}}(E_\mathrm{i},\theta_\mathrm{i},\theta_\mathrm{f})
\sin\theta_\mathrm{f}\,\dd\theta_\mathrm{f}.
\label{cumul}
\eeq 
At fixed polar angles $\theta_\mathrm{i,f}$, the cumulative distribution function
of photon azimuthal deflection by angle $\Delta\varphi \in [0,\varphi]$
(where $0 \leq \varphi < 2\pi$)
can be obtained from \req{sigma_dif} by integration over $\Delta\varphi$:
\begin{align}&
   \widetilde{f}_\mathrm{j_\mathrm{f}j_\mathrm{i}}
   (E_\mathrm{i},\theta_\mathrm{i},\theta_\mathrm{f},\varphi) =
  \frac{
  \sigma_{\mathrm{d0}}\,\varphi + 
  \sigma_{\mathrm{d1}} \sin\varphi
  +
  \sigma_{\mathrm{d2}} \sin(2\varphi) / 2
  }{2\pi\sigma_{\mathrm{d0}}}
  .
\label{cumul1}
\end{align}
}
Accordingly, we 
{determine the}
photon direction after scattering
in two steps: first we pick the final polar angle 
$\theta_\mathrm{f}$ from the cumulative distribution
{(\ref{cumul})}
and next we
take $\Delta\varphi$ from the cumulative
distribution
{(\ref{cumul1}).
Since the $\varphi$-dependence is analytical, we can use tables of 
$\sigma_{\mathrm{d0}}$, $\sigma_{\mathrm{d1}}$, $\sigma_{\mathrm{d2}}$ and $f$ with entries only 
in $E_\mathrm{i}$, $\theta_\mathrm{i}$ and $\theta_\mathrm{f}$ (for each of the four combinations of 
$j_\mathrm{i}$ and $j_\mathrm{f}$). The arrays of entries for $E_\mathrm{i}$ and $\theta_\mathrm{i}$ 
are the same as above, and the array for $\theta_\mathrm{f}$ is the same as for $\theta_\mathrm{i}$.
}

In Paper~I we found that in the case of the circle
geometry a typical scale of radiation penetration 
into the accretion channel is about 1--2~km for 
subcritical accretion regimes. 
This is $\sim10$--20\% of $R_\mathrm{NS}$, therefore the variation 
of the magnetic fields can be about a
few tens percent at this scale. Such variation could affect the position and width 
of the cyclotron line, as noted by \citet{2008ApJ...672.1127N} 
for supercritical pulsars. 
However, in our case of subcritical pulsars, 
resonance photons have much smaller penetration depths 
in the large-gradient plasma velocity zone 
near the NS
pole (see Section \ref{sec:plasma_dec}). 
Besides, in the case of the ring geometry, as we will see in
Section~\ref{sec:num_res}, 
{the total} height of the radiation-penetration zone
{in the channel
is only a few tens of metres}.
Therefore, the approximation 
of a constant $E_B$ 
should not introduce large errors in our case. On the other hand, $\rho$
and $T$, if determined self-consistently, are not
constant in the accretion channel,
but we apply this approximation for $f_\mathrm{i
\rightarrow f}$ to improve the numerical efficiency of the code.

\subsubsection{Energy-momentum exchange}
\label{sec:en_moment}

In Section~\ref{sec:hydro_rad} 
we neglected
the integral terms $Q_\mathrm{rad}$ and
$\bm{F}_\mathrm{rad}$ 
in equations (\ref{rhov2}) and
(\ref{rhov3}). Their contributions are restored 
at the third substep by solving the system of equations
\beq
\frac{\partial\rho}{\partial{t}}=0, 
\quad
\frac{\partial}{\partial{t}}\left(\rho\epsilon+
{\frac{\rho v^2}{2}}
\right)=Q_\mathrm{rad},
\quad
\frac{\partial\rho\bm{v}}{\partial{t}}=\bm{F}_\mathrm{rad},
\label{eq:rad_source}
\eeq
which is equivalent to neglecting contributions of gravity and spatial inhomogeneities
in equations (\ref{rhov1})--(\ref{rhov3}). We solve the system of equations (\ref{eq:rad_source})
using
the explicit Euler method, which is appropriate in the case of subcritical pulsars
(in the case of supercritical pulsars it would require too short time steps to be practical).

Instead of the integration in equations (\ref{Qrad}) and (\ref{Frad}),
we calculate the contributions for the energy and momentum exchange in
the course of Monte Carlo simulations at the preceding radiation substep
as follows. A simulated monochromatic packet of photons with
$\hbar\omega=E$ carries a total energy $E_\mathrm{p}$. If the photon
energy changes in a scattering from $E$ to $E^\prime$, the energy of the
packet changes to $E_\mathrm{p}^{\prime}=E_\mathrm{p}E^\prime/E$,
where $E'$ is calculated according to \req{E_f}, assuming that
$p_\mathrm{i}=p_\mathrm{bulk}$ (without thermal dispersion). The energy and
momentum transfer from radiation to matter in a
scattering event are, respectively, $\Delta E_\mathrm{p}=E_\mathrm{p}-E_\mathrm{p}^{\prime}$ and 
$\Delta p_\mathrm{p}= (E_\mathrm{p}\cos\theta_\mathrm{i}
-E_\mathrm{p}'\cos\theta_\mathrm{f})/c $.
Here we have taken into account that only the
longitudinal (along $\bm{B}$) momentum component 
is transferred to the plasma in a strongly quantizing magnetic field.
At each time step, $\Delta E_\mathrm{p}$ and $\Delta p_\mathrm{p}$
are summed up as cumulative totals in arrays for the respective spatial
grid cells, thus providing the total energy and momentum transferred from
radiation to matter
in each cell. 
Then $Q_\mathrm{rad}$ and $\bm{F}_\mathrm{rad}$ are
obtained by dividing these totals by the time step and the cell volume.

%%%%%%%%%%%%%%%%%%%%%%%%%%%%%%%%%%%%%%%%%%%%%%%%%%%
\subsection{Boundary conditions}
\label{sec:bounds}

We set the plasma inflow velocity at the top of the accretion channel 
$v_\mathrm{in}$ equal to the free-fall velocity
$v_\mathrm{ff}=
\sqrt{{2GM_\mathrm{NS}/(R_\mathrm{NS}+H)}}
\approx \sqrt{{2GM_\mathrm{NS}/R_\mathrm{NS}}}$. 
The plasma density at the top boundary
equals $\rho=\dot{M}/v_\mathrm{in}S_\mathrm{b}$, where $S_\mathrm{b}$ is the channel cross section
($S_\mathrm{b}=\pi R_\mathrm{c}^2$ in the circle geometry and $S_\mathrm{b}=
2\pi R_\mathrm{c}d_\mathrm{w} -\pi d_\mathrm{w}^2$ 
in the ring geometry).
The inflow plasma temperature $T_\mathrm{in}$
is taken to be effectively zero (for numerical reasons we set
$\kB T_\mathrm{in}=10^{-9}$~keV). 
Radiation can escape freely through the top
and side boundaries of the channel. 

The boundary conditions at the bottom of the channel are not so obvious.
In our model, the bottom is placed at the NS surface. Therefore the
proper bottom boundary conditions are determined by the physics of the
interaction of the accreting plasma and radiation with the NS surface or
the atmosphere, which is a complex, not completely resolved problem. The
details of plasma deceleration in the upper layers of a strongly
magnetized NS and the role of collective phenomena in this process are
not fully understood. The poorly known plasma behaviour in a strong
magnetic field at the base of the accretion channel affects the 
energy fraction radiated into the channel through this surface and the
characteristics of this radiation.

For example, collective modes in a two-temperature plasma can produce a
shock wave in the accretion channel
\citep{1969AZh....46..721B,1975ApJ...198..671S,1982ApJ...257..733L,2004AstL...30..309B}.
Besides, at sufficiently high accretion rates, thermal pressure can
exceed the magnetic pressure in the lower part of the accretion column
of a supercritical XRP, which results in a leakage of the plasma through
the cylinder walls \citep{1976MNRAS.175..395B,2023MNRAS.524.4148A}. We
will not consider such situations, which are unlikely to occur in the
subcritical XRPs.

We apply the zero-gradient conditions for hydrodynamical variables
$\bm{v}$, $\rho$ and $P$ at the bottom of the accretion channel and
assume that the total (kinetic and radiation) energy passed through its
lower boundary into the NS at a given time step is fully re-emitted from
the NS surface into the channel at the subsequent time step.

The next non-trivial problem is setting up the spectrum, angular
distribution and polarization of the radiation coming from the NS
surface into the accretion channel. 
The seed energy packets are emitted with equal energies
$E_\mathrm{p}$. We assume that each
packet is monochromatic and consists of $E_\mathrm{p}/E$ photons. 
We use the algorithm described in 
section 4.2.2 of \citet{2019LRCA....5....1N} to generate the 
blackbody spectrum of the seed radiation at the bottom of the accretion channel.
The effective temperature of this spectrum, $T_\mathrm{b}$ is determined 
by the energy balance condition: the energy emitted from the surface is equal to the sum of 
the kinetic energy of plasma at the surface and the energy 
of radiation that is carried into the surface by back-scattered photons.

In Paper~I we considered only
non-polarized radiation with black body spectrum. 
In the present work,
to test possible effects of polarization of the surface radiation in a
strong magnetic field, this model is supplemented by consideration of two 
limiting cases, where the surface radiation 
is generated entirely in one of the two NMs. 

In addition, 
we {also consider the possibility of using an arbitrary seed spectrum}.
For this purpose we
calculate in advance a table of a cumulative distribution function 
$F_\mathrm{ph}(E)$, which gives the probability that a seed photon has
an energy below a certain value $E$. The energy of a seed photon is
generated according to this distribution as 
$E=F_\mathrm{ph}^{-1}(\eta)$, where $\eta$ is a random number generated
with a uniform distribution in the interval $(0,1)$ and
$F_\mathrm{ph}^{-1}$ is the inverse function.
The power of the seed surface radiation is determined by the same energy 
balance condition as described above for the blackbody model.

In this way, in addition to the initial blackbody radiation spectrum, we test photon 
distribution functions $F_\mathrm{ph}(E)$
corresponding to the three best-fit spectral models for the
observed 
spectra of the subcritical X-ray pulsar 
1A~0535+262, presented by \citet{2019MNRAS.487L..30T}
for  three different accretion states of  this  XRP 
with luminosities 
$L_1 = 7\times10^{34}$~erg~s$^{-1}$,
$L_2 = 6\times10^{35}$~erg s$^{-1}$ and
$L_3 = 2\times10^{36}$~erg~s$^{-1}$,
with the cumulative photon distribution functions
$F_1 (E)$, $F_2(E)$ and $F_3(E)$, respectively.

Obviously, using observed spectra as initial ones
is not self-consistent, so we consider
it only as a test of possible alternatives to using 
the blackbody initial spectrum. In particular,
this test model includes cyclotron features 
(which are present in the spectra of XRP 1A~0535+262)
into the seed photon distribution. 

%%%%%%%%%%%%%%%%%%%%%%%%%%%%
\subsection{Remarks on mode conversion}

%%%%%%%%%%%%%%%%%
\begin{figure}
	\centering 
	 \includegraphics[width=.9\columnwidth]{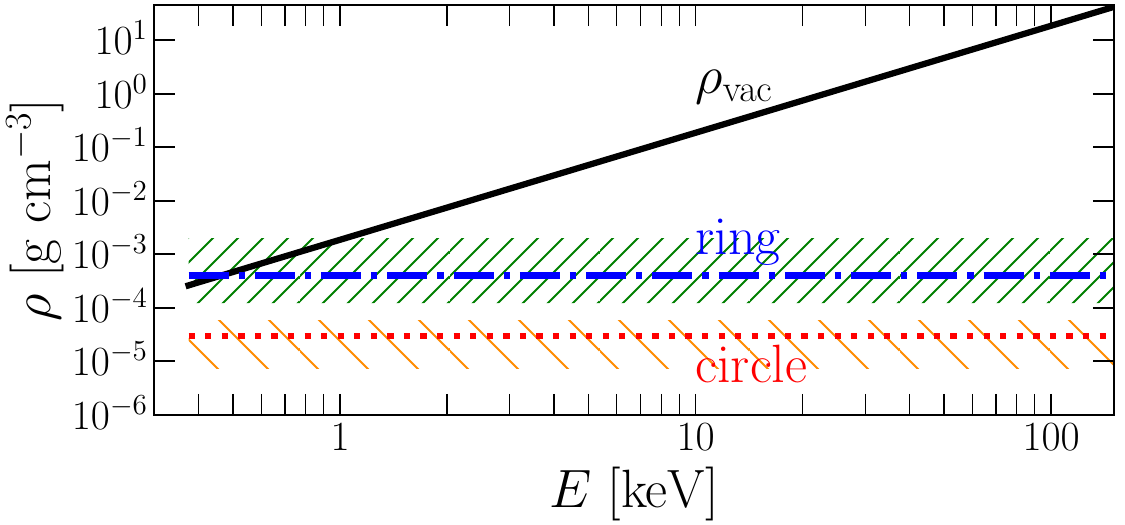}
 	\caption{{Dependence of the mass density $\rho_\mathrm{vac}$, corresponding to the vacuum resonance, 
on photon energy (solid line) compared with the characteristic plasma densities considered in this work. 
The hatched bands cover the densities encountered in our numerical calculations in the cases of circle 
(lower band) and ring (upper band) geometries.
The horizontal lines mark the densities 
$\rho = 4\times10^{-4}$ g cm$^{-3}$ and $3\times10^{-5}$ g cm$^{-3}$, 
at which the cross sections (Fig.~\ref{f:cr_sec}) have been calculated.}
        }
\label{fig:rho_E_beta}
\end{figure}
%%%%%%%%%%%%%%%%%%%%%%%%%%%%%%%%%%%%

As noticed by \citet{1979JETP...49..741P}, a 
linear transformation of NMs
can occur if NM photons propagate 
in an inhomogeneous medium. 
As noted in Section~\ref{sec:polarNM}, the plus-mode is ordinary, 
if the polarization parameter $\beta$ is positive, 
and it is extraordinary, if $\beta < 0$, and vice versa for the minus-mode. 
According to \req{beta-W}, the sign of $\beta$ is the same as the sign of $\cos\theta_B$ at $E > E_B$, whereas at $E < E_B$
it is so only if the NM polarization is dominated by the plasma, but the sign is opposite if it is dominated by the QED vacuum.

Since the vacuum resonance energy $E_\mathrm{vac}$ 
decreases
with 
decreasing
density, 
a photon with a given energy $E < E_\mathrm{vac}$ can encounter the vacuum resonance 
at a density $\rho_\mathrm{vac}$ where
$E\approx E_\mathrm{vac}$. 
If the density variation is sufficiently gentle, 
the polarization of a photon evolves adiabatically, preserving 
the electric vector rotation direction (plus-mode or minus-mode),
but changing the direction of the major axis of the rotation ellipse relative to the magnetic field, 
which means a conversion between the O-mode and X-mode \citep{2002ApJ...566..373L}.
According to equation (\ref{Evac}), $E=E_\mathrm{vac}$ at
$
  \rho_\mathrm{vac} \approx 10^{-4}B_{12}^2E_1^2\text{ g cm}^{-3}.
$

Although such densities are attained in our simulations
{at the lowest considered energies, as illustrated in Fig.~\ref{fig:rho_E_beta}}, we neglect the mode 
conversion effect, because the accretion channel 
is almost transparent for radiation at the relevant photon energies.
Let us demonstrate it with simple estimates.

{In all simulations presented below,}
we take $E_B =50$~keV, which corresponds to $B=4.319\times10^{12}$~G. 
Mass density can be calculated as $\rho=\dot{M}/(|v|S_\mathrm{b})$.
The typical subcritical accretion rates considered in our work are  $\dot{M}=(10^{15}-10^{16})$ g s$^{-1}$. 
As we will see in Section~\ref{sec:num_res}, the typical plasma velocity in accretion channels 
of subcritical pulsars is about $|v|=0.2-0.5c$, so we will use $|v|=0.5c$ for estimates.
A typical radius of an accretion channel is $R_\mathrm{c}\sim 0.5$~km (see, e.g., estimates 
in \citealt{2015MNRAS.454.2539M} and references therein) and its thickness 
is $d_\mathrm{w}\sim 10$~m in the case of ring geometry. Using these estimates, 
we obtain $\rho\lesssim3\times10^{-4}$ g cm$^{-3}$ for the circle geometry 
and $\rho\lesssim10^{-2}$ g cm$^{-3}$ for the ring geometry. 
In both cases $E_\mathrm{vac} \ll E_B $, therefore around the cyclotron 
line the normal mode polarization is dominated by vacuum.

\begin{figure}
	\centering 
	 \includegraphics[width=.8\columnwidth]{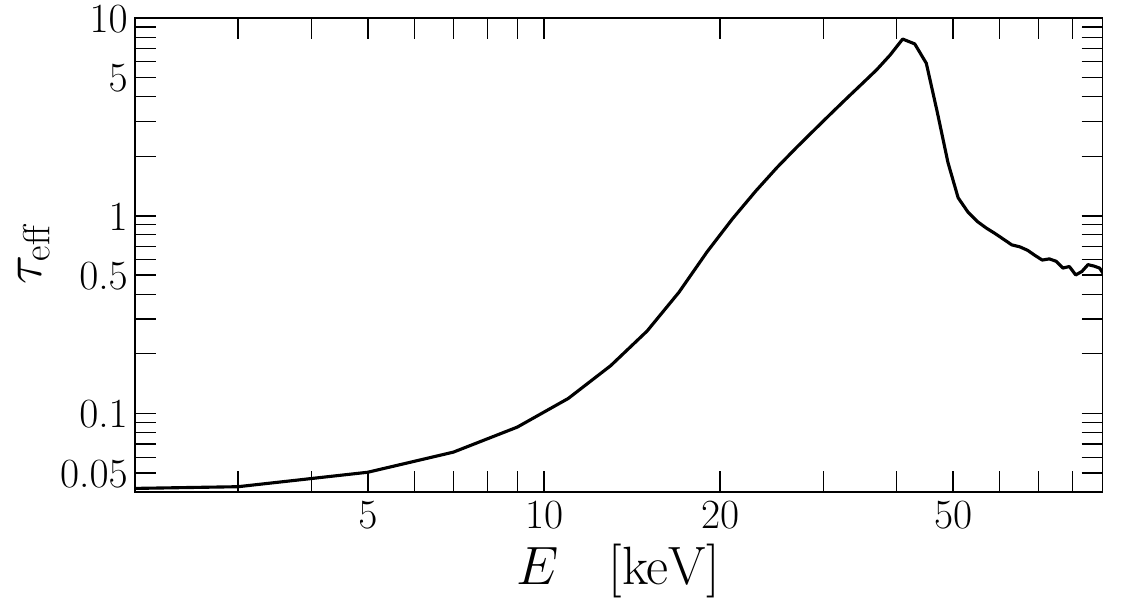}
 	\caption{Effective optical depth,
averaged over photon directions $0 < \theta_B < \theta_\mathrm{max}$
as a function of photon energy 
  for  the ring-shaped base of the accretion channel with 
  radius $R_\mathrm{c} = 0.5$~km and thickness $d_\mathrm{w} = 15$~m at
 the mass accretion  rate  $\dot{M} = {2.5}\times10^{15}$~g
 s$^{-1}$ and cyclotron energy
$E_B = 50$~keV.
The scattering cross sections have been computed at mass density 
$\rho=4\times10^{-4}$ g~cm$^{-3}$ and electron temperature $\kB T = 5$~keV.
}
\label{Pic:tau}
\end{figure}

Let us define an effective optical depth $\tau_\mathrm{eff}=-\ln(N_\mathrm{ns}/N_\mathrm{tot})$, 
where $N_\mathrm{tot}$ is the total number of photons with energy $E$, 
injected in the accretion channel, and $N_\mathrm{ns}$ is the number of 
injected photons with the same energy which leave the accretion channel 
without scattering. Thus $\tau_\mathrm{eff}$ is the characteristic of the scattering region.
Figure~\ref{Pic:tau} shows an example of $\tau_\mathrm{eff}(E)$ from our 
calculations (discussed in more detail in the next section) at 
$\dot{M}={2.5\times}10^{15}$ g s$^{-1}$,
$\kB T=5$~keV and $E_B=50$~keV, for the ring 
geometry with $R_\mathrm{c}= 0.5$~km an $d_\mathrm{w} = 15$~m. 
The height of the computational domain is
$H=40$~m; the underlying tables for 
interpolation of cross sections and photon redistribution functions  
have been calculated with account of the vacuum polarization, 
assuming $\rho=4\times10^{-4}$~g cm$^{-3}$
(the vacuum polarization parameter $W \simeq  10^4$).
{Note that the maximum of $\tau_\mathrm{eff}(E)$ is redshifted from $E_B$. 
This redshift is caused by the Doppler effect due to the bulk plasma 
velocity, discussed in more detail below.}
It can be seen from Fig.~\ref{Pic:tau} that 
$\tau_\mathrm{eff}\ll1$, if $E\ll E_B$.
Therefore, at $E\sim E_\mathrm{vac} \simeq E_B/\sqrt{W}$
most of the photons pass through the accretion 
channel without scattering.   
Under this condition, the mode conversion 
can only change the sign of the polarization but not its magnitude.
Moreover, it is clear that for an unpolarized initial radiation (which will be considered 
in all but one examples presented below) the mode conversion 
has no effect under the considered conditions.

%%%%%%%%%%%%%%%%%
%%%%%%%%%%%%%%%%%%%%%%%%%%%%%%%%%%%%

%%%%%%%%%%%%%%%%%%%%%%%%%%%%

\section{Numerical results}
\label{sec:num_res}

In this section we present results of numerical simulations. In all the
simulations, we have adopted the surface gravity $g=1.29\times10^{14}$
cm s$^{-2}$, which corresponds to $M_\mathrm{NS}=1.4 M_\odot$ and
$R_\mathrm{NS}=12$ km in the non-GR approach used here (or to
$M_\mathrm{NS}=1.4 M_\odot$ and $R_\mathrm{NS}=13.2$ km with the GR corrections). 
We will present steady-state results, to which the
time-dependent solutions converge at the end of each simulation. In
fact, our model does not allow us to trace the time evolution 
accurately, since we use the stationary radiative transfer equations
(\ref{IOm}) in order to accelerate the calculations. We have checked 
that our stationary solutions remain stable in the
non-stationary calculations with the term $c^{-1}{\partial I_E^j/\partial
t}$ added to the left-hand side of \req{IOm}.

\subsection{Plasma deceleration}
\label{sec:plasma_dec}

%%%%%%%%%%%%%%%%%
\begin{figure}
	\centering 
	\includegraphics[width=.9\columnwidth]{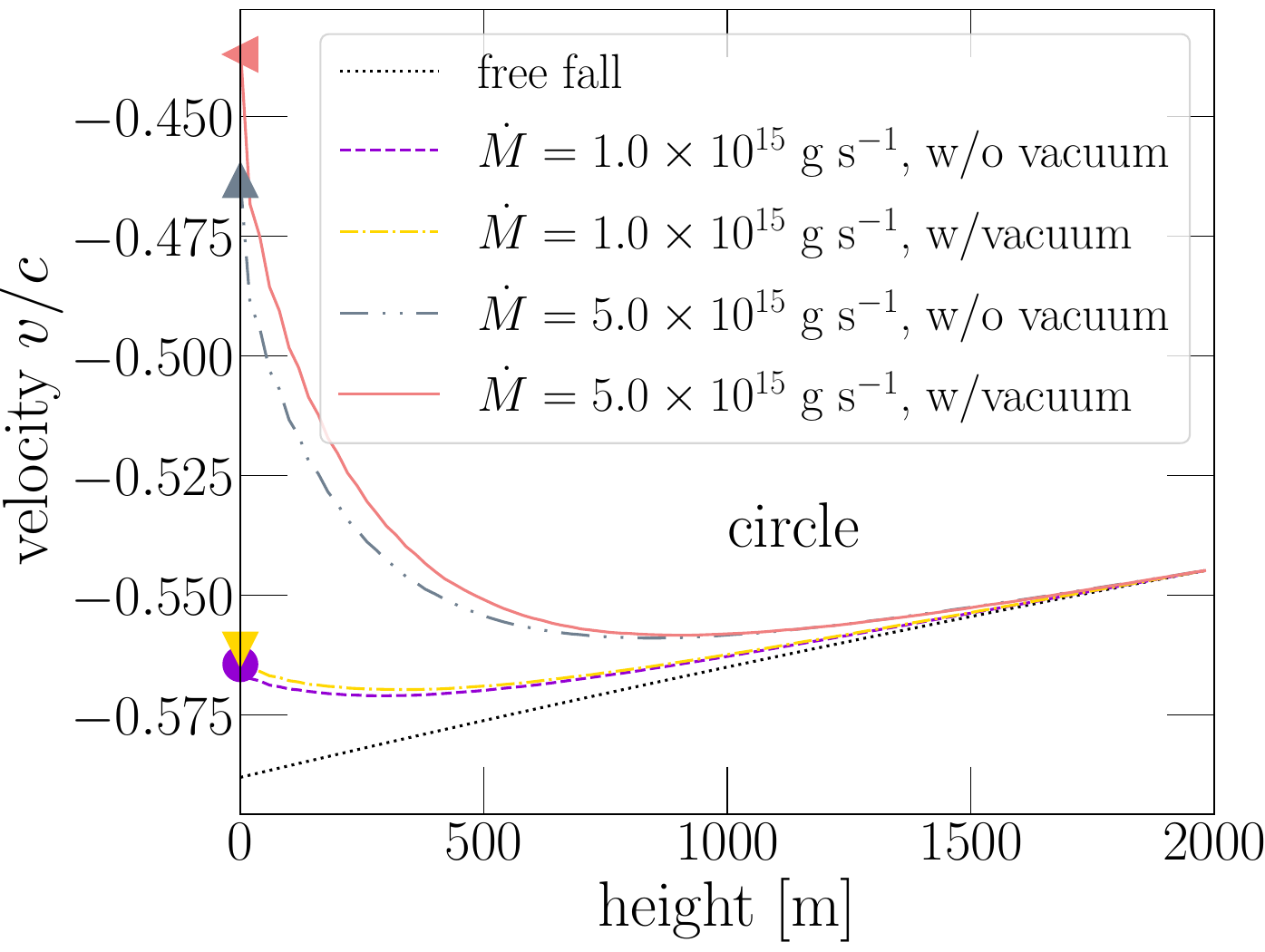} 	
  	\caption{Plasma velocity $v$ in units of $c$ as a function of
     height above the hot polar cap of a NS emitting the blackbody
     radiation 
in  the circle geometry with
$R_\mathrm{c} = 0.5$~km.
The negative sign of $v$ means motion toward the surface. 
The dotted 
line corresponds to the free-falling plasma; the dashed line corresponds to the accretion
rate $\dot{M}=10^{15}$ g s$^{-1}$ and electron
plasma birefringence; the dashed-dotted line corresponds to the same
$\dot{M}$ with vacuum polarization birefringence; 
the dash-double-dot line and the solid line show the cases
without and with the vacuum polarization effects, respectively, for
$\dot{M}=5\times10^{15}$~g s$^{-1}$.
The symbols of respective colours at the left vertical axis 
visualize the $v/c$ values attained at the NS surface
(see Table~\ref{tab:velo0}).
The boundary condition at the base of the accretion channel is the
non-polarized blackbody radiation with the effective temperature
$T_\mathrm{b}$ obtained from the energy balance 
condition
(see section~\protect{\ref{sec:bounds}}).
The scattering cross sections have been computed at 
$\rho={3\times10^{-5}}$ g~cm$^{-3}$ and $\kB T = 5$~keV.}
\label{f:velo_vac_pl}
\end{figure}
%%%%%%%%%%%%%%%%%%%%%%%%%%%%%%%%%%%%

%%%%%%%%%%%%% 
\begin{figure}    
  \centering
  \includegraphics[width=.9\columnwidth]{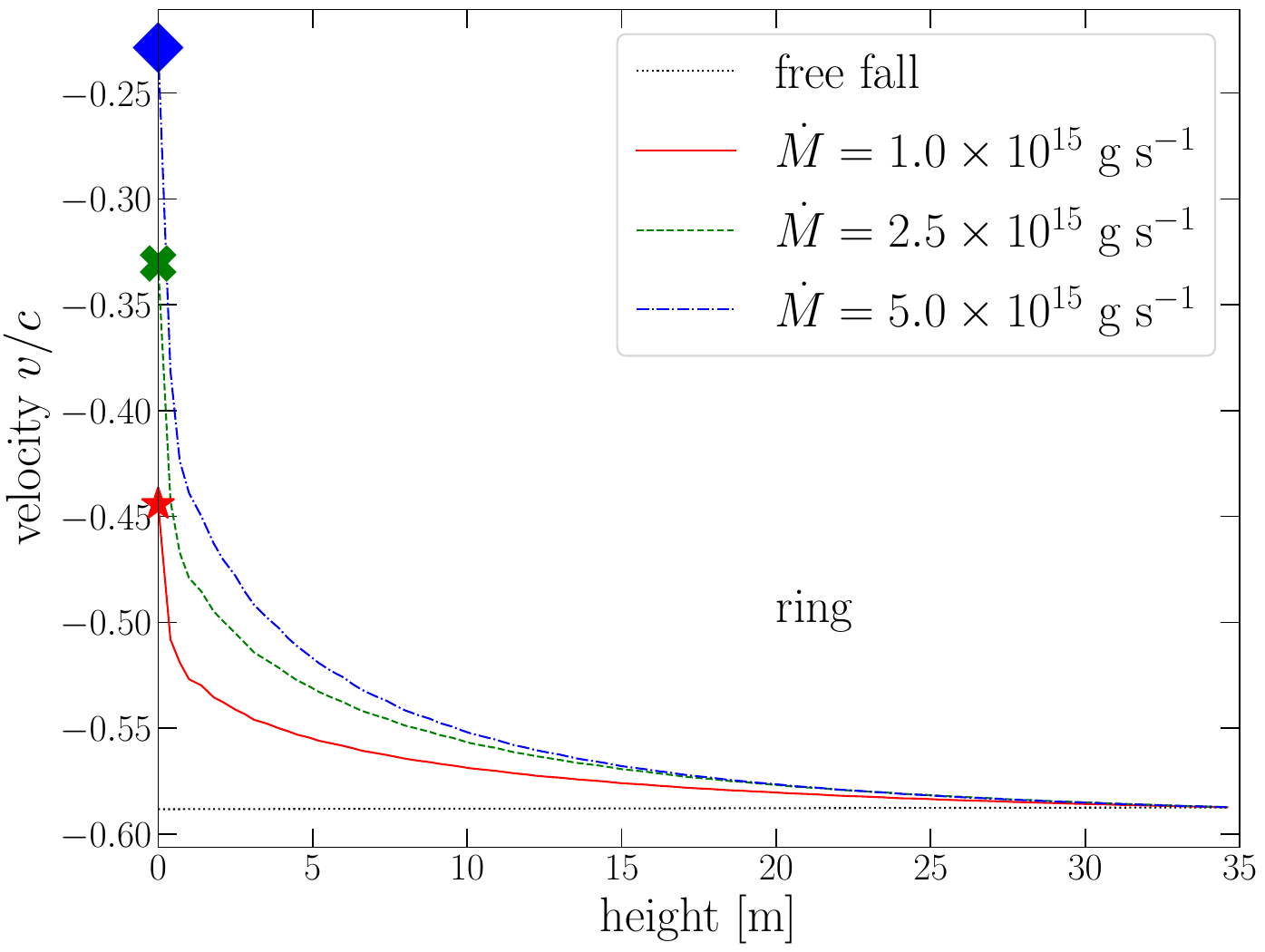}
  \caption{Same as in Fig.~\ref{f:velo_vac_pl} but for the ring
  geometry with
the same model setup
as in Fig.~\ref{Pic:tau}.
The solid line corresponds to $\dot{M}=10^{15}$~g s$^{-1}$, the 
dashed one to $\dot{M}=2.5\times10^{15}$~g s$^{-1}$, the 
dash-dotted line to $\dot{M}=5\times10^{15}$~g s$^{-1}$, the 
dotted one corresponds to the free fall.
Vacuum polarization is taken into account.
}
\label{f:velo_vs_dMd}
\end{figure}
%%%%%%%%%%%%%%%%%%%%%%%%%%%%%%%%%%%%%

\begin{figure}
  \centering
  \includegraphics[width=.9\columnwidth]{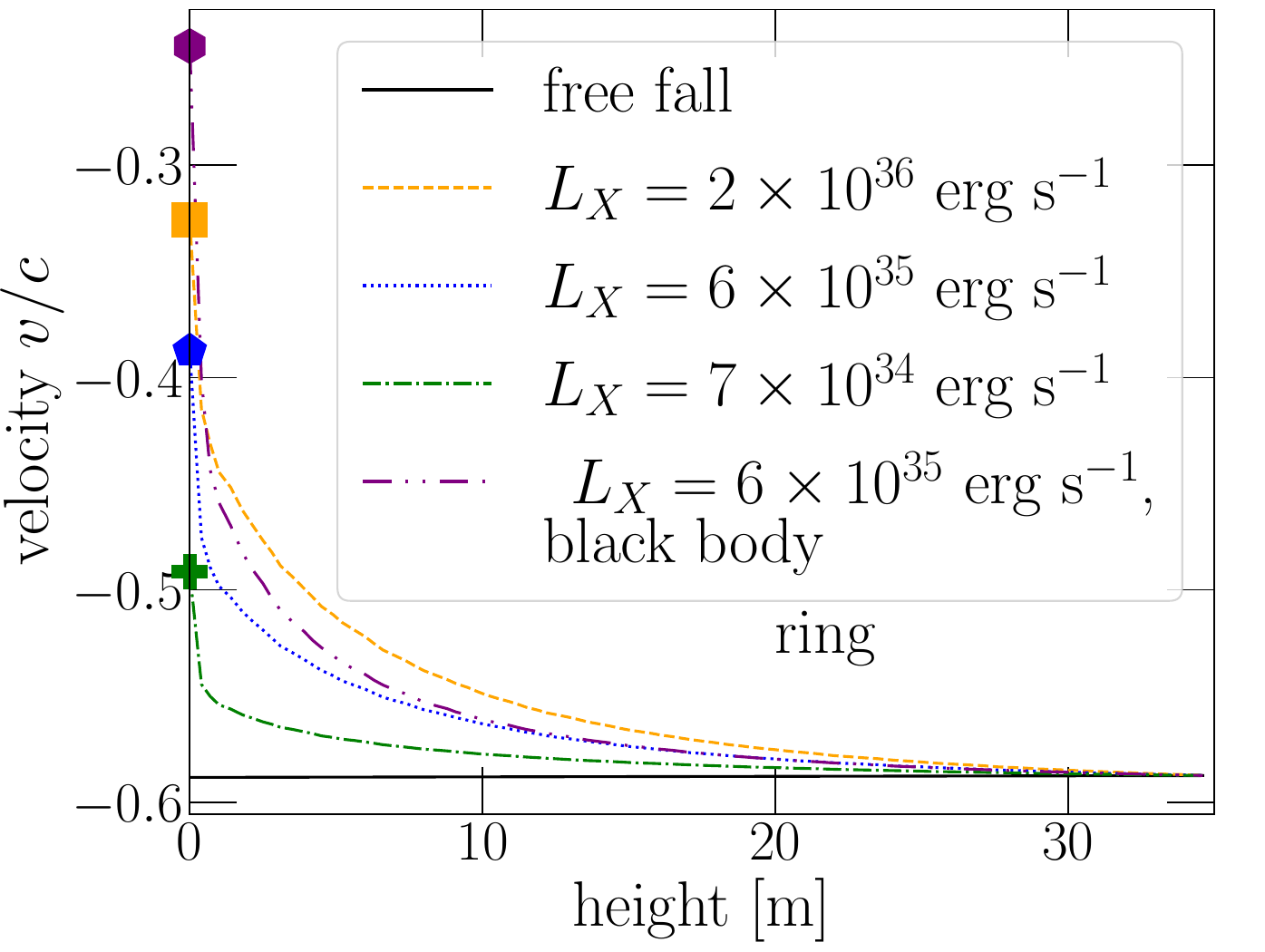} 
  \caption{Same as in Fig.~\ref{f:velo_vs_dMd} but for
  complex (non-blackbody) spectra of radiation from the NS surface, taken to be the fits 
  by \citet{2019MNRAS.487L..30T} to the observed spectra of XRP 1A~0535+262 
  for the states with X-ray luminosities 
$L_1=7\times10^{34}$~erg s$^{-1}$, $L_2=6\times10^{35}$~erg s$^{-1}$ and
$L_3=2\times10^{36}$~erg s$^{-1}$, with
 $R_\mathrm{c}$ and $d_\mathrm{w}$ determined from equations~(\ref{eq36}) and~(\ref{eq37}).
The dash-dotted, dotted and dashed lines correspond to
the simulated spectra from the accretion 
channel for these three cases, respectively. For comparison, the 
dash-double-dot
line corresponds to $L_\mathrm{X}=6\times10^{35}$~erg s$^{-1}$ with a
blackbody spectrum from the base of the channel and
the black solid line corresponds to the free fall.
}
\label{f:velo_A_true}
\end{figure}

\begin{figure}  
 	\centering 
	\includegraphics[width=.9\columnwidth]{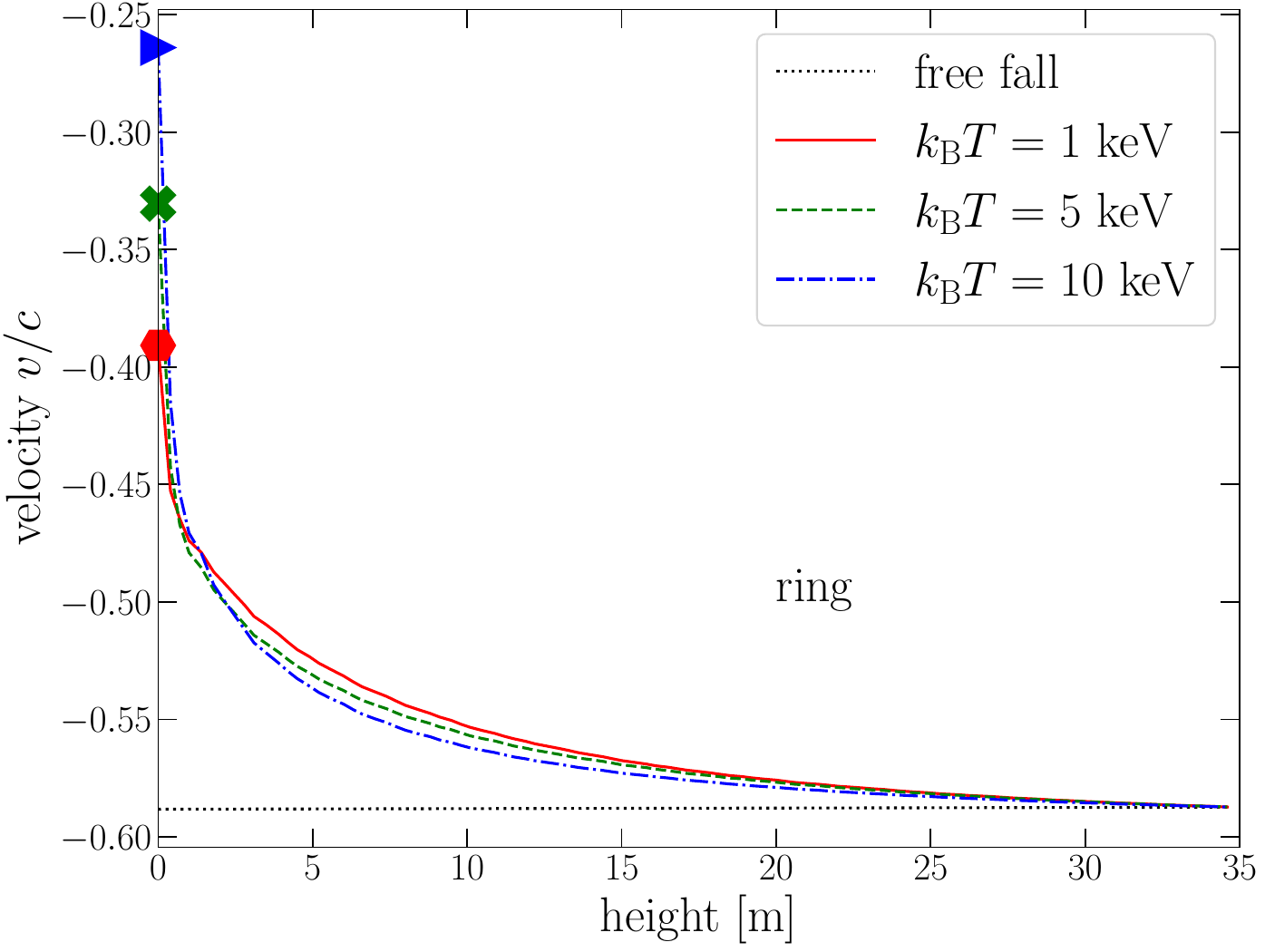} 	
      \caption{Same as in Fig.~\ref{f:velo_vs_dMd} at
$\dot{M}=2.5\times10^{15}$~g s$^{-1}$ but for different electron temperatures
$T$ used in the cross-section interpolation tables: $\kB T=1$~keV (solid line), $\kB T=5$~keV (dashed line) and $\kB T=10$~keV (dash-dotted line).
}
\label{f:velo_compare_T}
\end{figure}

In Figs.~\ref{f:velo_vac_pl}--\ref{f:velo_compare_T} 
we consider the influence of different factors on
plasma deceleration in an accretion channel. 
The higher the mass accretion rate, the higher the deceleration.

Figure~\ref{f:velo_vac_pl} displays bulk plasma velocity profiles in
the circle geometry with the same radius and magnetic field, with 
and without taking the vacuum
polarization effects
into account, for accretion rates
$10^{15}$ g s$^{-1}$ and $5 \times 10^{15}$~g s$^{-1}$.
{For comparison, the free-fall velocity is also shown.} 
The height of the accretion channel is higher 
for the circle geometry than for the ring geometry. 
Accordingly, in this case the computational domain has been increased to $H=2$~km 
and the interpolation tables have been calculated at a smaller density 
$\rho={3\times10^{-5}}$~g cm$^{-3}$. 
{Although the dipole field $B$ should vary by a factor $\sim 1.6$ 
along this height, we neglect this variation 
because of computational-resource limitations
(the resulting inaccuracy is not expected to be significant, as argued
in Section~\ref{sec:rad_transfer})}.
For both accretion rates, the plasma deceleration in the accretion channel is
stronger when the vacuum polarization is taken into account, although
the difference is not large. 

Figure~\ref{f:velo_vs_dMd} 
{provides a clearer illustration} of
the influence of the mass accretion rate on the deceleration of the
accreting plasma. 
Compared with Fig.~\ref{f:velo_vac_pl},
the filled (circle geometry) accretion
channel is replaced by the hollow one (i.e., ring geometry). In this
case, it is sufficient to use a relatively small height $H=35$~m. The
vacuum polarization effects are included in consideration. The plasma
velocity is shown as a function of the height above the NS surface for
accretion rates $\dot{M} = 10^{15}$~g s$^{-1}$, $2.5 \times 10^{15}$~g
s$^{-1}$ and $5 \times 10^{15}$~g s$^{-1}$. The free-fall velocity is
also shown as a reference line.  We see that the increase of 
$\dot{M}$ leads to a noticeable enhancement of the radiative braking
of the plasma.

%%%%%%%%%%%%%%%%%%%%%%%%%%%%
%%%%%%%%%%%%%%%%%%%%%%%%%%%%%%%%%%%%%%%

%%%%%%%%%%%%%%%%%%%%%%%%%%%%%%%%%%%%%%%%%%%%%%%%%%
\begin{table}
\centering
\caption{Plasma velocity (in units of speed of light) at the NS surface for the accretion models 
considered in Figs.~\ref{f:velo_vac_pl}--\ref{f:velo_A_true}: the circle or ring geometry of the accretion 
channel without (the `plasma' model) or with (the `vacuum' model) account of vacuum polarization 
effects. For Figs.~\ref{f:velo_vac_pl} and~\ref{f:velo_vs_dMd} (lines 2--8), 
the blackbody emission from the hot spot was assumed. 
For Fig.~\ref{f:velo_A_true} (lines 9--12), the ring geometry with vacuum polarization effects 
is assumed and either a complex model of the initial spectrum at the surface 
(`1A 0535+262', see text; lines 9--11) or the blackbody initial spectrum (line 12) is used.
The free-fall velocity is given in the first line for comparison.
}
\label{tab:velo0}.
\begin{tabular}{clccc}
\hline
Fig. & model &
$\dot{M}$ [$10^{15}$ &  $L_\mathrm{X}$  [$10^{35}$ & $v/c$  
\\
no. & & g\,s$^{-1}$] & erg\,s$^{-1}$]
\\
\hline
\ref{f:velo_vac_pl}--\ref{f:velo_A_true} & free fall & any  &  any  & $-0.587$
\\
\ref{f:velo_vac_pl} & circle, plasma  &  $1.0$  & $1.55$ & $-0.564$
\\
\ref{f:velo_vac_pl} & circle, vacuum  &  $1.0$  & $1.55$ &  $-0.561$
\\
\ref{f:velo_vac_pl} & circle, plasma  &  $5.0$  & $7.74$ & $-0.463$
\\
\ref{f:velo_vac_pl} & circle, vacuum  &  $5.0$  & $7.74$ & $-0.435$
\\
\ref{f:velo_vs_dMd} & ring, vacuum  &  $1.0$  & $1.55$ & $-0.444$
\\
\ref{f:velo_vs_dMd} & ring, vacuum  &  $2.5$  & $3.87$ & $-0.331$
\\
\ref{f:velo_vs_dMd} & ring, vacuum  &  $5.0$  & $7.74$ & $-0.229$
\\
\ref{f:velo_A_true} & 1A 0535+262  &  $0.452$  & $0.7$ & $-0.491$
\\
\ref{f:velo_A_true} & 1A 0535+262  &  $3.88$  & $6.0$ & $-0.387$
\\
\ref{f:velo_A_true} & 1A 0535+262  &  $12.9$  & $20$ & $-0.326$
\\
\ref{f:velo_A_true} & blackbody  &  $3.88$  & $6.0$ & $-0.244$
\\
\hline 
\end{tabular}
\end{table}
%%%%%%%%%%%%%%%%%%%%%%%%%%%%%%%%%%%%%%%%%%%%%%%%%%

Figure~\ref{f:velo_A_true} illustrates an influence of the radiative
lower boundary conditions on the plasma deceleration. For this purpose,
along with the blackbody radiation from the NS surface, we also consider
the models $F_1$, $F_2$ and $F_3$ that fit the spectrum of XRP
1A~0535+262 in three different states \citep{2019MNRAS.487L..30T},
mentioned in Section~\ref{sec:bounds}. For comparison, the free-fall
plasma velocity is also shown. In this set of simulations, the radius
$R_\mathrm{c}$ and thickness $d_\mathrm{w}$ of the accretion channel
have been determined self-consistently
from the relations for disc accretion with a
low rate \citep{2007ARep...51..549S}:
\begin{align}&
  \label{eq36}
  S_\mathrm{b}=6.6\times10^9\Lambda^{-7/8}L_{39}^{2/5}B_{12}^{-1/2}\left(\frac{M}{M_\odot}\right)^{\!\!-13/20}R_6^{19/10}\text{cm}^2,
\\&
  \label{eq37}
  2\pi\frac{R_\mathrm{c}}{d_\mathrm{w}}=95\Lambda^{-1/8}L_{39}^{-4/35}B_{12}^{-1/14}\left(\frac{M}{M_\odot}\right)^{71/140}R_6^{19/10},
\end{align}
where 
$L_{39}=L_\mathrm{X}/(10^{39}~\text{erg~s}^{-1}$,
$R_6=R_\mathrm{NS}/(10^6~\text{cm})$
and $\Lambda$ is a dimensionless
correction parameter, which we set equal to 0.5, as appropriate for the
ring geometry. In the calculations we have set $E_B=50$~keV and obtained the 
ring radius $R_\mathrm{c}=349$~m and width $d_\mathrm{w}=7$~m for
the case $F_1$; $R_\mathrm{c}=474$~m
and $d_\mathrm{w}=12$~m for $F_2$; $R_\mathrm{c}=563$~m and
$d_\mathrm{w}=16$~m for $F_3$.
The height of the computational domain was $H=35$~m.

The strong differences in the deceleration curves are primarily
caused by the different number of photons with energies close to the
cyclotron resonance. These photons mainly control the deceleration of
the accretion flow due to the resonant peak in the cross sections. For
a fixed luminosity, the presence of an absorption line in the seed
spectrum causes a lack of resonance photons and suppresses the braking,
as seen from a comparison of the curves in Fig.~\ref{f:velo_A_true},
obtained for $L=L_2$ with the blackbody and non-blackbody boundary
conditions.

Note the large difference of height scales in 
Figs.~\ref{f:velo_vac_pl} and~\ref{f:velo_A_true}. 
It corresponds to much higher density
in the ring-shaped geometry than in the circle geometry at similar
mass accretion rates. In the circle geometry, the typical height values, at
which the deceleration mostly occurs, are $z \sim R_\mathrm{c}$
while in the ring geometry they are typically $z\sim d_\mathrm{w}$, 
both values corresponding to approximately the same optical depth
\citep[cf.][]{2015MNRAS.447.1847M}.

The final values of the velocities at the NS surface for the model settings 
used in Figs.~\ref{f:velo_vac_pl}--\ref{f:velo_A_true} are listed in Table~\ref{tab:velo0}. 
We have verified the robustness of these values with respect 
to numerical code parameters (such as the mesh size).
In these simulations, we used the tables of the scattering cross sections 
calculated at typical densities $\rho={3\times10^{-5}}$~g cm$^{-3}$ for the circle geometry or
$\rho=4\times10^{-4}$~g cm$^{-3}$ for the ring geometry and at 
fixed electron temperature $\kB T=5$~keV.
Figure~\ref{f:velo_compare_T} demonstrates the influence of the value 
of the latter parameter on the velocity profiles at fixed
$\dot{M} = 2.5\times10^{15}$~g s$^{-1}$ in the ring geometry.
Although this dependence is not very strong
({when we vary $\kB T$ between 1 keV and 10 keV, the velocity at the NS surface varies within tens per cent})
we can see a stronger deceleration of the 
accretion flow near the NS surface at higher temperatures.
It is explained by a larger Doppler broadening of the cyclotron resonance and 
accordingly a wider involvement of photons in the resonant scattering. 

%%%%%%%%%%%%%%%%%%%%%%%%%%%%
%%%%%%%%%%%%%%%%%%%%%%%%%%%%%%%%%%%%%%

%%%%%%%%%%%%%%%%%%%%%%%%%%%%%%%%%%%%%%%%%%%%%%%%%
\begin{figure*}
  \centering
 
\includegraphics[width=\textwidth]{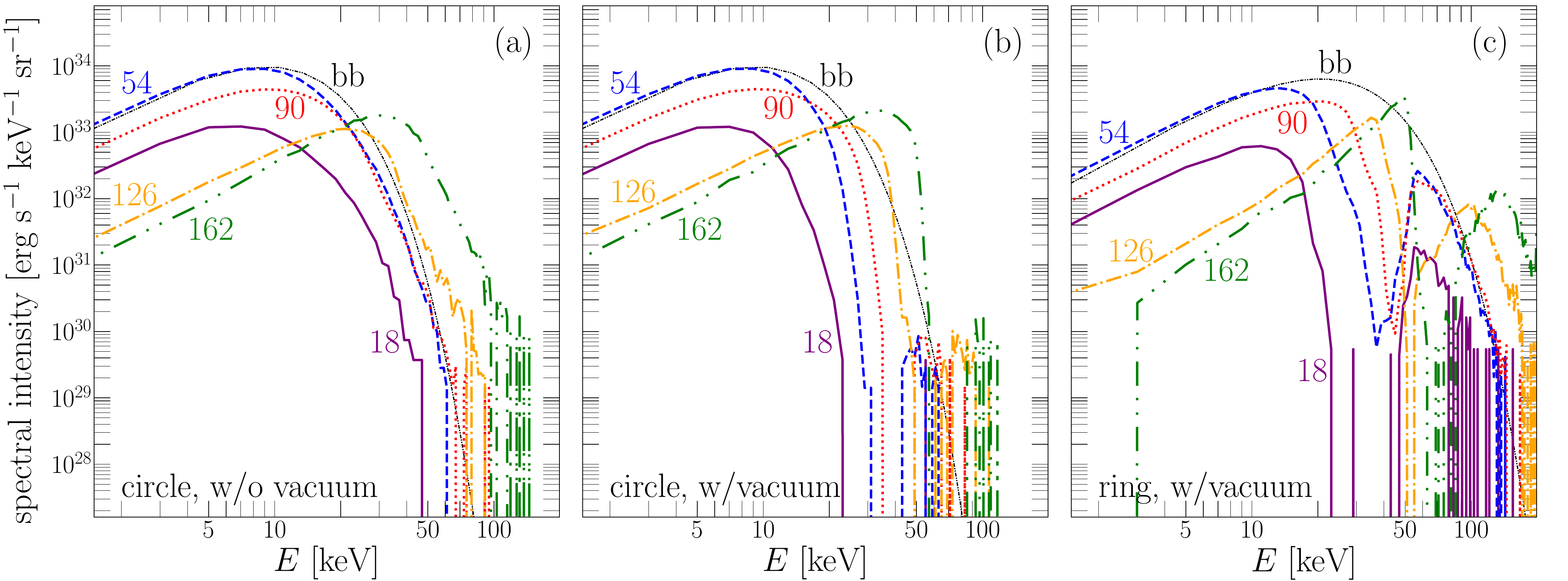}
  \caption{Integrated intensity (i.e., spectral flux per unit solid angle) of radiation outgoing from an accretion 
  channel at different angles $\vartheta$ to the channel axis
  as a function of photon energy $E$ for the mass accretion rate 
  $\dot{M}=5\times10^{15}$~g s$^{-1}$ at the electron cyclotron energy $E_B=50$~keV for three models: 
  (a) the circle geometry and the collisionless plasma approximation neglecting vacuum polarization, (b)
  the circle geometry and the more accurate description 
  of polarization according to \req{xi}, including the vacuum polarization effect
  and (c) the ring geometry with using 
  the same description of the vacuum polarization effects,
  as in panel (b).
  The values of $\vartheta$ are marked near the curves:
  $\vartheta=18^\circ$ (solid line), 
  $54^\circ$ (dashed line),
  $90^\circ$ (dotted line),
  $126^\circ$ (dot-dashed line)
  and $162^\circ$  (double-dot-dashed line).
  The thin 
  double-dot-dashed curve marked `bb' represents the blackbody integrated intensity
  emitted from the hot spot at the effective surface temperature 
  $\kB T_\mathrm{b} = 3.3$~keV and 7.3~keV in the cases of circle (a, b) and ring (c) geometries, respectively. 
{The model parameters are the same as in Fig.~\ref{f:velo_vac_pl} for the circle 
geometry (panels a,b) and the same as in Fig.~\ref{f:velo_vs_dMd} for the ring geometry (panel c).}
}
  \label{fig:spectra_angle}
\end{figure*}
%%%%%%%%%%%%%%%%%%%%%%%%%%%%%%%%%%%%%%%%%%%%%%%%%%

At high accretion rates (near the transition to the supercritical regime) we observe oscillations 
of the plasma density and velocity in the accretion channel, 
which could not be damped away in our calculation. These oscillations can be a sign of the incipient 
radiative-dominated shock, although we cannot exclude the possibility that they   
are an artifact due to the limited computational time. In the figures we have averaged plasma 
characteristics over the computational time in the cases where the oscillations are observed 
(the {dash-dotted}
line  in Fig.~\ref{f:velo_vs_dMd} and the {dash-double-dot} line in Fig.~\ref{f:velo_A_true}).

The increase of the deceleration with increasing temperature affects the cyclotron feature and polarization, 
which will be described in Sections \ref{sec:spectra}{\,--\,}\ref{sec:pol}. 
Therefore the variation of $T$ in the calculations of interpolation tables affect 
quantitatively the results of our calculations. At contrast, we have checked that a moderate 
variation of the plasma density $\rho$ in these calculations does not noticeably 
affect the results, which justifies using a constant $\rho$ in the tables.

%%%%%%%%%%%%%%%%%%%%%%%%%%%%%%%%%%%%%%%%%%%%%%%%%%%%%%
\subsection{Pulsar spectra}
\label{sec:spectra}

Let us consider how different factors affect the radiation from an accretion channel. 
First we consider the energy spectra for the sum of the two NMs.

Figure~\ref{fig:spectra_angle} shows examples of calculated spectral flux per unit solid angle 
{(equal to the specific intensity integrated over
a surface perpendicular to $\bm{k}$),
averaged over a small range $\delta\vartheta$
(we have set $\delta{\vartheta}=\pi/50$)
around several chosen angles $\vartheta$ between $\bm{k}$ and $\bm{n}$ for outgoing photons:}
\beq
J=\frac{1}{\delta{\vartheta}}\int\limits_{\vartheta-\delta\vartheta/2}^{\vartheta+\delta\vartheta/2}\frac{\mathrm{d}Q_E(\vartheta')}{2\pi\sin{\vartheta'}\dd\vartheta'\mathrm{d}E\mathrm{d}t}\,\mathrm{d}\vartheta',
\eeq
as functions of the photon energy $E$. 
Here, 
$\mathrm{d}Q_E(\vartheta)$
is the energy of radiation emitted by the accretion channel into the interval 
at angles $\vartheta\pm\dd\vartheta/2$ and photon energies $E\pm\mathrm{d}E/2$
{during time interval}
{$\mathrm{d}t$}.
The interpolation tables have been calculated at $\kB T=5$~keV and the typical densities 
$\rho={3\times10^{-5}}$ g cm$^{-3}$ for the circle geometry and $4\times10^{-4}$ g cm$^{-3}$ 
for the ring geometry. To evaluate the spectral intensities and energy fluxes as functions 
of photon energy, the energy range from 0 to 200~keV has been divided into 100 equal intervals, 
and the wave packets, which were traced in the Monte Carlo simulations, have been counted 
separately in each interval to give the flux estimate at its centre.

The left panel (a) of Fig.~\ref{fig:spectra_angle} presents the spectra calculated 
in the simplified approximation (\ref{xi0}), which does not take into account plasma
collisions and vacuum polarization, for the circle geometry, as in Paper~I.
The middle (b) and right (c) panels show the results of calculations 
with using the more accurate expression (\ref{xi}) for the circle 
and ring geometry, respectively.
{The increase of numerical noise towards high energies, 
which is observed in this and subsequent figures, 
is related to the sharp decrease of the Wien tail of the spectrum, 
which results in an insufficient number of wave packets in the Monte Carlo simulations 
at such energies.}

%%%%%%%%%%%%%%%%%%%%%%%%%%%%%%%%%%%%%%%%%%
\begin{figure}
	\centering 

\includegraphics[width=.9\columnwidth]{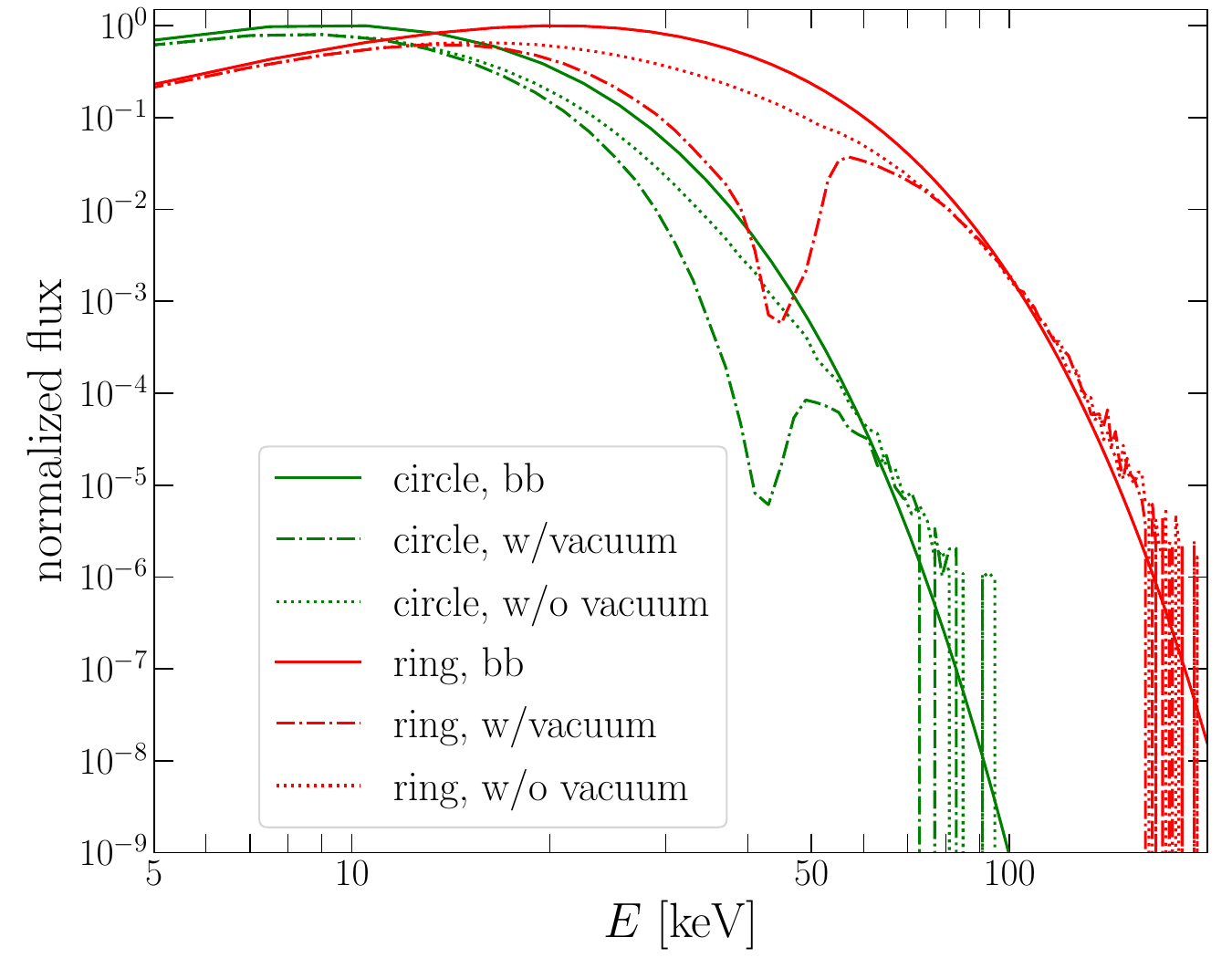}
     \caption{Normalized spectral flux into the upper 
     hemisphere ($\vartheta\leq90^\circ$) as a function of photon energy 
     for the circle (green
     curves with maxima at lower $E$) 
     and ring (red curves with maxima at higher $E$)
     geometries of the accretion channel with the same parameters as 
     in Fig.~\ref{fig:spectra_angle}, including
     the accretion rate $\dot{M}=5\times10^{15}$~g s$^{-1}$.
    The cases with (dot-dashed lines) and without (dotted lines) 
    account of the vacuum polarization are shown.
    For each channel geometry, the normalized blackbody  spectral flux from the surface
     is shown by the solid line.
\label{Pic:Spectra_fhcr}
}
\end{figure}
%%%%%%%%%%%%%%%%%%%%%%%%%%%%%%%%%%%%%%%%%%%%%%%%%%

It can be seen that the spectra in Figure~\ref{fig:spectra_angle} become progressively redshifted 
with decreasing $\vartheta$. In panels (b) and (c), the absorption feature is also shifted to $E < E_B$.
Qualitatively, this can be explained by the Doppler effect at the resonant 
scattering (in the electron's rest frame)  of softer photons ($E < E_B$) propagating
in the direction opposite to the flow movement
($\vartheta \leq \pi/2$),
and harder photons ($E > E_B$) propagating in the direction of the flow movement
($\vartheta > \pi/2$). The scattered photons predominantly leave 
their initial spectral distribution (in the NS frame), 
forming visible absorption features.
{Consequences of this effect for the beaming pattern are discussed in \citet{Markozov_26beaming}.}
When taking into account 
the vacuum polarization, the
cyclotron feature becomes more pronounced,%
\footnote{The opposite result in Paper I
was caused {by} an accidental numerical error.}
although the specific form of the features depends heavily on the chosen parameters.

Figure~\ref{Pic:Spectra_fhcr} highlights 
the difference between spectra for the circle and ring geometries. 
It shows the spectral flux into the upper hemisphere ($\vartheta < 90^\circ$). 
For each channel geometry, the curves are normalized to the maximum of the initial blackbody 
spectral flux from the surface, with effective temperatures 
$\kB T_\mathrm{b}=3.3$~keV in the case of circle geometry and {7.3}~keV in the case of ring geometry.
The same numerical parameters and interpolation tables 
as in Fig.~\ref{fig:spectra_angle} have been used.
The spectra obtained with taking the vacuum polarization into account 
are compared with the results obtained in the approximations of Paper~I, 
neglecting vacuum polarization. 
In the latter case, however, the hydrodynamic structure (the density and velocity profiles)
have been fixed the same as for the case where the vacuum polarization 
is taken into account, in order to highlight specifically the radiation transfer effects.
We see that the vacuum polarization leads to the appearance of a strong cyclotron feature.
{This} feature is located at a slightly higher
energy for the ring geometry than for the circle geometry.
{In general,  its position and shape depend heavily on
parameters of the seed radiation, as well as}
on the mass accretion rate, 
as will be discussed in Section~\ref{sec:cyclo}.

%%%%%%%%%%%%%%%%%%%%%%%%%%%%%%%%%%%%%%%%%%%%%%%%%%%%%%%%%%%%%%%
\subsection{Dependence of cyclotron feature on luminosity}
\label{sec:cyclo}

The position of the cyclotron absorption feature in the spectrum of some XRPs varies
with X-ray luminosity 
\citep[see][and references therein]{2022arXiv220414185M,2019A&A...622A..61S}.
In supercritical XRPs (including V0332+53, see \citealt{2006MNRAS.371...19T}, 
and 4U~0115+63, see \citealt{2007AstL...33..368T}), where the luminosity $L_\mathrm{X} \gtrsim 10^{37}\,{\rm erg\,s^{-1}}$, 
a cyclotron line centroid energy shows a negative correlation with $L_\mathrm{X}$.
In some subcritical XRPs (including Her~X-1, see \citealt{2007A&A...465L..25S}, 
and GX~304-1, see \citealt{2012A&A...542L..28K,2017MNRAS.466.2752R}), 
where $L_\mathrm{X} \lesssim 10^{37}\,{\rm erg\,s^{-1}}$, a positive correlation between 
the line centroid energy and luminosity was reported.
Recently, it was found that the negative correlation arises again 
at luminosity level well below the critical one \citep{2021ApJ...919...33C,2024MNRAS.528.7320S}.

To explain the relation between the accretion luminosity and cyclotron line centroid energy, 
different authors have proposed models accounting for changes in the geometry 
of a line forming region and physical condition{s} there 
(see Section~4.4.3 in \citealt{2022arXiv220414185M}, and references therein).
In the case of subcritical accretion, \citet{2015MNRAS.454.2714M} proposed 
a mechanism explaining the variation of the observed cyclotron-line energy with 
luminosity, attributing it to the Doppler effect 
within the accretion channel of a subcritical pulsar. 
The authors adopted an approximate plasma velocity profile dependent 
on the accretion rate and evaluated the magnetized Compton opacity 
averaged over the channel, together with the resulting centroid energy of the cyclotron feature. 
Their calculations showed that the cyclotron line centroid 
exhibits a systematic blueshift with increasing accretion rate.

However, the previous authors
did not conduct a full radiation hydrodynamical modelling and used many simplifications. 
Therefore, it is of interest to verify 
the  explanation
discussed above using 
the present calculations
of magnetic radiation transfer coupled to hydrodynamics.
The results of the calculations show that 
the
blueshift of the feature
for directions opposite
to the flow motion ($\vartheta < \pi/2$),
at $E < E_B$, 
can be qualitatively  explained by a decrease of the averaged Doppler effect
 with increasing the  deceleration of the accretion flow
under the influence of increasing  radiation  luminosity  
emerging from the base of the accretion channel.

\begin{figure}
\centering 
 \includegraphics[width=.9\columnwidth]{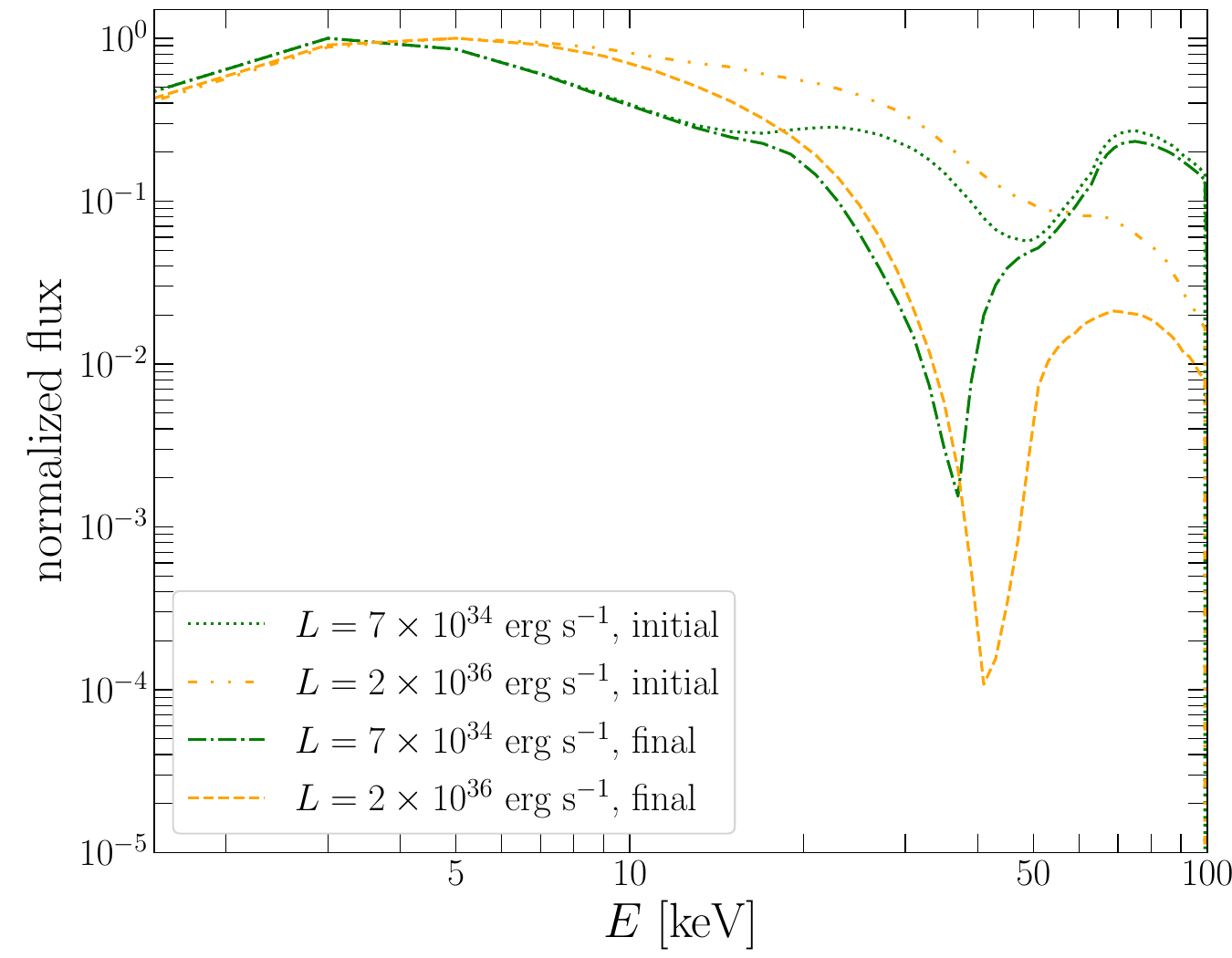} 
	\caption{Radiative flux into the upper hemisphere ($\vartheta\leq90^\circ$) as a function of photon energy 
    for the same model spectra of radiation from the NS surface as in Fig.~\ref{f:velo_A_true} 
    for the states with the lowest and highest luminosities: $L_\mathrm{X}=L_1$
    (dot-dashed curve) and $L_\mathrm{X}=L_3$
    (dashed curve).
  The initial model spectra from the NS surface are also shown 
  (dotted ($L_1$)
  and double-dot-dashed ($L_3$) curves, respectively).
{The parameters $R_\mathrm{c}$ and $d_\mathrm{w}$ are chosen consistently 
with the accretion luminosity according to equations (\ref{eq36}), (\ref{eq37}).}
\label{fig:flux_A_true}
}
\end{figure}

We check the influence of {the} Doppler effect on the observed 
cyclotron energy by applying our code to the model of a complex initial spectrum, 
corresponding to the subcritical XRP 1A~0535+262 as fitted by \citet{2019MNRAS.487L..30T}. 
Its accretion channel structures in the states $F_1$, $F_2$ and 
$F_3$ with luminosities $L_1=7\times10^{34}$~erg~s$^{-1}$, $L_2=6\times10^{35}$~erg~s$^{-1}$ and 
$L_3=2\times10^{36}$~erg~s$^{-1}$ have been shown in Fig.~\ref{f:velo_A_true}. 
Figure~\ref{fig:flux_A_true} shows the flux into the upper hemisphere 
as a function of photon energy, obtained by the Monte Carlo radiative transfer 
simulations for the models $F_1$ and $F_3$.
{The dot-dashed line}
corresponds 
to $L_\mathrm{X}=7\times10^{34}$~erg~s$^{-1}$,
{while the dashed line corresponds to 
$L_\mathrm{X}=2\times10^{36}$~erg~s$^{-1}$}
Other parameters of the simulations are $T=5$~keV, $\rho=4\times10^{-4}$ g cm$^{-3}$, $R_\mathrm{c}=500$~m, 
$d_\mathrm{w}=15$~m, $E_B=50$~keV. 
The initial model spectra of radiation from the hot spot 
(the seed spectra) are also shown in the figure.
This seed radiation has been assumed unpolarized.

%%%%%%%%%%%%%%%%%%%%%%%%%%%%%
%%%%%%%%%%%%%%%%%%%%%%%%%%%%%%%%%%%%%%%%%%%%

We see that the spectra outgoing from the accretion channel reveal 
sharp and deep cyclotron lines, unlike the shallow features in the seed spectra. 
The cyclotron line centroid energy is higher for the larger accretion rate.

Qualitatively similar results were obtained by \citet{Fotiadis_26}, who performed Monte Carlo unpolarized 
radiative-transfer simulations of cyclotron-line formation in subcritical X-ray pulsars 
in the electron plasma (without QED vacuum) approximation, using prescribed velocity 
and density profiles in the filled (circle geometry) accretion funnel. 
Here we have extended the treatment by coupling the radiative transfer to the hydrodynamic 
structure of the flow and by including polarized normal-mode transfer together with vacuum-polarization effects.
{In particular, in Fig.~\ref{Pic:Spectra_fhcr} we observe no significant absorption feature in the polarization-averaged 
outgoing spectrum in the case of the plasma approximation, 
but there appear strong cyclotron features when vacuum polarization effects are included. 
It is because only in the latter case both cross sections for the X- and O-modes display 
a strong cyclotron resonance (see Fig.~\ref{f:cr_sec}).}

Besides, it should  be noted
that the cyclotron line position and shape can be also affected 
by some other factors, which are not currently included in our model, such as variations 
of temperature and magnetic field over the height above the NS
surface (see, e.g., \citealt{2008ApJ...672.1127N}, for a discussion).

% ---------------------------------------------
\subsection{Linear polarization}
\label{sec:pol}

One of the key features of our model is its ability to provide not only the energy spectrum, 
but also the polarization by self-consistent radiation hydrodynamical simulation. 
The degree of linear polarization with respect to the magnetic field direction is
\begin{equation}
  \label{eq34}
  P_L=\frac{I_\mathrm{O}-I_\mathrm{X}}{I_\mathrm{O}+I_\mathrm{X}}P_\mathrm{O},
\end{equation}
where $I_\mathrm{O}$ is the O-mode intensity, $I_\mathrm{X}$ is the X-mode intensity and
\begin{equation}
  \label{eq35}
  P_\mathrm{O} = \frac{|K_\mathrm{O}|^2-1}{|K_\mathrm{O}|^2+1}
  = \frac{1-|K_\mathrm{X}|^2}{1+|K_\mathrm{X}|^2}
  = - P_\mathrm{X}
\end{equation}
is the linear polarization of the O-mode,
$K_\mathrm{O}=K_2$ and $K_\mathrm{X}=K_1$ being the O- and X-mode 
polarization parameters given by \req{xi}. 

\begin{figure}
	\centering 
  \includegraphics[width=.9\columnwidth]{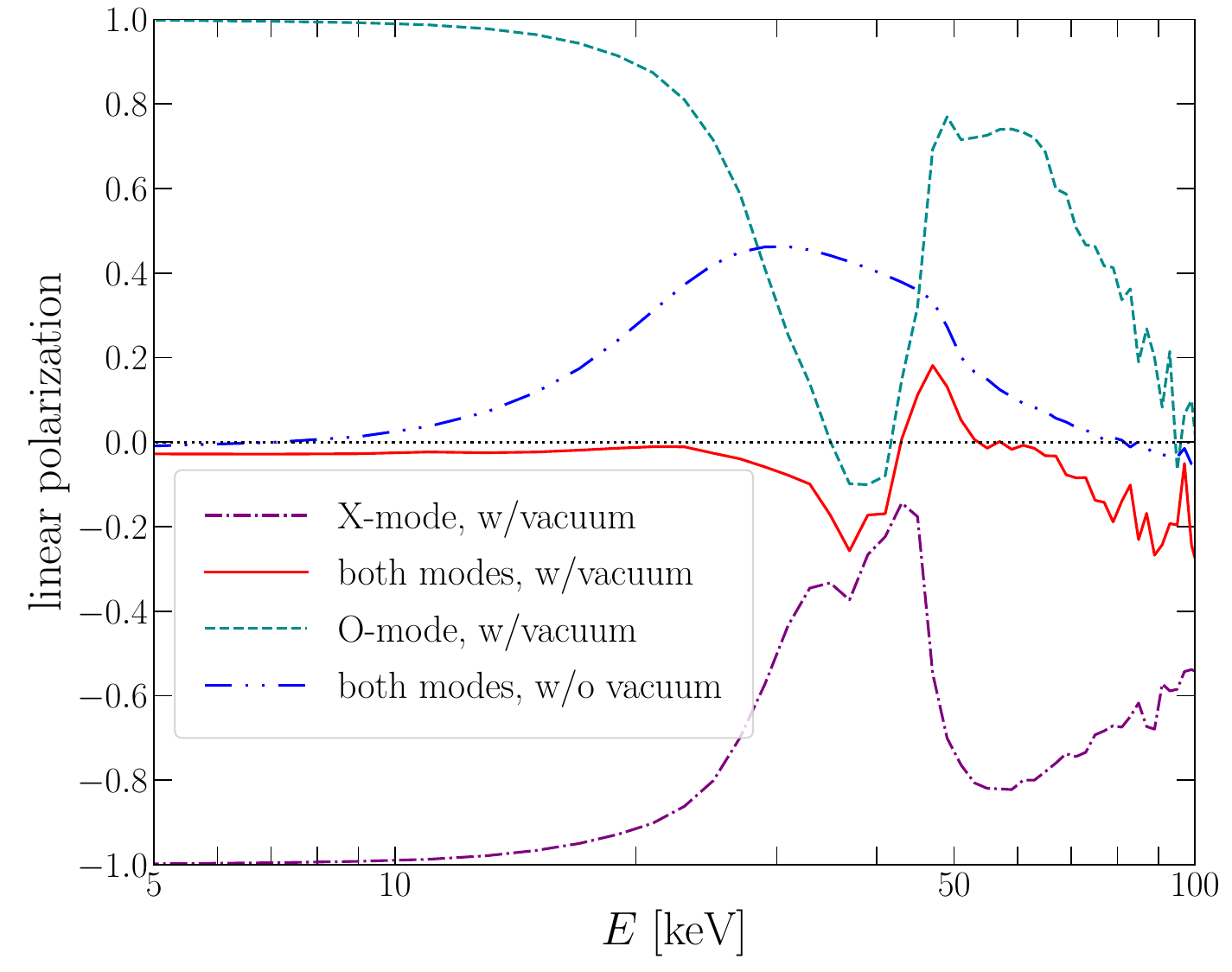} 	
	\caption{Linear polarization degree $P_L$ as a function of photon energy, 
    averaged for radiation emitted into the upper hemisphere ($\vartheta\leq90^\circ$)
 {(with weights proportional to the total flux in each direction $\vartheta$)
for the ring geometry with  $R_\mathrm{c}=500$~m, $d_\mathrm{w}=15$~m 
and accretion rate $\dot{M}=2.5\times10^{15}$~g s$^{-1}$.}
The calculation of $P_L$ including vacuum 
  polarization effects (solid line) is compared with the cold 
  collisionless plasma 
  approximation neglecting the vacuum polarization (double-dot-dashed line), {assuming} unpolarized radiation from the NS surface.
  The {dot-dashed} (X-mode) and {dashed} (O-mode) curves show the results 
  of analogous calculations in the cases of purely elliptical
  hot-spot surface polarization, where the seed photons are injected in the channel in only one of the NMs,
  {with the vacuum polarization effects taken into account}.
\label{Pic:P_l_sum_first}
}
\end{figure}

Figure~\ref{Pic:P_l_sum_first} presents $P_L$ as a function of photon energy 
$E$ in different model approximations.
{The simulations were performed for the ring 
geometry with $R_\mathrm{c}=0.5$~km, $d_\mathrm{w}=15$~m and accretion rate 
$\dot{M}=2.5 \times 10^{15}$ g s$^{-1}$.
Interpolation tables were calculated with
$\rho=4\times10^{-4}$~g cm$^{-3}$ and $\kB T=5$~keV.}

{First,}
we compare the results 
obtained in the approximation for the polarization parameter given by \req{xi0}, 
which does not take into account polarization of vacuum {(double-dot--dashed line)}, with more accurate results {(solid line)}
obtained with taking the vacuum polarization into account according to 
equations~(\ref{xi}), (\ref{beta}).
{These two sets of simulations were performed}
assuming that the non-polarized blackbody radiation is emitted from the NS surface. 
Despite the surface radiation has been assumed unpolarized, 
we observe
{significant}
linear polarization $P_L$ of the outgoing radiation
{at sufficiently high energies, $E \gtrsim 20$~keV}.
It means that the accretion channel substantially contributes 
into the polarization at such photon energies even in the subcritical XRPs.
{We also observe an important effect of the vacuum polarization. 
In the
approximation where the vacuum polarization is neglected,
 $P_L$ 
is positive and rather large (reaching $\sim{40}$\%) in a wide energy band
close to
$E_B$.
If the vacuum polarization effects are taken into account, then $P_L$ 
is also
non-zero
at $E\sim E_B$, but in a narrower band;
{in this case} it is smaller in magnitude
{and sign-alternating around $E_B$}
{due to the strong redistribution of photons between NMs near the resonant energy}.
At $E \ll E_B$, $P_L$ is negative and moderate (several percent) 
in both cases with and without the vacuum polarization effects.
Note that only the latter (lower) energy range is accessible with}
the current polarization measurements by \textit{IXPE} (2--8 keV).

%%%%%%%%%%%%%%%%%%%%%%%%%%%%%%%%%%%%%%%%%%%%%
%%%%%%%%%%%%%%%%%%%%%%%%%%%%%%%%%%%%%%%%%%%%%%%%%%%%5

%%%%%%%%%%%%%%%%%%%%%%%%%%%%%%%%%%%%%%%%%%%%%%%%%%%%%%%%%%%%%%%%%%%%%%%%%%
\begin{figure}
	\centering 
	\includegraphics[width=.9\columnwidth]{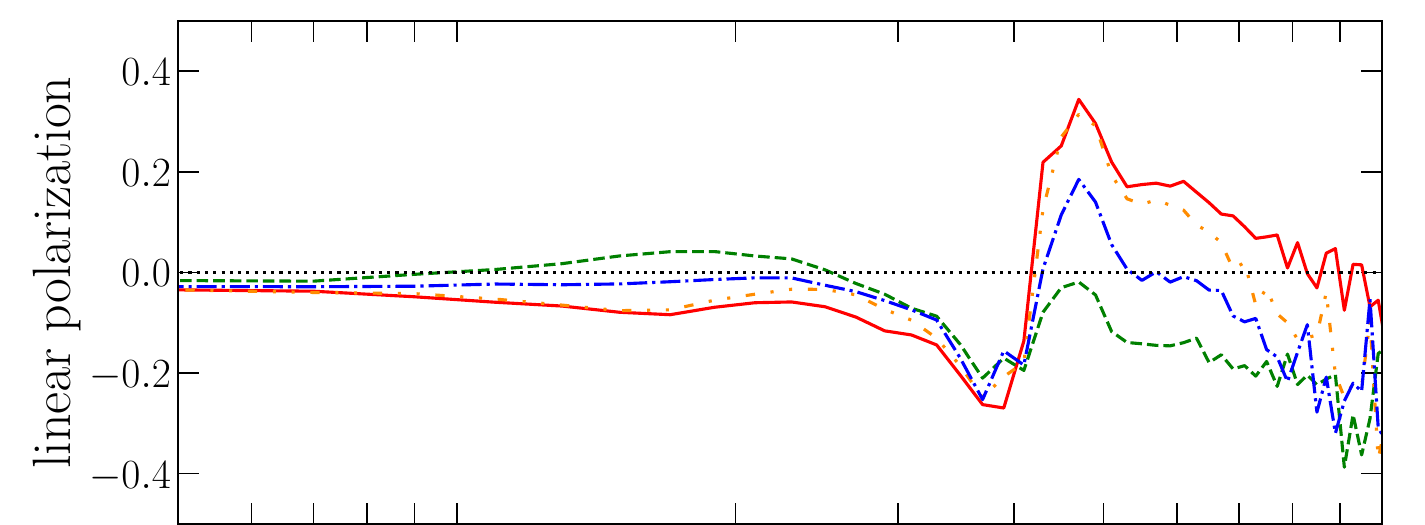} 
 
\includegraphics[width=.89\columnwidth]{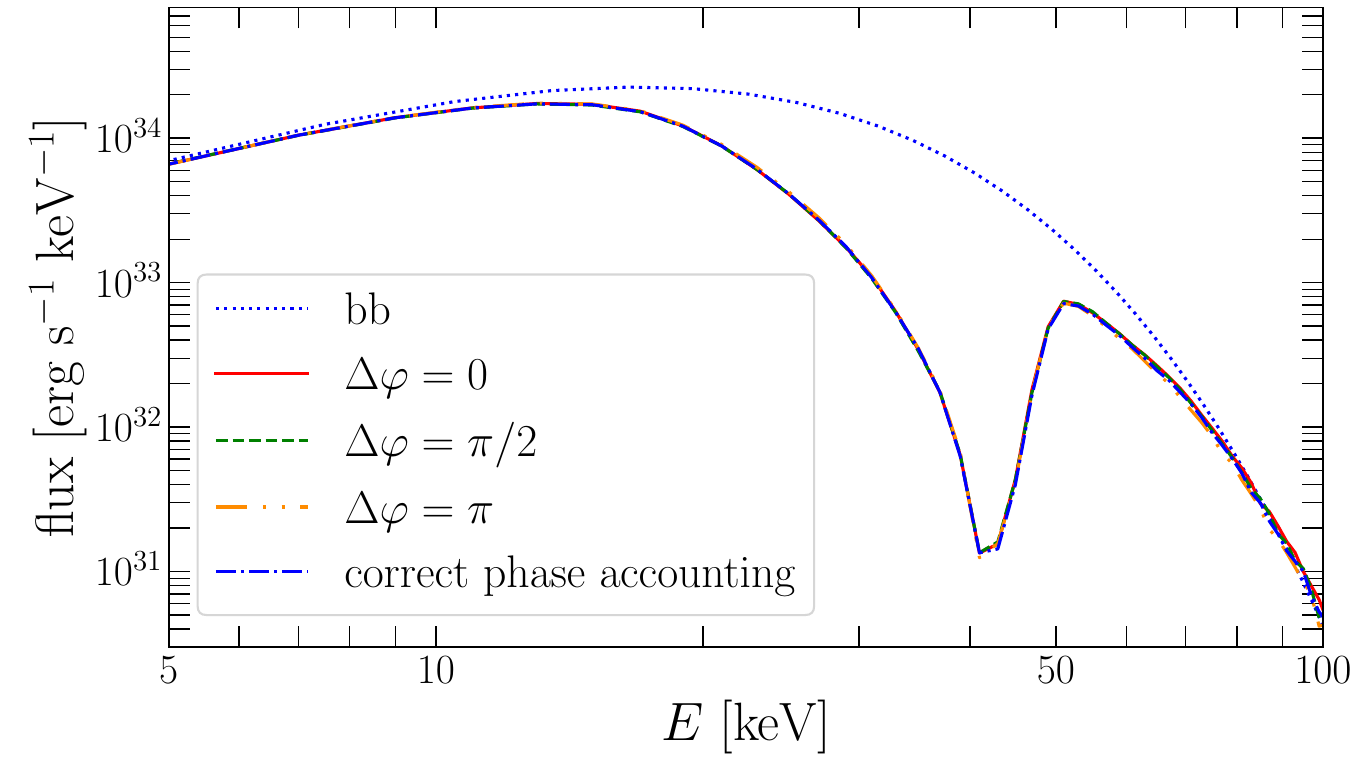} 
\caption{
Average linear polarization degree (\textit{upper panel})
and flux (\textit{lower panel}) of radiation emitted into the upper 
hemisphere 
in the ring geometry of the accretion channel 
with $R_\mathrm{c} = 500$~m and $d_\mathrm{w}=15$~m
as a function of photon energy for mass accretion rate
$\dot{M} = 2.5 \times 10^{15}$~g s$^{-1}$,
calculated using the cross sections with fixed azimuthal scattering angle $\Delta\varphi=0$ (solid line),
$\Delta\varphi=\pi/2$ (dashed line),
{$\Delta\varphi=\pi$ (double-dot--dashed line)}
and with the accurate $\Delta\varphi$ dependence according to equation
(\ref{sigma_dif}) (dot-dashed line). 
The dotted line in the lower panel 
represents the spectrum of the blackbody 
radiation with 
$\kB T_\mathrm{b}=5.9$~keV, which was 
used as the hot-spot boundary condition; the vacuum polarization effects are included.
\label{fig:azimuth}
}
\end{figure}
%%%%%%%%%%%%%%%%%%%%%%%%%%%%%%%%%%%%%%%%%%%%%%%%%%%%

{Second,}
we carried out
calculations for two extreme
cases, when  only one of the two  NMs is emitted  
from the hot-spot surface {(the dashed and dot-dashed lines in Fig.~\ref{Pic:P_l_sum_first})}. 
It is clearly seen that at $E\ll E_B$ (in the IXPE band) 
the degree of linear polarization strongly correlates with that of the seed radiation. 
This is expected,
because the accretion channel of a subcritical 
pulsar has a small optical thickness at such low photon energies. 
{However, in a band around $E \sim {40}$~keV, which corresponds to the cyclotron energy red-shifted due to the plasma motion,
$P_L$ is strongly reduced in magnitude. 
This means that the scattering in the accretion channel strongly mixes the two NMs 
in this energy band
and 
to lesser extent
redistributes the photons between them at other energies.}

{To summarize, the results shown in Fig.~\ref{Pic:P_l_sum_first} 
indicate}
that an accurate model of the hot spot surface radiation 
is needed to calculate the properties of the outgoing radiation at $E\ll E_B$.
On the other hand,
{these results demonstrate}
that the reprocessing of the NMs polarization
by the accretion channel 
{not only}
plays
{a key}
role in formation of the polarization of outgoing radiation
{near the resonance, but also may contribute appreciably to $P_L$ at lower and higher energies.}

We should note that most of the previous simulations of XRP radiation, 
including our Paper~I, neglected the azimuthal dependence of the scattering cross sections.
Here we find that the accurate $\varphi$-dependence of the cross sections according 
to equation (\ref{sigma_dif}) is not negligible. 
To illustrate it, in Fig.~\ref{fig:azimuth} the results of an accurate 
calculation of the polarization degree and the spectral flux
are compared with the results obtained using the scattering cross 
sections at fixed $\Delta\varphi=0$,
{$\pi/2$ or $\pi$. The spectrum in the lower panel is insensitive to the choices with fixed or variable $\Delta\varphi$. However, for polarization (the upper panel) we}
see a considerable difference between the results obtained 
with fixed $\Delta\varphi$ and  with using 
the cumulative distribution function {(\ref{cumul1}).

The upper panel of Fig.~\ref{Pic:Spectra_pol_dMd} presents
the degree of linear polarization as 
a function of photon energy for accretion rates
$\dot{M}=10^{15}$ g s$^{-1}$, 
$2.5\times10^{15}$ g s$^{-1}$ and 
$5\times10^{15}$ g s$^{-1}$ in the ring geometry, 
taking into account the vacuum polarization effects. 
The lower panel of Fig.~\ref{Pic:Spectra_pol_dMd} presents 
the respective radiation energy fluxes 
emitted into the upper hemisphere.
The unpolarized blackbody radiation was assumed as the boundary 
condition at the NS surface, with temperatures
determined by the energy balance, as described above: 
$\kB T_\mathrm{b}=4.5$~keV, 5.9~keV and {7.3}~keV, 
for the lowest, middle and highest accretion rate, respectively 
(the corresponding spectra are shown in the lower panel by dotted lines).
The underlying interpolation tables have been calculated at 
$E_B=50$~keV, $\rho=4\times10^{-4}$~g cm$^{-3}$ and $\kB T=5$~keV.

As in
Fig.~\ref{fig:azimuth},
the degree of linear polarization is substantial at photon energies 
around the cyclotron resonance, 
so we see the same changes of polarization signs 
near this energy.
{The change of sign of $P_L$ in the upper panels of Figs.~\ref{fig:azimuth} and \ref{Pic:Spectra_pol_dMd} occurs near the energy where the spectral flux in the lower panels has a minimum.}
The non-monotonic behaviour of the polarization degree 
around this resonance in the upper panel of Fig.~\ref{Pic:Spectra_pol_dMd} 
is an artifact related to our use of the tabulated cross sections to speed-up calculations: 
we have checked that this non-monotonic structure changes with a change 
of the grid of energies and angles for the tabulation. 
However,
the average trend of 
$P_L(E)$ is rather robust, so that its maxima and minima around the cyclotron resonance 
remain nearly the same within this wavy structure. 
Since the seed surface radiation is set unpolarized, the observed polarization 
of the outgoing radiation is produced by the radiative transfer in the accretion channel. 
The cyclotron feature in the spectral flux, seen in the lower panel, 
shifts to higher energies with increasing accretion rate, 
in accordance with observations of subcritical XRPs \citep{2022arXiv220414185M}. 
The growth of $P_L$ near the cyclotron resonance in the upper panel 
is also produced by scatterings in the accretion channel.

%%%%%%%%%%%%%%%%%%%%%%%%%%%%%%%%%%%%%%%%%%%%%%%%%%%%%%%%%%%%%%%%%%%%%%%%%%
\begin{figure}
	\centering 
	\includegraphics[width=.9\columnwidth]{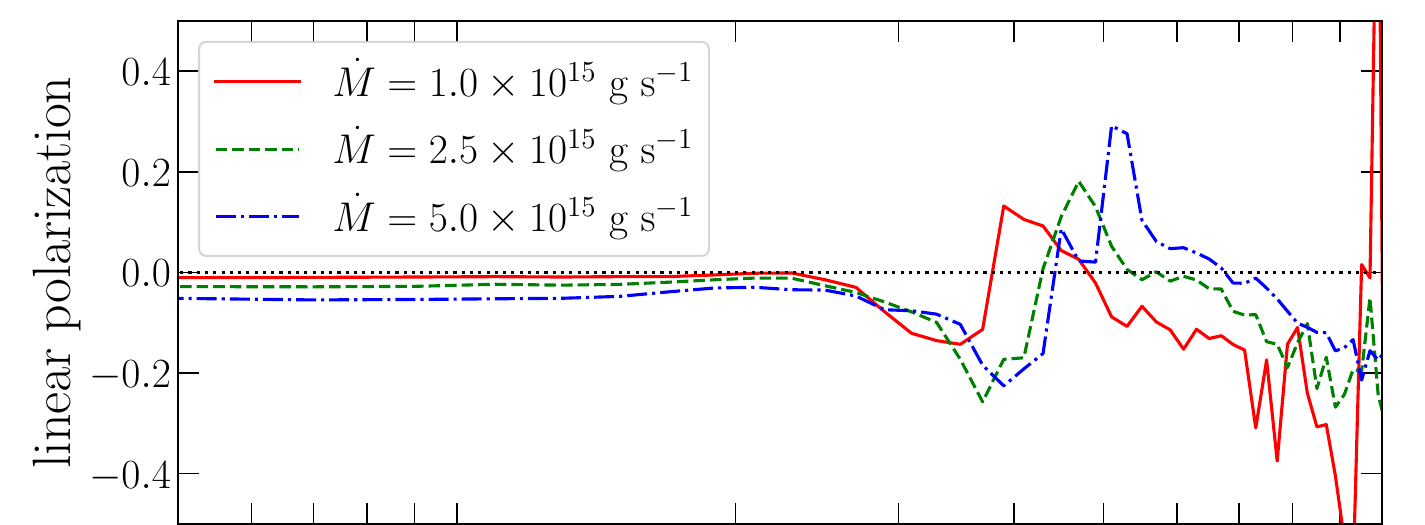} 
 
\includegraphics[width=.89\columnwidth]{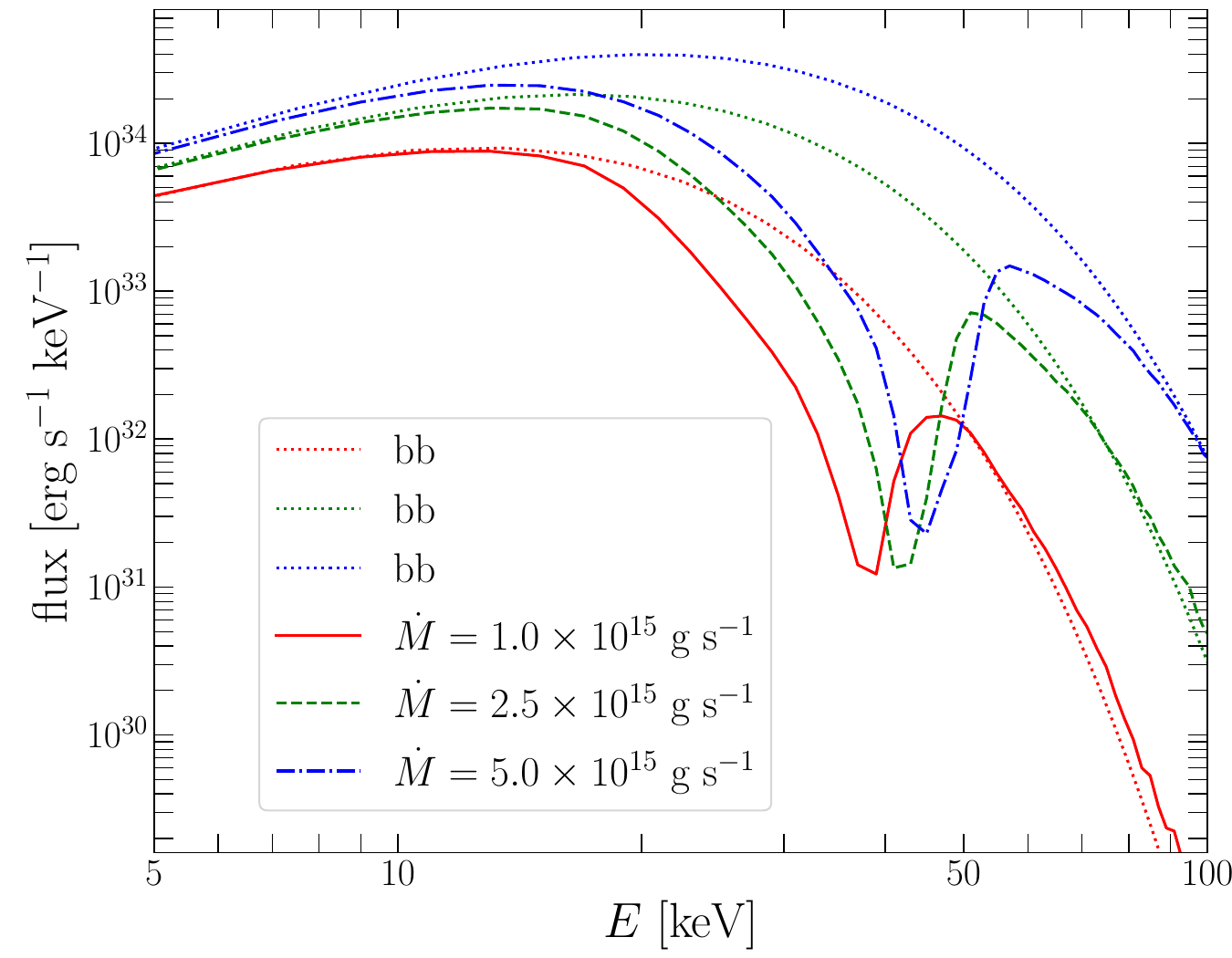} 
\caption{Same as in Fig.~\ref{fig:azimuth}
for mass accretion rates 
$\dot{M} = 10^{15}$~g s$^{-1}$ (thick solid line),
$\dot{M} = 2.5 \times 10^{15}$~g s$^{-1}$  (dashed line) and
$\dot{M} = 5.0 \times 10^{15}$~g s$^{-1}$ (dashed-dotted line).
The dotted lines in the lower panel 
represent the spectra of the blackbody 
radiation with 
{(from bottom to top)}
$\kB T_\mathrm{b}=4.5$~keV, 5.9~keV and {7.3}~keV, 
which were
used
as the hot-spot boundary conditions.
\label{Pic:Spectra_pol_dMd}
}
\end{figure}
%%%%%%%%%%%%%%%%%%%%%%%%%%%%%%%%%%%%%%%%%%%%%%%%%%%%

%%%%%%%%%%%%%%%%%%%%%%%%%%%%%%%%%%%%%%
\section{Summary and discussion}
\label{sec:summary}
%%%%%%%%%%%%%%%%%%%%%%%%%%%%%%%%%%%%%%

We have presented the results of self-consistent numerical simulations of the structure of 
an XRP accretion channel and radiation, emitted from a polar hot spot and reprocessed 
by Compton scattering in the strongly magnetized plasma, which falls vertically 
along the magnetic field lines, assuming that the electrons reside on the ground Landau level. 
The finite-{volume method of} solution of the system 
of equations of radiation hydrodynamics was coupled to the Monte Carlo simulation of two-mode 
radiative transfer in an accretion channel, which has the shape of either filled 
(circle geometry) or hollow (ring geometry) cylinder.
We have {jointly accounted for the effects of the}
strongly magnetized plasma and the QED vacuum
in the accretion channel. 

We have shown that in an accretion channel of a subcritical XRP
the polarization of the NMs is mostly determined  by the  
vacuum polarization  effects.
As a consequence, the polarization 
of both NMs is almost linear and their scattering cross sections have 
comparable cyclotron resonances.
It slightly increases the plasma deceleration in comparison to the pure
plasma approximation. 
The vacuum polarization changes the shape of the spectra formed in the accretion channel 
around the electron cyclotron energy.
The cyclotron absorption
{feature becomes quite pronounced in the outgoing spectra.
Without allowance for the vacuum polarization, this absorption is smeared away 
because of the mostly non-resonant character of the O-mode scattering
(see Fig.~\ref{Pic:Spectra_fhcr}; the absence of this 
smearing in \citealt{Fotiadis_26} 
is caused by averaging cross sections over polarizations, thus neglecting the difference between the X- and O-modes).
The vacuum polarization makes both NMs almost equally resonant (Fig.~\ref{f:cr_sec}), 
thus enhancing the photon depletion in the core of the line}.

However,
the quantitative results of our simulations
are only preliminary, 
because of simplifications inherent in our model. 
In particular, taking the vacuum polarization into account, we obtain a more pronounced 
cyclotron depression than is observed in real XRP spectra. 
A possible cause of this disagreement may be our neglect of population 
and radiative decay of higher Landau levels, which may cause the 
`spawning' of the cyclotron photons \citep{1999ApJ...517..334A}.
 
An important part of our model is the seed radiation that is emitted from the hot spot on the NS
surface and scattered by the plasma in the channel. 
The blackbody seed radiation does not contain the cyclotron absorption feature, 
which is expected in the radiation of underlying atmosphere 
(see, e.g., \citealt{2021MNRAS.503.5193M,2021A&A...651A..12S}).
To test the influence of such a feature,
we performed a set of simulations 
{in which the blackbody hot-spot spectrum was replaced}
by the spectrum of the subcritical pulsar 
{1A~0535+262} in different luminosity states. 
We found that this modification of the initial spectrum 
noticeably
reduces plasma deceleration by radiation,
{because}
the plasma 
interacts mostly with the resonant photons and remains transparent 
for photons below the cyclotron resonance.

To test an effect of a possible polarization of the seed radiation, we performed 
{the simulations assuming that} 
the hot-spot radiation consists {entirely} of one of the two NMs.
As noted in Paper~I, at $E\ll E_B$ 
the polarization of the outgoing radiation is almost completely determined by {that} of the seed radiation
because the channel is transparent to radiation at these energies. 
Near the resonance the opposite situation occurs and the channel is highly opaque. 
We demonstrate that 
the initial polarization state {is completely erased}
at $E \sim E_B$
because of the multiple scattering in the accretion channel. 
It means that at such photon energies it is necessary to treat 
accurately the radiation transfer in the falling plasma even at low accretion rates.

A number of subcritical XRPs demonstrate a positive correlation between the energy 
of the cyclotron absorption feature in the spectrum and luminosity. 
One of 
various hypotheses
to explain this property of the resonance absorption features proposes the Doppler shift 
of Comptonized photons in accretion channels.
In order to check it, we 
carried out
simulations with different accretion rates
and found that our radiation-hydrodynamical model agrees with
the proposed mechanism: 
the cyclotron feature energy increases with increasing accretion rate
although this effect is not very strong. 
In addition, we have found a positive correlation between the luminosity 
and the degree of linear polarization at the energies near and above the cyclotron resonance. 
Such energies are not observed by the IXPE telescope, so this result cannot yet be verified experimentally.

The angular dependence of the outgoing flux (Fig.~\ref{fig:spectra_angle}) 
controls the formation of the beaming {patterns} and light curves
at different photon energies. 
The polarization degree also shows such angular dependence, which results in
formation of the \emph{X-ray polarization light curves} of the XRPs.
These beaming patterns and light curves are studied 
in more detail in another paper \citep{Markozov_26beaming}.

%%%%%%%%%%%%%%%%%%%%%%%%%%%%%
\section*{Acknowledgements}
%%%%%%%%%%%%%%%%%%%%%%%%%%%%%%

The work of IDM, AYP and ADK was supported by the Russian Science Foundation Grant No.\,24-12-00320. 
AAM acknowledges support from the UKRI Stephen Hawking fellowship.
This research was supported by the International Space Science Institute (ISSI) in Bern, through International Team project 25-657 `Polarimetric Insights into Extreme Magnetism'.

%%%%%%%%%%%%%%%%%%%%%%%%%%%%%%%%%%%%%
\section*{Data availability}
%%%%%%%%%%%%%%%%%%%%%%%%%%%%%%%%%%%%%

The calculations presented in this paper were performed using a private code 
developed and owned by the corresponding author. All the data appearing 
in the figures are available upon request. 

%%%%%%%%%%%%%%%%%%%%%%%%%%%%%%%%%%%%%%%%%%%%%%%%%%%%%%%%%%%%%%%%%%%%%%%%%%%%%%
%% Bibliography %%
%%%%%%%%%%%%%%%%%%%%%%%%%%%%%%%%%%%%%%%%%%%%%%%%%%%%%%%%%%%%%%%%%%%%%%%%%%%%%%
%\bibliographystyle{mn2e}
\bibliographystyle{mnras}
\bibliography{allbib}

\appendix

%%%%%%%%%%%%%%%%%%%%%%%%%%%%%%%%%%%%%%%%%%%%%%%%%%%%%%%%%%%%%

\section{Scattering of the normal modes}
\label{app:scattering}

Equation~(\ref{MaM}) explicitly reads
\beq
   a_{j'j} = \sum_{l'=1,2}\sum_{l=1,2} M'_{j'l'}a^\ell_{l'l}M^\ast_{jl},
\label{ampli}
\eeq
where the matrix elements $M_{jl}$ are given by any of Eqs.~(\ref{M1}) or (\ref{M}), the amplitudes for linear polarizations $a^\ell_{l'l}$ are given by Eqs.~(\ref{Herold1})--(\ref{Herold3}) and the primed quantities are taken for the final state with $E' = \hat{E}_\mathrm{f}(p)$ according to Eq.~(\ref{E_f}).
As mentioned above, we fix the phases in Eqs.~(\ref{M1}), (\ref{M}) as $\phi_1=0$ and $\phi_2=\pi$. The result can be written in the form
\beq
   a_{j'j} =  C^{(0)}_{j'j} + \frac{C^{(1)}_{j'j}\,E\,\ee^{\ii(\varphi'-\varphi)}}{E-E_B + \ii\hbar\nu_\mathrm{eff,e}} ,
\label{a_via_C}
\eeq
where
\begin{align}&
   C^{(0)}_{11} = \left( 1+{K\phantom{'}\!\!}_1^2 \right)^{-1/2} \left(1+{K'_1}^2 \right)^{-1/2} \big[
     2K'_1 K_1 \sin\theta'\sin\theta
\nonumber\\&\,
   + A_+ \,\big(1 + K'_1 K_1 \cos\theta'\cos\theta
   + K'_1 \cos\theta' + K_1 \cos\theta \big)
   \big],
\label{C0_11}
\\&
   C^{(1)}_{11} = \left( 1+{K\phantom{'}\!\!}_1^2 \right)^{-1/2} \left(1+{K'_1}^2 \right)^{-1/2} \big[
    1 + K'_1 K_1 \cos\theta'\cos\theta 
\nonumber\\&\qquad
   - K'_1 \cos\theta' - K_1 \cos\theta
   \big] ,
\\&
   C^{(0)}_{22} = \left( 1+{K\phantom{'}\!\!}_1^2 \right)^{-1/2} \left(1+{K'_1}^2 \right)^{-1/2} 
    \mathrm{sign}(K'_1)\,\mathrm{sign}(K_1)
\nonumber\\&\qquad
 \times\big[
       2\sin\theta'\sin\theta +A_+ (K'_1 K_1 +
       \cos\theta'\cos\theta
\nonumber\\&\qquad
   - K_1 \cos\theta' - K'_1 \cos\theta )\,\big] ,
\label{C0_22}
\\&
   C^{(1)}_{22} = \left( 1+{K\phantom{'}\!\!}_1^2 \right)^{-1/2} \left(1+{K'_1}^2 \right)^{-1/2} 
    \mathrm{sign}(K'_1)\,\mathrm{sign}(K_1)
\nonumber\\&\qquad
\times\big[
    K'_1 K_1  + \cos\theta'\cos\theta
  + K_1\,  \cos\theta'
   + K'_1\,\cos\theta
   \big] ,
\\&
   C^{(0)}_{12} = \left( 1+{K\phantom{'}\!\!}_1^2 \right)^{-1/2} \left(1+{K'_1}^2 \right)^{-1/2}
     \mathrm{sign}\,K_1 \,\big[
       2K'_1 \sin\theta'\sin\theta 
\nonumber\\&\quad
 + A_+ (K'_1 \cos\theta'\cos\theta - K_1
   - K'_1 K_1 \cos\theta' + \cos\theta )\,\big] ,
\\&
   C^{(1)}_{12} = \left( 1+{K\phantom{'}\!\!}_1^2 \right)^{-1/2} \left(1+{K'_1}^2 \right)^{-1/2} \,\mathrm{sign}(K_1)
\nonumber\\&\qquad
   \times\big[
     K'_1 \cos\theta'\cos\theta - K_1 + K'_1 K_1\, \cos\theta' - \cos\theta
   \big] ,
\\&
   C^{(0)}_{21} = \left( 1+{K\phantom{'}\!\!}_1^2 \right)^{-1/2} \left(1+{K'_1}^2 \right)^{-1/2}
    \mathrm{sign}(K'_1)
\nonumber\\&\qquad\times\big[
       2 K_1\sin\theta'\sin\theta + A_+ (K_1 \cos\theta'\cos\theta - K'_1
\nonumber\\&\qquad
   - K'_1 K_1 \cos\theta + \cos\theta' )\,\big] ,
\\&
   C^{(1)}_{21} = \left( 1+{K\phantom{'}\!\!}_1^2 \right)^{-1/2} \left(1+{K'_1}^2 \right)^{-1/2} \mathrm{sign}(K'_1)
\nonumber\\&\qquad
    \times\big[\, K_1 \cos\theta'\cos\theta
    - K'_1 + K'_1 K_1 \, \cos\theta - \cos\theta'
   \big]
\label{C1_21}
\end{align}
and $A_+$ is defined by Eq.~(\ref{A+-}).
Here we have separated resonant contributions with coefficients $C^{(1)}_{j'j}$ from non-resonant terms 
with coefficients $C^{(0)}_{j'j}$, which is convenient for numerical 
implementation -- for example, to optimize the integration in Eq.~(\ref{sigma_mean}).

Making use of Eqs.~(\ref{dsigma_a_fi}) and (\ref{a_via_C}), 
and taking into account that $C^{(0)}_{j'j}$ and $C^{(1)}_{j'j}$ in Eqs.~(\ref{C0_11})--(\ref{C1_21}) 
are real, we can write the differential cross sections in the right-hand side of 
Eq.~(\ref{dsigma_transform}) as
\beq
   \frac{\mathrm{d}\sigma_{j'j}^0}{\mathrm{d}\bm{\Omega}_\mathrm{f}^0}
   =
   \frac{3\sigma_\mathrm{T}}{32\pi}
   \Bigg[
   |C_{j'j}^{(0)}|^2 +
   \frac{
    |C_{j'j}^{(1)}|^2
   + 2C_{j'j}^{(0)}C_{j'j}^{(1)}(1-u_\mathrm{e})
   }{(1-u_\mathrm{e})^2+(\hbar\nu_\mathrm{e,eff}/E)^2} 
   \Bigg],
\label{sigma_explicit}
\eeq
where $u_\mathrm{e}=(E_B/E)^2$ as defined by Eq.~(\ref{u_v_def}). We see, in particular, that the factors sign$\,K_1$ and sign$\,K'_1$ that are present in Eqs.~(\ref{C0_22})--(\ref{C1_21}) are squared in Eq.~(\ref{sigma_explicit}). Therefore the change of these signs at $\theta_B=\pi/2$ or $\theta'_B=\pi/2$ may only lead to a change of sign of the amplitudes $a_{j'j}$, but $\mathrm{d}\sigma_{j'j}^0/\mathrm{d}\bm{\Omega}_\mathrm{f}^0$ remains continuous when crossing these angles.

We have checked that equations (\ref{ampli})--(\ref{sigma_explicit}) can also be derived from equations (1), (10) and (11) of \citet{Ventura79} using an appropriate coordinate transformation.

%%%%%%%%%%%%%%%%%%%%%%%%%%%%%%%%%%%%%%%%%%%%%%%%%%%%%%%%%%%%%

\section{Code verification}
\label{app:code_verif}

%%%%%%%%%%%%%%%%%%%%%%%%%%
\begin{figure}
	\centering 
	\includegraphics[width=\columnwidth]{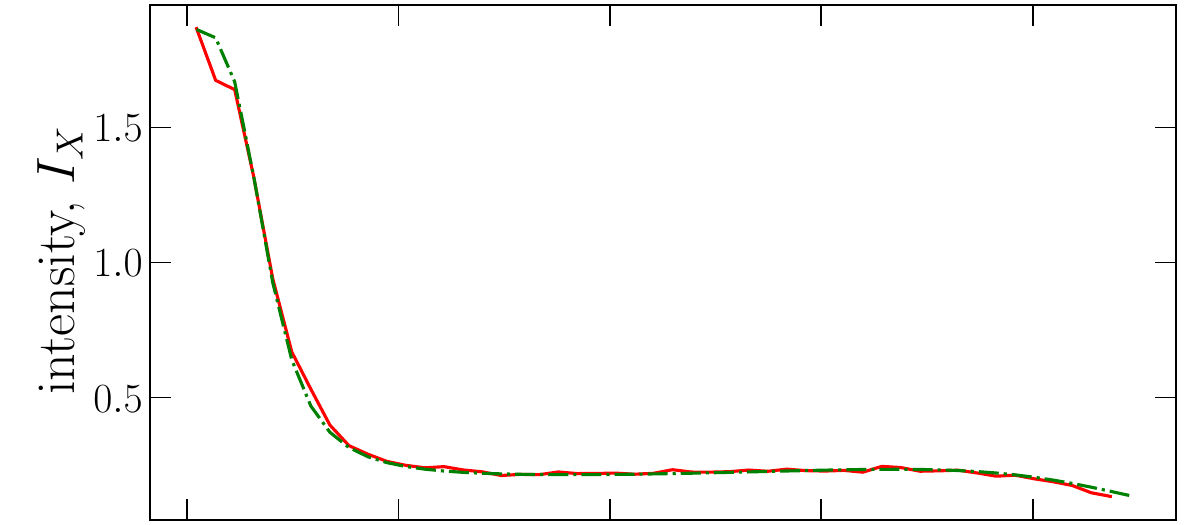} 
  \includegraphics[width=\columnwidth]{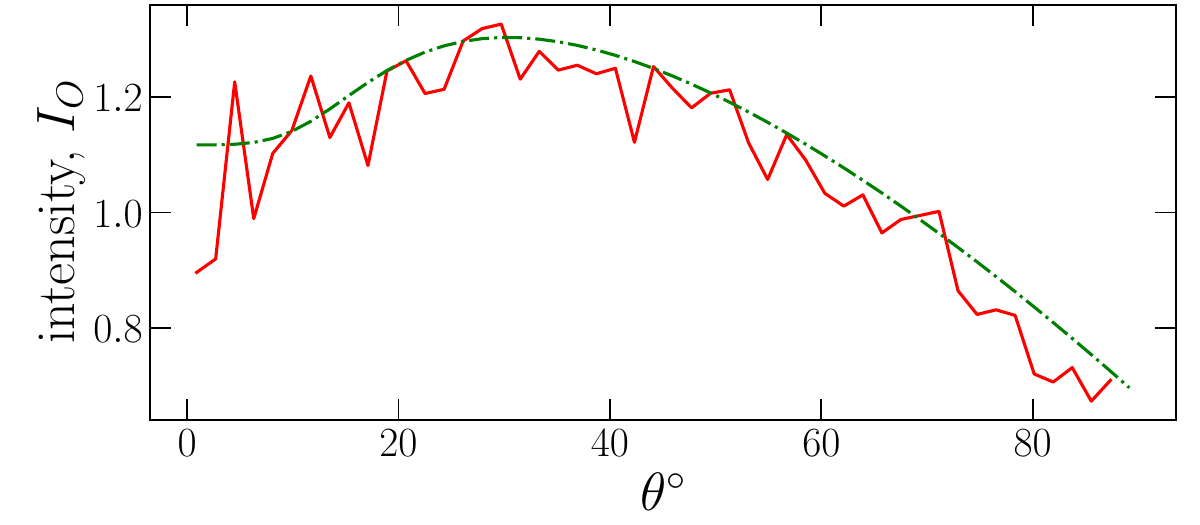}
	\caption{Integrated intensity in X-mode  
    (the upper panel)
 and O-mode (the lower panel)
 in relative units (normalized to the average)
  as a function of
    the angle $\vartheta$ to the outer normal to the plane parallel semi-infinite magnetized NS
atmosphere. At both panels the 
    dashed-dotted line corresponds to the 
    analytical formula and the 
    %% AK
    %red 
    solid one corresponds to our program results. 
    Computations were performed for the photon energy $E=70$~keV, the cyclotron energy 
    $E_B =51.1$~keV and the vacuum parameter $W=29.3$.
}
\label{Pic:Intensity_ver}
\end{figure}

In this Appendix we describe some of the tests of the code used for its verification.

\subsection{Magnetized atmosphere with coherent scattering}
\label{app:RT_coherent}

We have tested our Monte Carlo radiation transfer code on several
 radiative transfer problems with known analytical solutions. The most complicated of these tests
involves almost all parts of our Monte Carlo code:
simulation of the radiation emitted by a semi-infinite uniform 
atmosphere with a strong magnetic field, assuming a coherent resonant scattering. 
This problem has an analytical solution in the diffusion approximation for two NMs \citep{1982Ap&SS..86..249K}.
To test our Monte Carlo code, we compute intensities of the two NMs emitted by the 
semi-infinite magnetized atmosphere in the coherent scattering approximation 
and compare the results with equation~(19) of \citet{1982Ap&SS..86..249K}.

Figure \ref{Pic:Intensity_ver} shows the angle dependence of the X- and O-mode intensities, 
calculated numerically and analytically for a monochromatic 
radiation with photon energy $E=70$~keV, which experiences coherent scattering
in a semi-infinite atmosphere
with the cyclotron energy $E_B =51.1$~keV and vacuum parameter $W=29.3$. 
These values correspond to the case (Bb) (very large vacuum polarization, 
near the cyclotron resonance) of \citet{1982Ap&SS..86..249K}. 
The number of seed photons was $N_\mathrm{ph}=10^7$. 
The numerical noise seen in the lower panel of Fig.~\ref{Pic:Intensity_ver} is higher 
than in the upper one, because most of the photons come out of the atmosphere in the X-mode. 
We see that our numerical solution satisfactorily reproduces the analytical one within the numerical noise.
Thus we can conclude that our Monte Carlo code is free of a systematic bias (within the model assumptions).

\subsection{Radiation hydrodynamics: forward-backward scattering}
\label{app:RH_fb}

The second part of our tests deals with the coupling of radiative transfer with hydrodynamics. 
For this purpose, we consider a tube filled with a plasma, interacting with radiation, 
which is injected from one side of the tube (the bottom) and emitted from the other side 
(the top of the tube). To obtain an analytical solution, 
we consider one-dimensional propagation of photons along the tube
and neglect the gravity.

First we consider the case where the radiation interacts with the matter only via coherent 
forward-backward scattering. For simplicity, we assume that the probabilities of photon scattering 
in the forward or backward directions are equal in the reference frame of the tube.
A constant radiative flux from the bottom of the tube is treated 
as a parameter of the simulation.

This model setup is described by equations
\begin{equation}
\frac{\mathrm{d}\rho v}{\mathrm{d}x}=0 ,
\quad
\frac{\mathrm{d}}{\mathrm{d}x}\left(\rho v^2\right)+\frac{\mathrm{d}P}{\mathrm{d}x}=\rho a_\mathrm{r},
\quad
\frac{\mathrm{d}}{\mathrm{d}x}\left(\frac{Pv\gamma}{\gamma-1}+\frac{\rho v^3}{2}\right)=0,
\label{eq:app_RH_fb}
\end{equation}
where 
$\gamma$ is the adiabatic index of the plasma,
$a_\mathrm{r}={\sigma_\mathrm{T}F_0/\mion c}$ is the acceleration by the radiation force, 
$\mion$ is a proton mass and $F_0$ is a flux from the bottom boundary,
which can be expressed through the tube cross section $S_\mathrm{b}$ 
and bottom luminosity $L_\mathrm{b}$ as $F_0={L_\mathrm{b}/S_\mathrm{b}}$. 
The zero right-hand side of equation (\ref{eq:app_RH_fb}) is due to the absence 
of energy exchange at the coherent scattering.

%%%%%%%%%%%%%%%%%%%%%%%%%%%%%%%%%%%%%%%%%%%%%%%%%%%%%%%%%%%
\begin{figure}
	\centering 
	\includegraphics[width=\columnwidth]{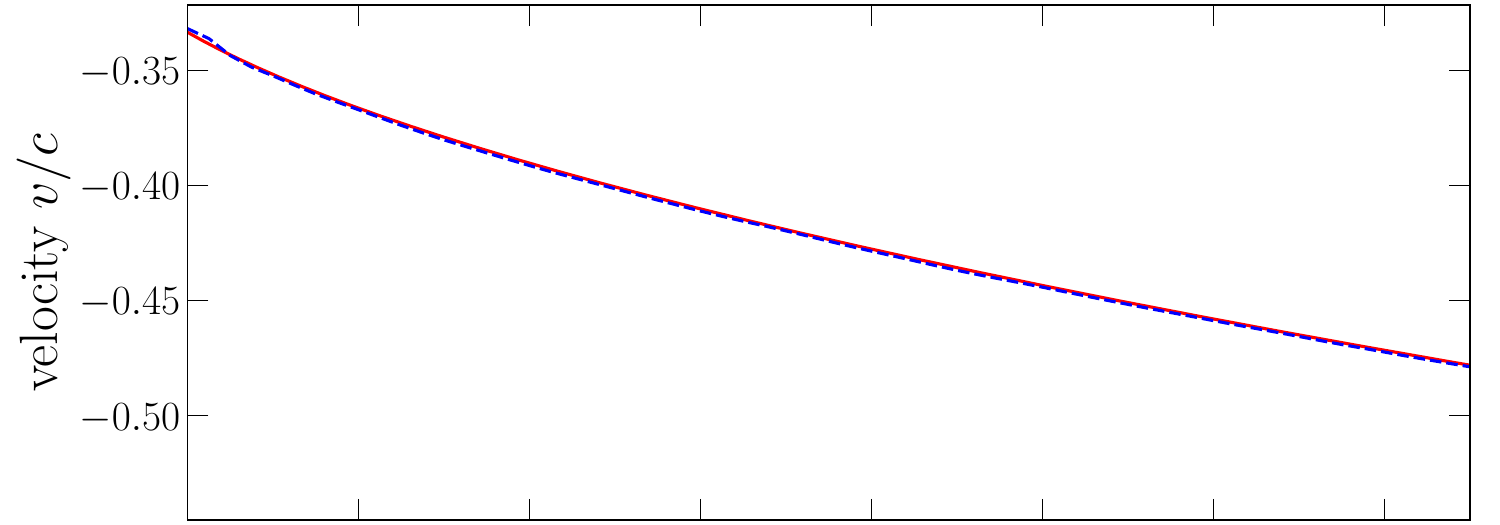} 	
\includegraphics[width=\columnwidth]{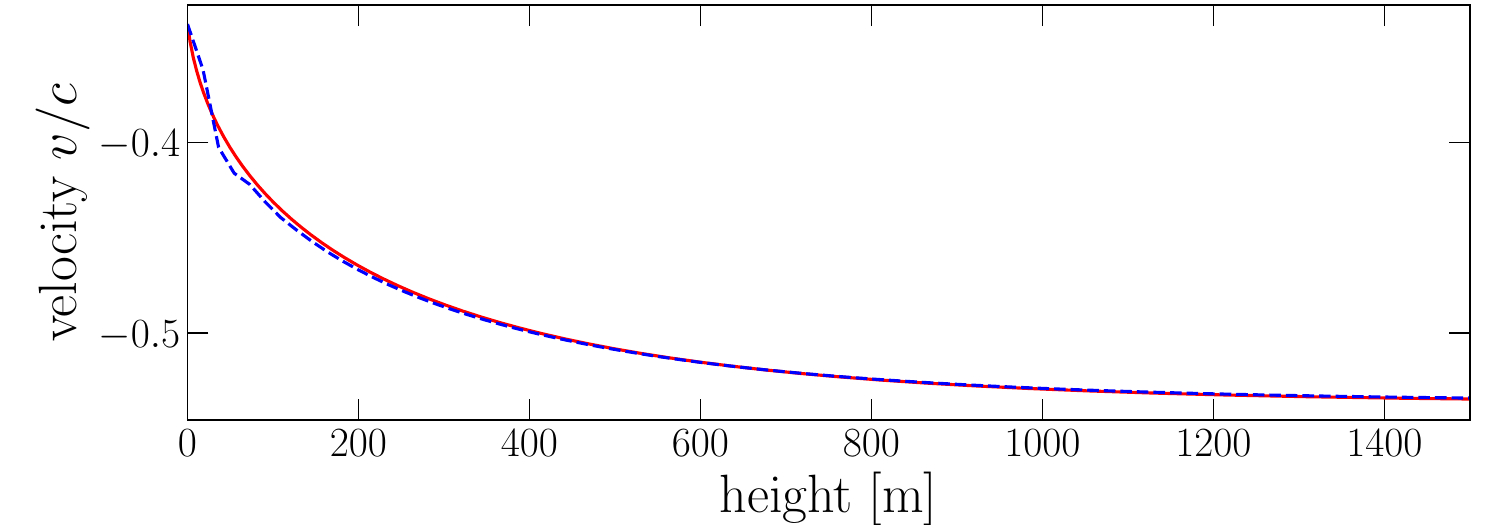} 	
	\caption{Velocity profiles in the one-dimensional tube for the forward-backward 
    radiation scattering (upper panel, see Section~\ref{app:RH_fb}{; scattering cross section $\sigma=\sigma_\mathrm{T}=6.65\times10^{-25}$~cm$^2$, 
   radiation acceleration at the lower boundary $a_\mathrm{r}=1.43\times10^{14}$~cm s$^{-2}$}) and 
    pure absorption (lower panel, see Section~\ref{app:RH_abs}{;
absorption cross section $\sigma=\sigma_\mathrm{T}$,  $a_\mathrm{r}=5.1\times10^{14}$~cm s$^{-2}$}). 
  The matter is treated as an ideal gas with an adiabatic index $\gamma=5/3$. 
  In both panels, conditions at the upper boundary of the simulation domain are 
  $v_0=0.54c$, $P_0=4\times10^{14}$~dyn cm$^{-2}$, $\rho_0=7\times10^{-5}$~g cm$^{-3}$. 
  At the lower boundary, we set zero gradient for hydrodynamical variables and 
  a constant flux of the radiation propagating along the normal to the surface.
}
\label{Pic:Velocity_ver}
\end{figure}
%%%%%%%%%%%%%%%%%%%%%%%%%%%%%%%%%%%%%%%%%%%%%%%%%%%%%%%

This system of equations has an analytical solution:
\begin{align}&
   \rho v=\rho_0 v_0, 
\qquad % \\&
   P(v)=\frac{c_1}{v}-\rho_0 v_0 v\,\frac{\gamma-1}{2\gamma}, 
\\&
   \frac{(v^2-v_0^2)(\gamma+1)}{4\gamma}-\frac{c_1}{\rho_0 v_0}
   \ln\left(\frac{v}{v_0}\right)=a_\mathrm{r}(x-x_0).
\label{eq:app_RH_fb_an_sol}
\end{align}
Here $\rho_0$ and $v_0$ are the inflow mass density and velocity 
at the top of the tube, placed at the coordinate $x=x_0$,
\begin{equation}\label{eq:app_c_1}
  c_1=P_0v_0+\rho_0 v_0^3\,\frac{\gamma-1}{2\gamma}
\end{equation}
is the integration constant and $P_0$ is the top-side pressure.
We use the following method to verify our numerical model. For 
certain values of $F_0$, $P_0$, $v_0$ and $\rho_0$ 
we compute hydrodynamical profile of the tube with a fixed length $H$
 according to \req{eq:app_RH_fb_an_sol}. 
 After that, we perform a simulation with the same boundary conditions 
 and constant values of $P=P_0$, $v=v_0$, $\rho=\rho_0$ over the tube 
 as the initial condition. 
{The system relaxes} to a steady-state solution, which is {then compared with the analytical solution}.
 
The upper panel of Fig. \ref{Pic:Velocity_ver} demonstrates the results 
of comparison of the numerical solution with the semi-analytical one. 
We see that these solutions almost coincide, 
which means that the test has been passed successfully.

\subsection{Radiation hydrodynamics: pure absorption}
\label{app:RH_abs}

Now let us consider the same model tube as in the previous section, but
with pure absorption of radiation by the plasma instead of scattering, with the absorption cross section $\sigma$ set equal to the Thomson cross section
$\sigma_\mathrm{T}$. 
In addition, 
we assume that photons {are emitted} from the lower boundary
only in the direction strictly normal to the boundary surface. 
{This test is useful for examining the} explicit Euler method {implemented in our code, because absorption produces a steeper solution profile} than scattering.
The system of radiation hydrodynamical equations in this case has the form:
\begin{align}&
\frac{\mathrm{d}\rho v}{\mathrm{d}x}=0 ,
\label{continuity0}
\\&
\frac{\mathrm{d}}{\mathrm{d}\tau}\left(\rho v^2\right)+\frac{\mathrm{d}P}{\mathrm{d}\tau}=-\frac{\sigma F_0}{c\mion}\mathrm{e}^{-\tau(x)},
\\&
\frac{\mathrm{d}}{\mathrm{d}\tau}\left(\frac{Pv\gamma}{\gamma-1}+\frac{\rho v^3}{2}\right)=\frac{\sigma F_0}{c\mion}v\mathrm{e}^{-\tau(x)},  	
\label{eq:app_RH_abs}
\end{align}
where $\tau=(\rho/\mion)\sigma x$ is the optical depth. Unlike the pure scattering case, 
in the present case the radiative flux is not conserved 
in the computational volume, but exponentially decreases. 

We obtained the profiles of hydrodynamical quantities as functions of $\tau$ by solving the system 
of equations (\ref{continuity0})--(\ref{eq:app_RH_abs}) and restore the $x(\tau)$ profile as 
\begin{equation}\label{eq:app_x_lambda}
  x(\tau)=\int\limits_0^\tau\frac{\mathrm{d}\tau'}{\rho(\tau')}
\end{equation} 

To compare the results of our numerical model with a steady-state solution we proceed as follows. 
Firstly, the flux $F_0$ and hydrodynamical quantities 
$P_0$, $v_0$ and $\rho_0$ at the bottom boundary 
are 
specified. After that, 
the system (\ref{continuity0})--(\ref{eq:app_RH_abs}) is being solved until 
the optical depth $\tau$ reaches a certain value $\tau_s$, which is the additional 
parameter of the calculation. After that the total length $H$ of the computational 
domain that corresponds to the optical depth $\tau_s$ is obtained from equation 
(\ref{eq:app_x_lambda}). This length $H$ as well as the flux from the bottom boundary 
$F_0$ and values of pressure $p_1$, density $\rho_1$ and velocity $v_1$ at the optical 
depth $\tau_s$ 
are {supplied to} our radiation-hydrodynamics code as boundary conditions.
The simulation with these conditions starts with $p_1$, $\rho_1$ and $v_1$ constant 
over all tube. Then it evolves to the steady-state distribution which can be compared 
with the solution of the system (\ref{continuity0})--(\ref{eq:app_RH_abs}).  

The results of the comparison of our code calculations for this model with the semi-analytical solution are presented in the lower panel of Fig.~\ref{Pic:Velocity_ver}.  
Numerical and semi-analytical velocity profiles are in good agreement, 
which means that our code works accurately for this model.

% do not change these lines
\bsp % typesetting comment
\label{lastpage}
\end{document}